\documentclass[preprint,12pt]{elsarticle}

\usepackage{amsmath,amssymb,amsfonts}
\usepackage{amsthm}
\usepackage{algorithmic}
\usepackage{subcaption}
\usepackage{graphicx}
\usepackage{textcomp}
\usepackage{mathtools}
\usepackage[nice]{nicefrac}
\usepackage{blkarray, bigstrut}

\usepackage{enumitem}
\usepackage{multirow}
\usepackage{xcolor}
\newcommand{\HP}{\textit{HingePlace} }
\DeclareMathOperator*{\argmin}{arg\,min}
\DeclareMathOperator*{\argmax}{arg\,max}
\newcommand\myeq{\mkern1.2mu{=}\mkern1.2mu}
\newcommand\myplus{\mkern1.2mu{+}\mkern1.2mu}
\newcommand\myminus{\mkern1.2mu{-}\mkern1.2mu}

\newcommand\mylesseq{\mkern1.2mu{\leq}\mkern1.2mu}

\newcommand{\myVec}[1]{\mathbf{#1}}

\newcommand{\R}{\mathbb{R}}

\newcommand{\N}{\mathbb{N}}

\newcommand{\1}{\mathbf{1}}

\newtheorem{theorem}{Theorem}
\newtheorem{lemma}{Lemma}   

\newtheorem{claim}{Claim}

\newtheorem{property}{Property}
\newtheorem{conclusion}{Conclusion}

\newenvironment{subproof}[1][\proofname]{%
  \begin{proof}[#1]%
}{%
  \end{proof}%
}

\journal{Brain Stimulation}
\begin{document}

\begin{frontmatter}



\title{\textit{HingePlace}: Harnessing the neural thresholding behavior to optimize Transcranial Electrical Stimulation}

\author[label1]{Chaitanya Goswami} 
\author[label1]{Pulkit Grover}
\affiliation[label1]{organization={Electrical and Computer Engineering, Carnegie Mellon University},
            addressline={5000 Forbes Avenue}, 
            city={Pittsburgh},
            postcode={15213}, 
            state={PA},
            country={USA}}

\begin{abstract}
\textbf{Background}: Transcranial Electrical Stimulation (tES) is a neuromodulation technique that utilizes electrodes on the scalp to stimulate target brain regions. tES has shown promise in treating many neurological conditions, such as stroke rehabilitation and chronic pain. The efficacy of tES-based therapies can be significantly enhanced by optimizing multi-electrode montages to achieve more focal neural responses.

\noindent\textbf{Objective}: 
In this work, we identify a common restriction among existing electrode placement algorithms that design multi-electrode montages for tES: they do not harness the thresholding behavior of neural response. To overcome this restriction, we propose a new algorithm, \textit{HingePlace}, which fully harnesses the neural thresholding behavior to achieve more focal stimulation.

\noindent\textbf{Method}: We extend a previous unification result by Fernández-Corazza et al. to unify all major electrode placement algorithms. Using this unification result, we demonstrate mathematically that existing electrode placement algorithms do not harness the thresholding behavior of neural response. To explicitly account for the neural thresholding behavior, we employ a symmetrized hinge loss in the \HP optimization framework. We extensively compare the performance of \HP with existing electrode placement algorithms in two simulation platforms.

\noindent\textbf{Result}: 
In both simulation platforms, montages designed by \textit{HingePlace} consistently achieved significantly more focal neural responses—improving focality by up to 60\% compared to traditional algorithms. Additionally, we show that all existing electrode placement algorithms can be interpreted as special cases of \textit{HingePlace}.

\noindent\textbf{Conclusion}: We unify the major electrode placement algorithms, and show that these algorithms only partially harness the neural thresholding behavior. By explicitly accounting for neural thresholding, our proposed algorithm--\textit{HingePlace}--induces more focal neural stimulation, offering a promising advancement for improving the efficacy of tES-based treatments.
\end{abstract}

\begin{keyword}
convex optimization, transcranial electrical stimulation, non-invasive stimulation, electrode design, hinge loss, transcranial direct current stimulation, transcranial pulse stimulation


\end{keyword}

\end{frontmatter}




\section{Introduction}\label{sec:introduction}
Transcranial Electrical Stimulation (tES) refers to stimulating or modulating the neural activity in the brain non-invasively by injecting current through electrodes placed at the scalp. In recent years, tES has shown promise in treating clinical depression~\cite{kalu2012transcranial}, chronic pain~\cite{mori2010effects}, epilepsy treatment~\cite{yook2011suppression}, and many more conditions~\cite{brunoni2013transcranial}. Furthermore, tES has also emerged as a valuable tool in neuroscience research~\cite{parkin2015non}. 
{Two important limitations of traditional tES-based therapies are} significant off-target neural stimulation and considerable inter-subject variability. In tES, electrical current travels through the scalp, skull, and cerebrospinal fluid (CSF) to reach the brain, which causes the resultant electric fields to be diffused. These diffused electric fields still cause considerable undesired off-target stimulation. Improving the focality of tES can improve the clinical efficacy of tES-based therapies~\cite{kuo2013comparing,fischer2017multifocal} and help develop more precise brain-machine interfaces~\cite{dmochowski2011optimized}. Inter-subject variability in tES-based therapies is also a significant issue, leading to debates over tES's clinical efficacy~\cite{horvath2014transcranial,horvath2015evidence,antal2015conceptual}. Some experimental evidence suggests that the differences in the induced transcranial electric fields across subjects can explain the inter-subject variability in tES studies~\cite{laakso2019can}. Hence, the inter-subject variability in tES studies can be potentially reduced by personalizing the stimulation to each subject~\cite{fernandez2020unification}. 
 
{In order to address the above two limitations of traditional tES-based therapies, considerable research has been done on developing methods that design multi-electrode montages in the past decade~\cite{im2008determination,park2011novel,dmochowski2011optimized,sadleir2012target,dutta2013using,ruffini2014optimization,guler2016optimization,salman2016concurrency,cancelli2016simple,wagner2016optimization,fernandez2016transcranial,dmochowski2017optimal,guler2018computationally,saturnino2019accessibility,khan2022individually,prieto2022l1,wang2023multi}}. These methods use optimization techniques to design multi-electrode montages that minimize the off-target electric field, {thereby reducing off-target stimulation}. Furthermore, these methods use subject-specific head models built using magnetic resonance imaging (MRI) techniques to personalize tES-based therapies. We refer to these methods as \emph{electrode placement algorithms} and the task of designing multi-electrode montages as \emph{electrode placement}.
 
\begin{figure*}[!ht]
    \centering
    \includegraphics[width=\textwidth]{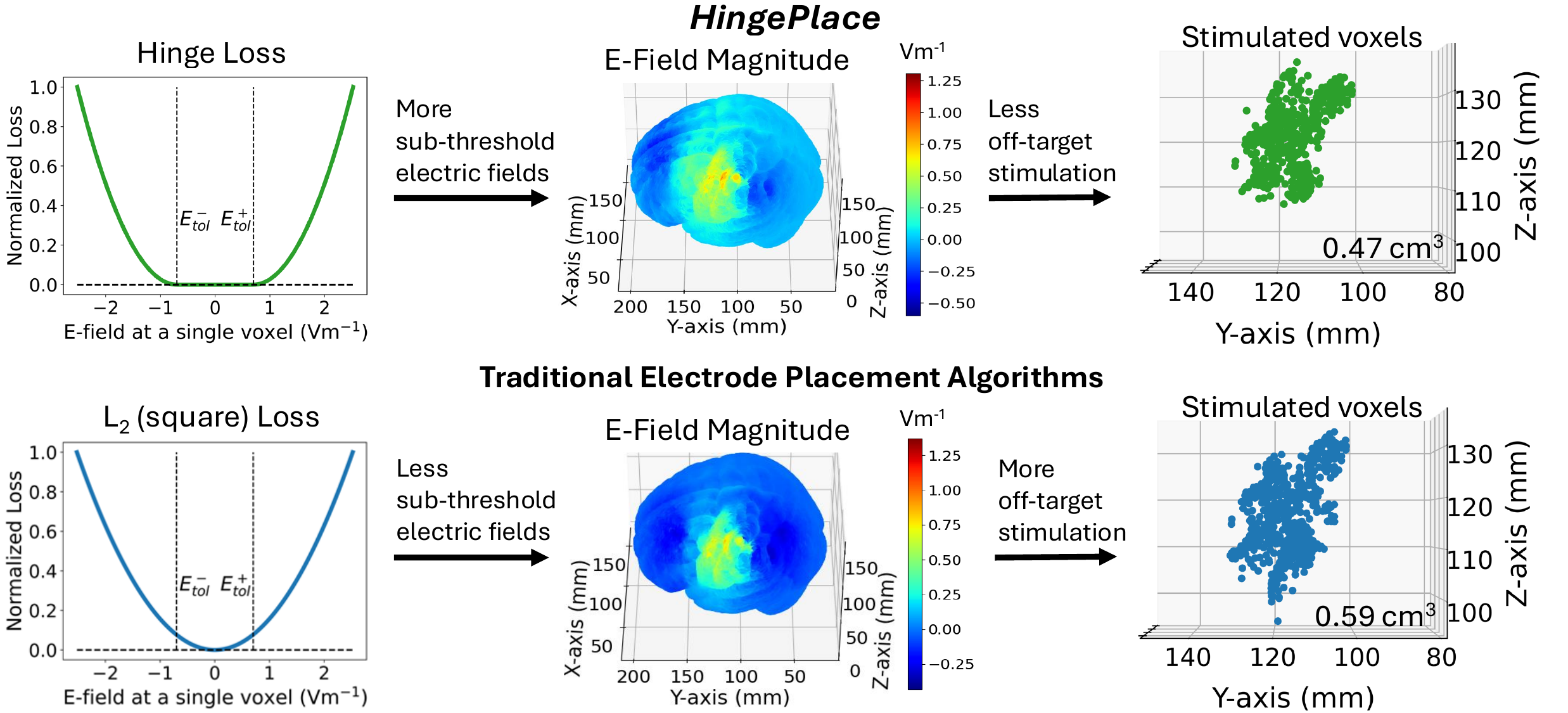}
    \caption{A representation of the \HP algorithm utilizing the hinge loss {(top-left)} to allow sub-threshold electric field amplitudes in off-target regions {(shown in the top-middle)} to reduce off-target stimulation volume {(represented by the smaller stimulated region depicted in top-right)}. In contrast, traditional electrode placement algorithms utilize the $l_2$ loss {(bottom-left)} that minimizes the electric field amplitude in the off-target regions {(shown in the bottom-middle)}, which counter-intuitively leads to a larger off-target stimulation volume than the \HP algorithm {(represented by the larger stimulated region depicted in the bottom-right)}.}
    \label{fig:graph-abstract}
\end{figure*}
{In this work, we observe that all major traditional electrode placement algorithms exhibit a common restriction, namely, they minimize off-target stimulation by \emph{forcing the off-target electric field magnitude to be as close to zero as possible}. Our key observation is as follows: neural response exhibits a thresholding behavior under tES. It is well known that the stimulating field needs to be above a certain threshold to \emph{directly} stimulate neurons~\cite{kandel2000principles}. Even in {``low-amplitude stimulation''} therapies with \emph{no direct neural activation}, such as transcranial direct current stimulation (tDCS), consistent neural effects (e.g., increases in plasticity) require the stimulating field to have a sufficiently high amplitude~\cite{hsu2023robust}. In either case, there are non-zero amplitudes of the stimulating field that do not have significant off-target effects. Hence, we do not require the off-target electric field amplitudes to be \emph{zero}, but only to be below a \emph{non-zero threshold} to create focal neural responses. Consequently, forcing the off-target electric field magnitude to be close to zero is overly restrictive, as we only require the off-target magnitude to be below a non-zero threshold.}

{We theoretically show this restriction in traditional electrode placement algorithms by extending the unification result by Fern\'{a}ndez-Corazza et al.~\cite{fernandez2020unification} (in Sec.~\ref{sec:equiv-LCMV-EDM}) to unify the four major categories of electrode placement algorithms (see Sec.~\ref{sec:prob-description:background}). This unification result enables us to theoretically establish that all major electrode placement algorithms obtain electrode montages by minimizing the off-target electric field magnitude, thereby revealing their restriction.}

To {overcome this restriction of traditional electrode placement algorithms}, we propose the \HP algorithm in Sec.~\ref{sec:algo:HingePlace}. \HP utilizes a symmetrized hinge loss {(see Fig.~\ref{fig:graph-abstract})} to penalize only the off-target electric field amplitudes above user-specified tolerance levels. The hinge loss enables the \HP algorithm to design electrode montages that allow \emph{non-zero} {but below-threshold} off-target electric field amplitude. Consequently, the \HP algorithm better harnesses the {thresholding behavior of neural response to improve the focality of tES} compared to traditional electrode placement algorithms. Fig.~\ref{fig:graph-abstract} visually depicts the difference in the design philosophies of the \HP and traditional electrode placement algorithms.

We compare the performance of the \HP algorithm with traditional electrode placement algorithms (represented by the LCMV-E algorithm) in two simulation platforms: a standard MRI head model built using the ROAST platform~\cite{huang2019realistic} and a sea of neurons model that combines realistic bio-physical neurons with spherical head models used in~\cite{Caldas-Martinez2024}. Across both platforms, we find that \HP montages produce more focal neural responses than the traditional electrode placement montages, with as much as $\sim30\%$ (in the MRI-head model) and $\sim60\%$ (in the sea of neurons model) reduction in the off-target stimulation region in some cases. Finally, we conclude in Sec.~\ref{sec:discussion}, where we briefly discuss the limitations of our work and future directions. 

\section{Background}\label{sec:prob-description}
We provide a brief overview of the background pertaining to the electrode placement problem. Sec.~\ref{sec:prob-description:sys-model} describes the mathematical system model of non-invasive electrical stimulation used by existing electrode placement algorithms for designing electrode montages. Finally, we briefly discuss existing work on the electrode placement problem in Sec.~\ref{sec:prob-description:background}. We use standard mathematical notation in this paper, which is detailed in {Appendix Table~\ref{tab:math-not}}. 
\subsection{System Model}\label{sec:prob-description:sys-model}
{We utilize the system model proposed in previous works~\cite{dmochowski2011optimized,fernandez2020unification,khan2022individually,dmochowski2017optimal}. We assume there are only $N$ possible locations where the electrodes can be placed. The current flowing through each location is mathematically represented as an $N$-dimensional vector $\myVec{I}=\begin{bmatrix}
    i_1&\hdots&i_N
\end{bmatrix}^T\in\R^N$, where $i_k$ represents the current flowing through the $k$-th electrode location. Consequently, the vector $\myVec{I}$ mathematically represents all possible electrode montages in this system model. Under the assumptions of quasi-static Maxwell equations~\cite{griffiths2005introduction}, we can numerically solve the Laplace equation to obtain the electric field induced in a discretized head model. Furthermore, under these quasi-static assumptions, the electric field in the head is related to the electrode montage $\myVec{I}$ through the following linear systems:}
\begin{align}
    \myVec{T}^x\myVec{I}=\myVec{E}_I^x,        \myVec{T}^y\myVec{I}=\myVec{E}_I^y,\text{ and }        \myVec{T}^z\myVec{I}=\myVec{E}_I^z;\ \ {\myVec{T}^x,\myVec{T}^y,\myVec{T}^z\in\myVec{R}^{M\times N}}\label{eq:prob-descrip:sys-model:1}
\end{align}
where $\myVec{E}_I^x$, $\myVec{E}_I^y$, $\myVec{E}_I^z\in\R^{M}$, respectively, represent the $x$, $y$, and $z$ components of the electric field induced in the head model by {the} electrode montage $\myVec{I}$, and $M$ (typically much bigger than $N$) is the number of voxels into which the head model is discretized. The matrices {$\myVec{T}^x,\myVec{T}^y,\text{ and }\myVec{T}^z$ are} the forward models relating $\myVec{I}$ to the corresponding $x$, $y$, and $z$ components of the induced electric fields. We can combine the three linear systems in~\eqref{eq:prob-descrip:sys-model:1} into a single system, as follows:
\begin{align}
\myVec{T}\myVec{I}=\myVec{E}_{I};\text{ where }  \myVec{T} \myeq \begin{bmatrix}
        \myVec{T}^x\\
        \myVec{T}^y\\
        \myVec{T}^z
    \end{bmatrix}\in\R^{3M\times N}\text{ and }\myVec{E}_{I}\myeq\begin{bmatrix}
        \myVec{E}_I^x\\
        \myVec{E}_I^y\\
        \myVec{E}_I^z
    \end{bmatrix} \in\mathbb{R}^{3M}.\label{eq:prob-description:sys-model:forward-model}
\end{align}
The matrix $\myVec{T}$ only depends on the conductivities and geometry of the head model. We refer the reader to~\cite{dmochowski2011optimized} for a comprehensive derivation of~\eqref{eq:prob-description:sys-model:forward-model}. Most electrode placement algorithms also divide the head model into {two distinct regions: a} target/focus region and an off-target/cancel region. We denote the collections of all voxels present in the target and off-target regions as $\mathcal{F}$ and $\mathcal{C}$, respectively. We further denote the matrices constructed by sub-sampling the rows of the forward matrix ($\myVec{T}$) corresponding to the voxels in the target and off-target regions as $\myVec{T}_f\in\mathbb{R}^{3|\mathcal{F}|\times N}$ and $\myVec{T}_c\in\R^{3|\mathcal{C}|\times N}$, respectively. Therefore, $\myVec{T}_f\myVec{I}$ and $\myVec{T}_c\myVec{I}$ give the value of the electric field at each voxel in the target and off-target regions, respectively. We also further sub-divide $\myVec{T}_f$ and $\myVec{T}_c$ into \{$\myVec{T}_f^x$, $\myVec{T}_f^y$, $\myVec{T}_f^z$\} and $\{\myVec{T}_c^x,\myVec{T}_c^y,\myVec{T}_c^z\}$ to obtain the respective $x$, $y$, and $z$ components of the target and off-target electric fields. 

The linear system relating the electrode montages $\myVec{I}$ and the induced electric field $\myVec{E}_I$ is extremely crucial for designing electrode montages as it simplifies the problem of electrode placement into two sub-problems:
    \begin{enumerate}
        \item Specifying some desired property of the induced electric field $\myVec{E}_I$. For example, a common desired property of the induced electric field is to reduce off-target stimulation.
        \item Solve the linear system described in~\eqref{eq:prob-description:sys-model:forward-model} to find the electrode montage $\myVec{I}$ whose induced electric field best satisfies the specified desired properties.     
    \end{enumerate}
All existing electrode placement algorithms conceptually follow the above two steps to design electrode montages with differing implementations. The next section, i.e., Sec.~\ref{sec:prob-description:background} provides a brief discussion on the different algorithms proposed to design electrode montages.
\subsection{Existing Approaches for Electrode Placement}\label{sec:prob-description:background}
The problem of electrode placement has received significant attention with numerous works~\cite{im2008determination,park2011novel,dmochowski2011optimized,sadleir2012target,dutta2013using,ruffini2014optimization,guler2016optimization,salman2016concurrency,cancelli2016simple,wagner2016optimization,fernandez2016transcranial,dmochowski2017optimal,guler2018computationally,saturnino2019accessibility,goswami2021hingeplace,khan2022individually,prieto2022l1,wang2023multi} proposing different algorithms for placing electrodes for tES. We refer the reader to~\cite{gomez2024perspectives} for a broad overview of existing work on tES electrode placement algorithms and tES forward modeling. All the proposed electrode placement algorithms can be thought of as different ways for approximately inverting the linear system in~\eqref{eq:prob-description:sys-model:forward-model} such that the resultant electrode field satisfies some desired properties. Some examples of desired properties are ensuring safe stimulation, high electric field intensity at the target region, and low electric field intensity across the off-target regions. 

Most of the proposed electrode placement algorithms can be broadly classified into four categories: the weighted least squares (WLS) approach~\cite{dmochowski2011optimized,ruffini2014optimization,salman2016concurrency,dmochowski2017optimal}, the constrained directional maximization (CDM) approach~\cite{dmochowski2011optimized,guler2016optimization,guler2018computationally,wagner2016optimization,khan2022individually,im2008determination,saturnino2019accessibility}, the reciprocity-based approach~\cite{salman2016concurrency,fernandez2016transcranial,cancelli2016simple,dutta2013using,fernandez2020unification}, and the extended linearly constrained minimum variance (LCMV-E) approach~\cite{fernandez2016transcranial,dmochowski2011optimized,saturnino2019accessibility,wang2023multi}. An important unification result from~\cite{fernandez2020unification} shows that the WLS and reciprocity-based approaches are {special cases of} the CDM approach. {Consequently, we do not discuss the WLS and reciprocity-based approaches in this work and refer the reader to~\cite{fernandez2020unification} (and the references therein) for a more detailed discussion on these two approaches}. Sec.~\ref{sec:equiv-LCMV-EDM} discusses the CDM and LCMV-E algorithms in detail, in addition to providing a (previously not known) equivalence result unifying these two algorithms. Hence, we defer the discussion on the LCMV-E and CDM algorithms to Sec.~\ref{sec:equiv-LCMV-EDM}. 

We now discuss the few electrode placement algorithms that do not fall into the above four classes. The algorithms proposed in~\cite{park2011novel,sadleir2012target} design electrode montages that aim to maximize the electric field magnitude at the target region by solving a non-convex optimization framework~\cite{gomez2024perspectives}. These algorithms have no guarantee of converging to the global optimum due to their non-convexity~\cite{fernandez2020unification}, making them unattractive from an optimization perspective. Furthermore, it is well known that neurons preferentially respond to certain electric field directions~\cite{aberra2018biophysically,rattay1986analysis,rattay1999basic,fellner2022finite}. Hence, in many applications, it is more fruitful to maximize electric field intensity along these directions rather than the electric field magnitude, which fortuitously results in a convex optimization problem (providing a guarantee of converging to global optimum). Consequently, we elected not to compare \HP with~\cite{sadleir2012target,park2011novel}. We presented a special case of the \HP algorithm at the $43^{\text{rd}}$ International Conference of the IEEE Engineering in Medicine \& Biology Society (EMBC)~\cite{goswami2021hingeplace}. This special case of the \HP algorithm was also independently proposed (later) in~\cite{prieto2022l1}, albeit with a different goal. We discuss and compare these algorithms in Sec.~\ref{sec:algo:equiv-HP-L1L1}.

\section{Equivalence of CDM and LCMV-E}\label{sec:equiv-LCMV-EDM}
We present our theoretical results connecting the CDM and the LCMV-E approaches. These results extend the previous unification result of Fern\'{a}ndez-Corazza et al.~\cite{fernandez2020unification}, unifying the four major classes of electrode placement algorithms, namely, CDM, LCMV-E, WLS, and reciprocity-based approaches. We first describe the CDM and LCMV-E formulations and then state Theorem~\ref{theorem:CDM-LCMV-equiv}, showing the equivalence between these approaches.

\textbf{CDM Approach}: We use the optimization formulation proposed in Fern\'{a}ndez-Corazza et al.~\cite{fernandez2020unification} as the representative CDM approach. The goal of the CDM approach is to maximize the sum of electric field intensity along chosen directions in the target region ($\mathcal{F}$) while ensuring that the off-target electric field magnitude is below a pre-specified threshold $\sqrt{\alpha}$. Mathematically, CDM achieves this goal through the following optimization problem: 
\begin{align}
    \myVec{I}^{*} \myeq&\argmax_{\myVec{I}\in\R^N}\myVec{D}^T
\myVec{\Gamma}_F\myVec{T}\myVec{I}    
    ,\text{ s.t. }(\myVec{T}_c\myVec{I})^T\boldsymbol{\Gamma}_C\myVec{T}_c\myVec{I}\mylesseq \alpha, \|\myVec{I}\|_1\mylesseq 2I_{tot}, \|\myVec{I}\|_{\infty}\mylesseq I_{safe},\nonumber\\
    &\text{ and }\myVec{1}_N^T\myVec{I} \myeq 0.\label{eq:cdm:original}
\end{align}
{The objective $\myVec{D}^{T}\myVec{\Gamma}_F\myVec{T}_f\myVec{I}$ is a discrete approximation of the aggregate electric field projected along the direction $\myVec{d}(\myVec{r})$ (at the location $\myVec{r}$) in the target region, i.e., $\int_{\myVec{r}\in \mathcal{F}}\myVec{d}(\myVec{r})\cdot\myVec{E}_I(\myVec{r})dV$. Here, $\myVec{D}=[\myVec{D}_x^T,\myVec{D}_y^T,\myVec{D}_z^T]^T$ is the vector containing the directions along which the electric field should be projected at each target voxel, with $\myVec{D}_x$, $\myVec{D}_y$, $\myVec{D}_z\in\R^{|\mathcal{F}|}$ containing the respective $x$, $y$, and $z$ components. $\boldsymbol{\Gamma}_F$ is a diagonal matrix whose elements are the volumes of each target voxel. Hence, the goal of~\eqref{eq:cdm:original} is to maximize the aggregate electric field intensity (projected along the directions $\myVec{D}$) in the target region.}

{The quadratic term $(\myVec{T}_c\myVec{I})^T\myVec{\Gamma}_C\myVec{T}_c\myVec{I}$ is the discrete approximation of the integral $\int_{\myVec{r}\in \mathcal{C}} \left\|\myVec{E}_I(\myVec{r})\right\|_2^2dV$, representing the aggregate off-target electrical field magnitude. The elements of the diagonal matrix $\myVec{\Gamma}_C$ are the volumes of the off-target voxels. Therefore, the constraint $(\myVec{T}_c\myVec{I})^T\boldsymbol{\Gamma}_C\myVec{T}_c\myVec{I}\mylesseq \alpha$ bounds the aggregate off-target electric field magnitude to be below $\sqrt{\alpha}$. The constraints $\|\myVec{I}\|_1\leq 2I_{tot}$ and $\|\myVec{I}\|_{\infty}\mylesseq I_{safe}$ restrict the total current and the current per electrode to be less than $I_{tot}$ and $I_{safe}$, respectively, to ensure safe stimulation. The constraint $\myVec{1}_N^T\myVec{I}=0$ ensures Kirchhoff's law is satisfied. We denote $\myVec{D}^T\myVec{\Gamma}_F\myVec{T}_f$ as $\myVec{A}_f$, and $(\myVec{T}_c\myVec{I})^T\myVec{\Gamma}_C\myVec{T}_c\myVec{I}$ as $\|\myVec{A}_c\myVec{I}\|_2^2$ (which also defines $\myVec{A}_c$) for brevity.} We restate~\eqref{eq:cdm:original} using our updated notation for convenience:
\begin{align}
    \myVec{I}^{*} =& \argmax_{\myVec{I}\in\R^N}\myVec{A}_f\myVec{I},\text{ s.t. } \|\myVec{A}_c\myVec{I}\|_2^2\mylesseq \alpha,\|\myVec{I}\|_1\mylesseq 2I_{tot}, \|\myVec{I}\|_{\infty}\mylesseq I_{safe},\text{ and }\myVec{1}_N^T\myVec{I} \myeq 0.\label{eq:cdm:update-not}
\end{align}

\textbf{LCMV-E Approach}: We use the umbrella term LCMV-E to group several related but subtly different electrode placement algorithms (e.g.~\cite{saturnino2019accessibility,dmochowski2011optimized}). We broadly interpret the LCMV-E approach as the set of electrode placement algorithms that design electrode montages by minimizing the aggregate off-target electric field magnitude while constraining the target electric field by a linear constraint. The canonical LCMV-E optimization formulation is described in~\eqref{eq:lcmv-e}:
    \begin{align}
        \myVec{I}^{*} =& \argmin_{\myVec{I}\in\R^N} \|\myVec{A}_c\myVec{I}\|_2^2,\text{ s.t. }\myVec{A}_f\myVec{I}=\myVec{E}_{des},\|\myVec{I}\|_1\leq 2I_{tot},\|\myVec{I}\|_{\infty}\leq I_{safe},\nonumber\\
        &\text{ and }\myVec{1}_N^T\myVec{I} = 0,\label{eq:lcmv-e}
    \end{align}
where the term $\|\myVec{A}_c\myVec{I}\|_2^2$ is the same term as described in~\eqref{eq:cdm:update-not}. The linear constraint\footnote{We slightly abuse notation here to show the similarity between LCMV-E and CDM approaches. Under LCMV-E and later in \textit{HingePlace}, we more broadly interpret the matrix $\myVec{A}_f$ as any linear constraint used to ensure that the target electric field has some desired property in contrast to the specific meaning assigned to $\myVec{A}_f$ under CDM.} $\myVec{A}_f\myVec{I}=\myVec{E}_{des}$ is used to constrain different aspects of the target electric field. For example, Saturnino et al.~\cite{saturnino2019accessibility} use this linear constraint to ensure that the average target electric field along a chosen direction is at a desired intensity. The rest of the constraints in~\eqref{eq:lcmv-e} are the same as that in~\eqref{eq:cdm:update-not}, and serve the same purpose.  

\textbf{Equivalence}: We present Theorem~\ref{theorem:CDM-LCMV-equiv}, which shows that the LCMV-E and the CDM optimization problems (stated in~\eqref{eq:lcmv-e} and~\eqref{eq:cdm:update-not}, respectively) are essentially equivalent. We first introduce a few terms required to state Theorem~\ref{theorem:CDM-LCMV-equiv} formally. Consider the following optimization problem:
\begin{align}
&\argmin_{\myVec{I}\in\R^N}\myVec{A}_f^T\myVec{I},\text{ s.t. }\|\myVec{I}\|_1\mylesseq2I_{tot},\|\myVec{I}\|_{\infty}\mylesseq I_{safe},\myVec{1}_N^T\myVec{I}\myeq0.\label{eq:max-intenstity}
\intertext{Let $\mathcal{I}^*$ denote the set of all solutions of~\eqref{eq:max-intenstity}. Then,}
&\alpha_{MAX}=\argmin_{\myVec{I}^*\in\mathcal{I}^*}\|\myVec{A}_{c}\myVec{I}^*\|_2^2.\label{eq:alpha:defn}\\
&E_{MAX}= \myVec{A}_f\myVec{I}^*\text{ for some }\myVec{I}^*\in\mathcal{I}^*.\label{eq:emax:defn}
\end{align}
The variables $\alpha_{MAX}$ and $E_{MAX}$ respectively denote the highest possible aggregate target electric field intensity (along a desired direction), and the square of the corresponding smallest aggregate off-target magnitude, that can be induced under the safety limits, $I_{tot}$ and $I_{safe}$. Note that in~\eqref{eq:emax:defn}, it does not matter what $\myVec{I}^*$ is chosen from the set $\mathcal{I}^*$, as all elements of $\mathcal{I}^*$ produce the same value of $\myVec{A}_f\myVec{I}^*$.
\begin{theorem}\label{theorem:CDM-LCMV-equiv}
Assume that $I_{tot}$, $I_{safe}$, $\myVec{A}_c$, and $\myVec{A}_f$ are the same in~\eqref{eq:cdm:update-not} and~\eqref{eq:lcmv-e}. Denote the solution of~\eqref{eq:cdm:update-not} and~\eqref{eq:lcmv-e} as $\myVec{I}^*_1$ and $\myVec{I}_2^*$, respectively. 
\begin{itemize}
    \item If $\alpha<\alpha_{MAX}$, then for $E_{des}=\myVec{A}_f\myVec{I}_1^*$, we have $\myVec{I}_1^*=\myVec{I}_2^*$, i.e., the solutions of~\eqref{eq:lcmv-e} and~\eqref{eq:cdm:update-not} are same. Conversely, if $E_{des}<E_{MAX}$, then for $\alpha=\myVec{A}_f\myVec{I}_2^*$ we have $\myVec{I}_1^*=\myVec{I}_2^*$. Furthermore,~\eqref{eq:cdm:update-not} and~\eqref{eq:lcmv-e} have unique solutions. 
    \item If $\alpha\geq \alpha_{MAX}$, then for $E_{des}\myeq E_{MAX}$ we have $\myVec{I}_2^*=\argmin_{\myVec{I}_1^*\in\mathcal{I}_1^*}\|\myVec{A}_c\myVec{I}_1^*\|_2^2$, where $\mathcal{I}_1^*$ is the set of all solutions of~\eqref{eq:cdm:update-not}. Conversely, if $E_{des}\myeq E_{MAX}$, then for any $\alpha\geq\|\myVec{A}_c\myVec{I}_2^*\|_2^2$, we have $\myVec{I}_2^*\myeq\argmin_{\myVec{I}_1^*\in\mathcal{I}_1^*}\|\myVec{A}_c\myVec{I}_1^*\|_2^2$. For this case, only~\eqref{eq:lcmv-e} is guaranteed to have a unique solution. 
\end{itemize}     
\end{theorem}
\begin{proof}
    See Appendix A.
\end{proof}
{ Theorem~\ref{theorem:CDM-LCMV-equiv} shows that the LCMV-E and CDM optimization problems are exactly equivalent and are each other's dual formulation for $\alpha<\alpha_{MAX}$. This exact equivalence between LCMV-E and CDM does not hold for $\alpha\geq \alpha_{MAX}$. The CDM optimization problem has no guarantee of a unique solution for $\alpha\geq \alpha_{MAX}$, but the LCMV-E optimization always has a unique solution due to a strictly convex objective. In this case, the LCMV-E solution is the same as the most ``relevant'' CDM solution having the least aggregate off-target magnitude ($\alpha_{MAX}$). Note that a CDM solution with a larger off-target magnitude offers no additional advantage as the intensity at the target region remains the same. 
Hence, we argue that LCMV-E and CDM approaches are essentially equivalent. Combined with the unification result of~\cite{fernandez2020unification}, Theorem~\ref{theorem:CDM-LCMV-equiv} unifies the four major classes of electrode placements, namely, WLS, reciprocity-based, CDM, and LCMV-E approaches discussed in Sec.~\ref{sec:prob-description:background}.}  

\section{HingePlace}\label{sec:algo:HingePlace}
{In this section, we present our proposed algorithm, \textit{HingePlace}. We first show in Sec.~\ref{sec:algo:weakness} that existing electrode placement algorithms over-penalize the aggregate off-target electric field magnitude by utilizing the equivalence result of Sec.~\ref{sec:equiv-LCMV-EDM}. To harness the thresholding properties of neural response, we propose the \HP algorithm in Sec.~\ref{sec:algo:description}. Finally, Sec.~\ref{sec:algo:equiv-HP-L1L1} discusses the equivalence between ``L1L1-norm optimization'' proposed in Prieto et al.~\cite{prieto2022l1} and a special case of \HP algorithm.}
\subsection{Limitations of traditional electrode placement algorithms}\label{sec:algo:weakness}
{In Sec.~\ref{sec:equiv-LCMV-EDM}, we demonstrated that the WLS, reciprocity-based, and CDM approaches are equivalent to the LCMV-E approaches.} Hence, the behavior of these approaches can be understood by only analyzing the LCMV-E approach, specifically~\eqref{eq:lcmv-e}. Analyzing the objective of~\eqref{eq:lcmv-e}, we observe that the optimal $\myVec{I}^*$ for LCMV-E would be the one with the least possible off-target electric field magnitude.  {Furthermore, Appendix B shows that even the magnitude maximization approaches proposed in~\cite{park2011novel,sadleir2012target} minimize off-target electric field magnitude.} Consequently, all {major electrode placement algorithms} force the off-target electric field to be as close to \emph{zero} as possible. {As discussed earlier in Sec.~\ref{sec:introduction}, in both suprathreshold and subthreshold tES, we only require off-target electric fields to below a non-zero threshold.} Therefore, traditional electrode placement algorithms \emph{over-penalize} the off-target electric field {and do not harness the thresholding behavior of neural response}. 

\subsection{Algorithm Description}\label{sec:algo:description}
Mathematically, \HP constitutes a family of optimization algorithms {(stated in~\eqref{eq:algo:HP-main-opt})} that use a modified \emph{hinge loss} to avoid penalizing electric fields with  {off-target amplitudes below a threshold}. Consequently, \HP designs electrode montages that allow \emph{non-zero} but {below-threshold off-target electric field amplitudes}. We first mathematically state the \HP optimization formulation, followed by an intuitive description of this formulation.

\noindent\textbf{Mathematical Description}: We first define the loss vector $\myVec{L}_{\textrm{HP}}(\myVec{I})$ as follows:
\begin{align}
&\myVec{L}_{\textrm{HP}}\left(\myVec{I}\right) \myeq \begin{bmatrix}
    \myVec{L}_{\textrm{HP}}^x(\myVec{I})\\
    \myVec{L}_{\textrm{HP}}^y(\myVec{I})\\
    \myVec{L}_{\textrm{HP}}^z(\myVec{I})
\end{bmatrix},\text{ where }\text{ for }i=x,y\text{ and }z:\nonumber\\
&\myVec{L}_{\textrm{HP}}^i(\myVec{I})\myeq\max\left(\myVec{0}_{|\mathcal{C}|},\myVec{T}_c^i\myVec{I}\myminus\myVec{E}_{tol}^{i+}\right)\myplus\max\left(\myVec{0}_{|\mathcal{C}|},\myminus\myVec{T}_c^i\myVec{I}\myminus\myVec{E}_{tol}^{i-}\right)\label{eq:algo:loss-vector}\end{align}
In~\eqref{eq:algo:loss-vector}, $\max(\cdot,\cdot)$ computes the element-wise maximum between two vectors.  The vectors $\myVec{E}_{tol}^{x+},\myVec{E}_{tol}^{y+},$ and $\myVec{E}_{tol}^{z+}$ represent the upper limits of the off-target electric field amplitudes in $x$, $y$, and $z$ directions, respectively. Similarly, $\myVec{E}_{tol}^{x-}$, $\myVec{E}_{tol}^{y-}$, and $\myVec{E}_{tol}^{z-}$ represent the corresponding lower limits. We collectively refer to the threshold vectors as $\myVec{E}_{tol}^+=[(\myVec{E}_{tol}^{x+})^T,(\myVec{E}_{tol}^{y+})^T,(\myVec{E}_{tol}^{z+})^T]^T$ and $\myVec{E}_{tol}^-=[(\myVec{E}_{tol}^{x-})^T,(\myVec{E}_{tol}^{y-})^T,(\myVec{E}_{tol}^{z-})^T]^T$. We now state the \HP optimization problem as follows: 
\begin{align}
\myVec{I}^* \myeq &\argmin_{\myVec{I}\in\mathbb{R}^N} \underbrace{\left\|\myVec{\Gamma}_{C}\myVec{L}_{\textrm{HP}}\right\|_{p}^p}_{\mathcal{L}_{hinge}(\myVec{I})},   \text{ s.t. } \myVec{A}_f\myVec{I}= \myVec{E}_{des},\|\myVec{I}\|_1\leq 2\myVec{I}_{total},\|\myVec{I}\|_{\infty}\leq \myVec{I}_{safe},\nonumber\\
&\text{ and }
\myVec{1}_N^T\myVec{I} = 0. \label{eq:algo:HP-main-opt}
\end{align}
{\textbf{Intuitive Description}:} The \HP optimization formulation utilizes the same constraints employed by the LCMV-E optimization formulation (see~\eqref{eq:lcmv-e}). The constraints serve the same purposes described earlier in Sec.~\ref{sec:prob-description:background}. The main innovation in the \HP algorithm is the loss function $\mathcal{L}_{hinge}(\myVec{I})$ that only penalizes the off-target electric field with amplitudes above a user-specified threshold. We now describe the loss function $\mathcal{L}_{hinge}(\myVec{I})$. 

The matrix $\myVec{\Gamma}_{C}$ in $\mathcal{L}_{hinge}(\myVec{I})$ accounts for the different volumes of the voxels, similar to the CDM approach. The loss vector $\myVec{L}_{\textrm{HP}}(\myVec{I})$ assigns three loss values (corresponding to the three electric field components) for each voxel in the off-target region. Note that these loss values are only non-zero if the corresponding electric field component {violates} the upper limit (specified through $\myVec{E}_{tol}^+$) or the lower limit (specified through $\myVec{E}_{tol}^-$). Hence, the loss vector $\myVec{L}_{\textrm{HP}}(\myVec{I})$ only penalizes an off-target field whose amplitude exceeds pre-specified subthresholds. Consequently, $\myVec{L}_{\textrm{HP}}(\myVec{I})$ allows \HP to find electrode montages that induce more focused neural response by allowing {below-threshold but non-zero off-target amplitudes.}

\begin{figure}[ht]
    \centering
    \includegraphics[width=0.4\textwidth]{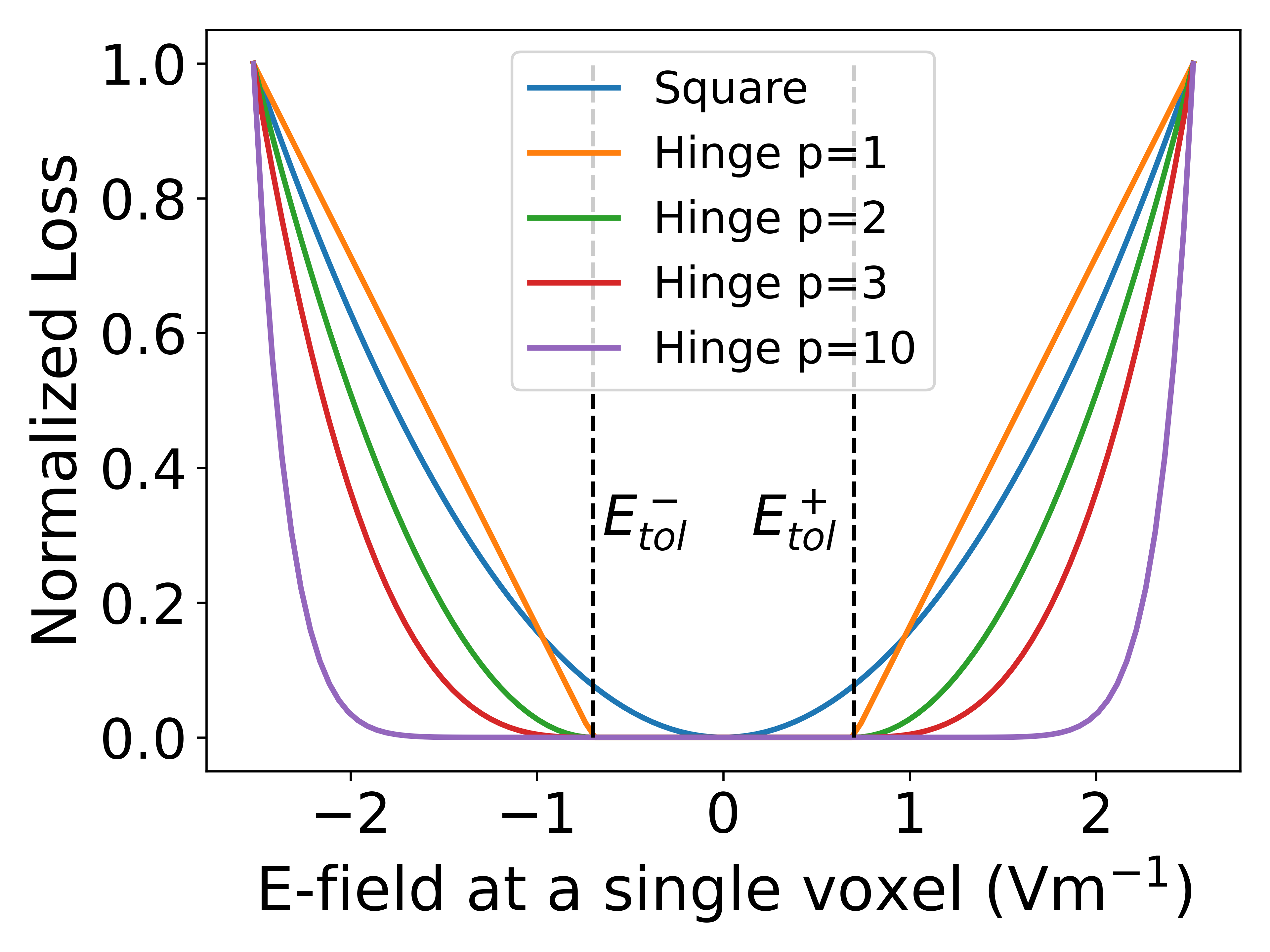}
    \caption{A visualization of the $l_2$ (square) loss and the \HP loss with different values of $p$ as a function of electric field amplitude at a single voxel in the off-target region.}
    \label{fig:-hp-loss}
\end{figure}

Finally, the $\|\cdot\|_{p}$ aggregates the loss vector into a single number $\mathcal{L}_{hinge}(\myVec{I})$, which can be minimized to obtain the desired electrode montage. Fig.~\ref{fig:-hp-loss} provides a visualization of the different \HP losses in comparison to the standard square loss utilized in the LCMV-E approach. From Fig.~\ref{fig:-hp-loss}, we can visually observe increasing $p$ causes $\mathcal{L}_{hinge}(\myVec{I})$ to sharply penalize off-target amplitudes much larger than the pre-specified thresholds ($\myVec{E}_{tol}^+$ and $\myVec{E}_{tol}^-$) while allowing off-target field amplitudes with smaller violations of the same thresholds. {In our simulation studies, we find that $p=1$ and $p=2$ generally yield the best performance. Furthermore, the \HP algorithm is a convex optimization problem, benefiting from all the computational advantage associated with solving convex optimization problem.} Lastly, for $p=2$ and $\myVec{E}_{tol}^+=\myVec{E}_{tol}^{-}=\myVec{0}_{|\mathcal{C}|}$, the \HP algorithm reduces to the LCMV-E algorithm. Consequently, the \HP algorithm also encompasses existing electrode placement algorithms due to the unifying results of Sec.~\ref{sec:equiv-LCMV-EDM}.
\subsection{L1L1-norm as a special case of \HP}\label{sec:algo:equiv-HP-L1L1}
Prieto et al.~\cite{prieto2022l1} proposed an ``L1L1-norm'' optimization problem that utilizes an $l_1$ loss and an $l_1$ regularizer to approximately solve the following system:
\begin{align}
    \begin{bmatrix}
        \myVec{T}_f\\
        \myVec{T}_c
    \end{bmatrix}\myVec{I}=\begin{bmatrix}
        \myVec{E}_{des}\\
        \myVec{0}_{|\mathcal{C}|}
    \end{bmatrix}.\label{eq:l1l1:sys}
\end{align}
As can be seen from~\eqref{eq:l1l1:sys}, the conceptual goal of L1L1-norm optimization is different from the \HP algorithm. The L1L1-norm optimization aims to obtain electrode montages that induce minimum (ideally zero) off-target electric field amplitudes. The loss of L1L1-norm utilizes a relaxation function $\Psi_{\epsilon}[\cdot]$ that is equivalent to $\mathcal{L}_{hinge}(\cdot)$ for $p=1$ and $\myVec{E}_{tol}^{+}=\myVec{E}_{tol}^-=E_{tol}\myVec{1}_{|\mathcal{C}|}$ for some $E_{tol}>0$ (see Appendix C). Consequently, L1L1-norm optimization is \emph{functionally} equivalent to a special case of \HP optimization ($p=1$ and $\myVec{E}_{tol}^{+}=\myVec{E}_{tol}^-=E_{tol}\myVec{1}_{|\mathcal{C}|}$ with an appropriately modified $l_{\infty}$ norm constraint\footnote{The last paragraph of Appendix C elaborates further on the difference of $l_{\infty}$ constraint in L1L1-norm optimization and \textit{HingePlace}.}) despite different conceptual goals (formally shown in Appendix C).

The difference in the conceptual goals of the \HP algorithm and the L1L1-norm optimization leads to different interpretations of $\myVec{E}_{tol}^+$ and $\myVec{E}_{tol}^{-}$ in these algorithms, namely, neural thresholds in the \HP algorithm and numerical tolerance in the 
 L1L1-norm optimization. This difference in interpretations of $\myVec{E}_{tol}^+$ and $\myVec{E}_{tol}^{-}$ is crucial for fully harnessing the potential of the \HP algorithm. Neurons are known to respond preferentially to electric fields oriented along specific directions. Hence, forcing $\myVec{E}_{tol}^+$ and $\myVec{E}_{tol}^{-}$ to be the same for all directions (as is done in the L1L1-norm optimization) is restrictive. Choosing different tolerances for different directions respects the underlying physiology more accurately, typically leading to more focal neural responses. We empirically demonstrate this point through a simulation study in Sec.~\ref{sec:results:sea-of-neurons:c3-c4}, observing that using the same tolerance for each direction leads to more diffused neural responses (see Fig.~\ref{fig:neuron:results}a-c). 

\section{Results}\label{sec:results}
\begin{figure*}[ht]
    \centering
    \includegraphics[width=\textwidth]{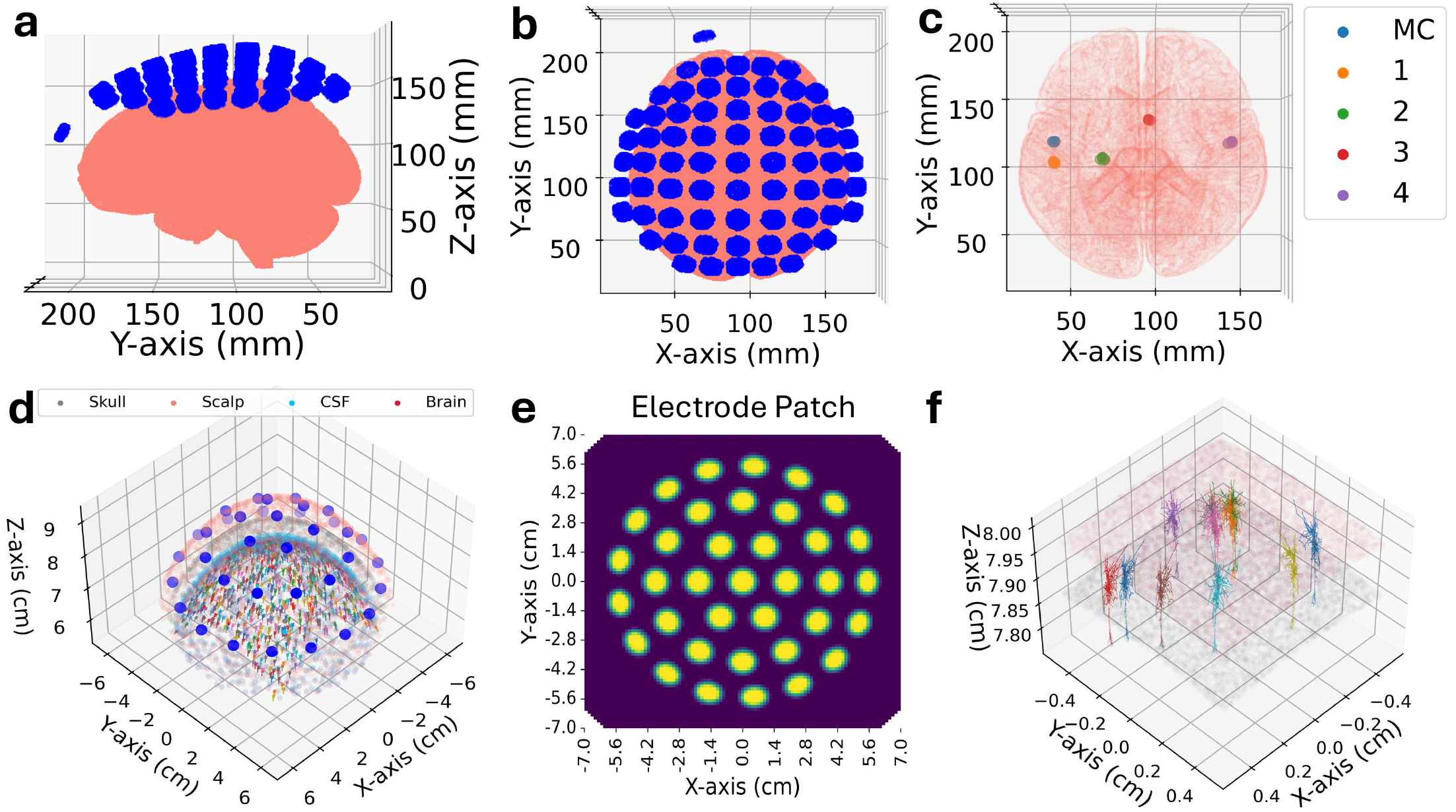}
    \caption{\textbf{a} and \textbf{b} present a sagittal and horizontal view of the realistic MRI head model discussed in Sec.~\ref{sec:results:mri}. The electrodes are shown in blue, and the brain is represented using the salmon color. \textbf{c} shows the five different target locations used in the simulation studies described in Sec.~\ref{sec:results:mri}. MC denotes a target location in the motor cortex, and the MNI coordinates of target locations $1$, $2$, $3$, and $4$ are provided in Appendix D. \textbf{d} provides a schematic of the sea of neurons model discussed in Sec.~\ref{sec:results:sea-of-neurons}. The electrode locations are denoted by blue points. \textbf{e} provides a 2-D representation of the electrode locations shown in \textbf{d}. \textbf{f} is a zoomed-in version of \textbf{d} to show the neuron morphology.}
    \label{fig:setup}
\end{figure*}
{Simulation platforms are widely used to validate the performance of various electrode placement algorithms across numerous studies~\cite{im2008determination,park2011novel,dmochowski2011optimized,sadleir2012target,dutta2013using,ruffini2014optimization,guler2016optimization,salman2016concurrency,cancelli2016simple,wagner2016optimization,fernandez2016transcranial,dmochowski2017optimal,guler2018computationally,saturnino2019accessibility,khan2022individually,prieto2022l1,wang2023multi}. In this work, we use two simulation platforms, namely, an MRI head model with neural response approximated through thresholding of the electric field, and a ``sea of neurons'' model in a spherical head geometry to test the \HP algorithm's performance. Across both simulation platforms, we compared the performance of the \HP algorithm against the LCMV-E algorithm (see~\eqref{eq:lcmv-e}) as a baseline electrode placement algorithm. We chose the LCMV-E algorithm for two reasons: (i) Sec.~\ref{sec:equiv-LCMV-EDM} shows that the LCMV-E algorithm is equivalent to all the other major electrode placement algorithms, namely, CDM, WLS, and reciprocity-based algorithms; (ii) The hyperparameters (e.g., $I_{safe}$, $I_{tot}$, and $E_{des}$) are the same for \HP and LCMV-E algorithms (except additional hyperparameters $\myVec{E}_{tol}^{+}$ and $\myVec{E}_{tol}^{-}$ for the \HP algorithm), allowing us to compare the performance of \HP and LCMV-E algorithms in a principled manner across different hyperparameters.} 

{All simulations studies compared \HP and LCMV-E across different values of $I_{safe}$ and $I_{tot}$. The smallest value for $I_{safe}$ was selected as the minimum value at which the LCMV-E and \HP optimizations remain feasible. We parameterized $I_{tot}=I_{tot}^{mul}I_{safe}$ in both~\eqref{eq:algo:HP-main-opt} and~\eqref{eq:lcmv-e}. This particular parameterization allowed us to interpret $I_{tot}^{mul}$ as the maximum number of active electrodes that can pass $I_{safe}$ amount of current. We considered three different values of $I_{tot}^{mul}$: \{$2$, $4$, $8$\} for MRI head model studies and \{$2$, $4$, $6$\} for sea of neurons model studies. We only consider $p=1,2,3$ for the \HP algorithm based on an initial study described in Appendix D. The rest of implementation details are discussed in Appendix D.}
 \begin{figure*}[ht]
    \centering
\includegraphics[width=\textwidth]{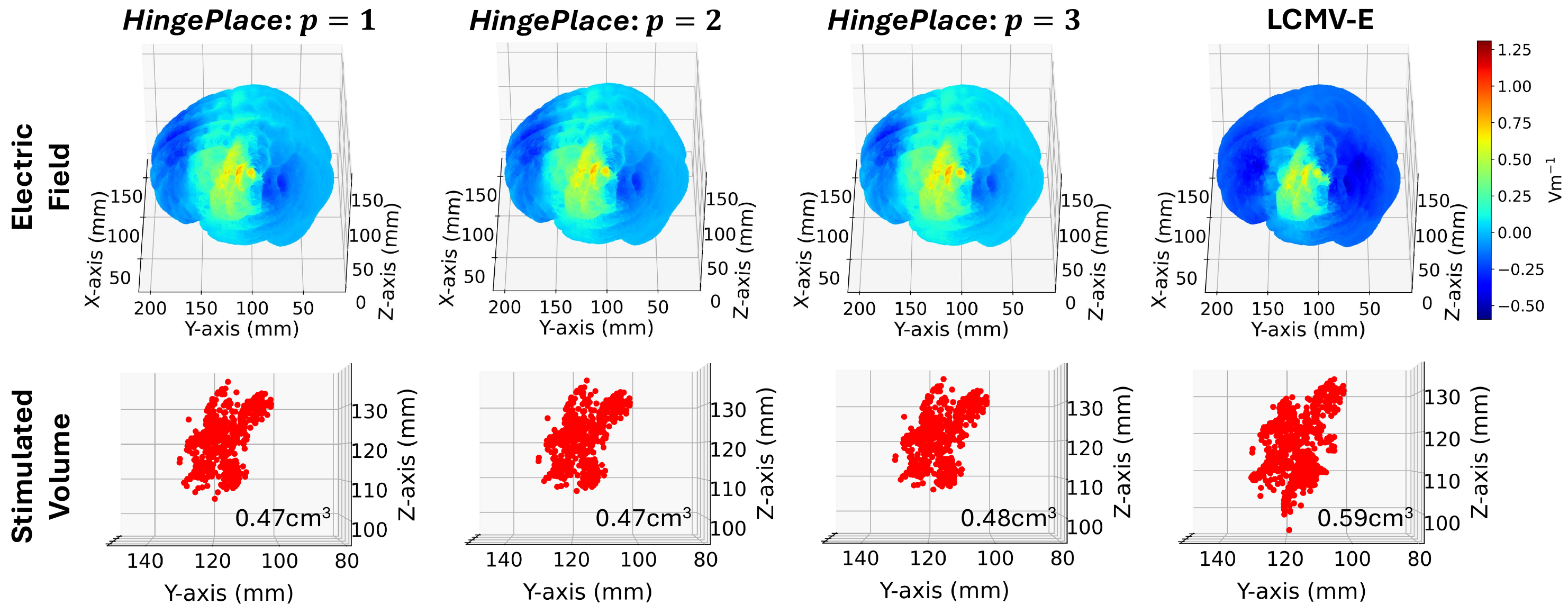}
    \caption{A representative plot of the electric field along the radial-in direction and the corresponding region where the radial-in electric field intensity above $0.8\text{Vm}^{-1}$ generated by electrode montages designed using the \HP and LCMV-E algorithms in Sec.~\ref{sec:results:mri:MC}. The target location was the motor cortex and the values of $I_{safe}$ and $I_{tot}^{mul}$ were $4.5$mA and $2$, respectively. We observe that LCMV-E produces the least off-target electric field, but has a larger volume above the stimulation threshold compared to \textit{HingePlace}. Additional representative plots are provided in Appendix E.}
    \label{fig:mri-visualisation}
\end{figure*}
\subsection{MRI Head Model Results}\label{sec:results:mri}
{MRI head models have been frequently used to validate the performance of the electrode placement algorithms in previous studies~\cite{im2008determination,park2011novel,dmochowski2011optimized,sadleir2012target,dutta2013using,ruffini2014optimization,guler2016optimization,salman2016concurrency,cancelli2016simple,wagner2016optimization,fernandez2016transcranial,dmochowski2017optimal,guler2018computationally,saturnino2019accessibility,khan2022individually,prieto2022l1,wang2023multi,Hermann2020,farahani2024transcranial}. We used a standard software ROAST~\cite{huang2019realistic} to simulate the standard MNI152 MRI head model with $1$mm resolution (see Fig.~\ref{fig:setup}). To compare neural activation in the MRI head model, we use a simplistic metric $\mathcal{V}_{\mathrm{Th}}$, which measures the volume of brain tissue having the ``radial-in'' electric field intensity  (modeling the electric field perpendicular to the cortical surface) above $80\%$ of the desired target intensity, namely, $E_{des}$ (based on our experimental work~\cite{forssellner2021}). We have previously used this metric to quantify neural activation in~\cite{goswami2021hingeplace}. All the implementation details regarding the MRI head models and $\mathcal{V}_{\mathrm{Th}}$ are provided in Appendix D.} 
We conducted four different simulation studies comparing the performance of the \HP and LCMV-E algorithms. {The \HP montages consistently produced the same or lower levels of off-target stimulation (measured using $\mathcal{V}_{\mathrm{Th}}$) than the LCMV-E montages across all the studies.}
\subsubsection{Creating focal fields at the motor cortex}\label{sec:results:mri:MC}

\begin{figure*}[ht]
    \centering
\includegraphics[width=\textwidth]{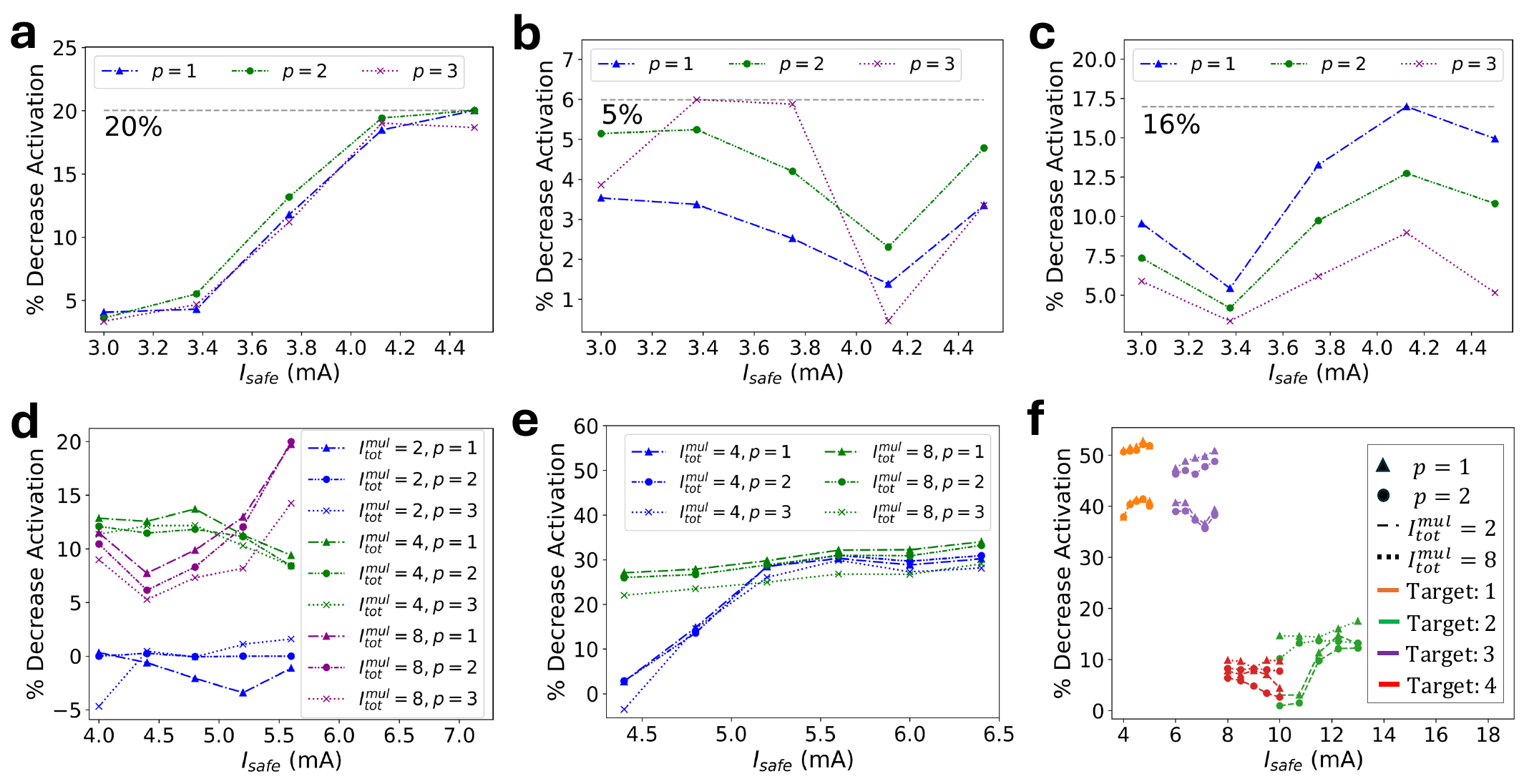}
    \caption{\textbf{a}-\textbf{c} plot the relative decrease in the stimulated volume between the \HP and LCMV-E montages for the simulation study described in Sec.~\ref{sec:results:mri:MC}. \textbf{a}, \textbf{b}, and \textbf{c} correspond to different values of $I_{tot}^{mul}$, namely, $2$, $4$, and $8$, respectively. Similarly, \textbf{d}, \textbf{e}, and \textbf{f} plot the relative decrease in stimulated volume between \HP and LCMV-E montages for the studies described in Sec.~\ref{sec:results:mri:larger-focus}, Sec.~\ref{sec:results:mri:multi-site}, and Sec.~\ref{sec:results:mri:diff-targets}, respectively. }
    \label{fig:mri-results}
\end{figure*}

The motor cortex is an important cortical region frequently targeted in non-invasive stimulation studies. Stimulation of the motor cortex has shown promise in treating neurological conditions such as stroke rehabilitation~\cite{ayache2012stroke} and fibromyalgia~\cite{sun2022repetitive}. In this study, we compare the focality of the neural responses generated at the motor cortex by the \HP algorithm (for $p=1,2,$ and $3$) with the LCMV-E algorithm.   
Fig.~\ref{fig:mri-visualisation} provides a visualization of the electric fields and the corresponding stimulated region generated by montages obtained using the \HP and LCMV-E algorithms. The LCMV-E montage generates the least off-target electric field amplitude but has the higher volume above the stimulation threshold. 
Fig.~\ref{fig:mri-results}a-c plot the relative decrease in the stimulated volume generated by the \HP montages ($\mathcal{V}_{\mathrm{Th}}^{HP}$) compared to stimulated volume generated by the LCMV-E montages ($\mathcal{V}_{\mathrm{Th}}^{LCMV-E}$), calculated as $\nicefrac{\mathcal{V}_{\mathrm{Th}}^{LCMV-E}-\mathcal{V}_{\mathrm{Th}}^{HP}}{\mathcal{V}_{\mathrm{Th}}^{LCMV-E}}\times 100$. We observe that the \HP montages always generate less stimulated volume than the LCMV-E montages, with relative decrease being as much as $20\%$ for some values of $I_{safe}$ and $I_{tot}^{mul}$. Generally, the most focal neural response was generated at $p=1$ or $p=2$ for the \HP algorithm.
\subsubsection{Effect of using a Larger Target Area}\label{sec:results:mri:larger-focus}
We analyze the behavior of the \HP algorithm for large target regions. The structure of this study is the same as the study performed in Sec.~\ref{sec:results:mri:MC}, except the volume of the target region was increased roughly $1000\times$. Fig.~\ref{fig:mri-results}d shows the relative decrease in the stimulated volume between the \HP and LCMV-E montages. We observe that results remain largely consistent with the results of Sec.~\ref{sec:results:mri:MC}, i.e., the \HP  montages stimulate less off-target volume than the LCMV-E montages. We again find that the \HP algorithm at $p=3$ continues to perform worse than the corresponding $p=1$ and $p=2$ cases.
\subsubsection{Effect of using multi-site stimulation}\label{sec:results:mri:multi-site}
We compare the \HP and LCMV-E algorithms when stimulating multiple target regions simultaneously. For this study, we chose two target regions: the left and right motor cortex {(denoted by MC and $4$ in Fig.~\ref{fig:setup}c)}. The study structure is the same as the previous studies described in Sec.~\ref{sec:results:mri:MC} and Sec.~\ref{sec:results:mri:larger-focus}. We only considered $I_{tot}^{mul}$ to be $4$ and $8$, as the case of $I_{tot}^{mul}=2$ required extremely high values of $I_{safe}$ to achieve the desired target intensity at both targets simultaneously. Fig.~\ref{fig:mri-results}e shows the relative decrease in the stimulated volume between the \HP and LCMV-E montages. The results remain consistent with our earlier studies in Sec.~\ref{sec:results:mri:MC} and~\ref{sec:results:mri:larger-focus}, showing that the \HP montages stimulate less off-target volume than the LCMV-E montages.

\subsubsection{Effect of choosing different targets}\label{sec:results:mri:diff-targets}
In this simulation study, we test the sensitivity of the \HP algorithm to different targets. We run the same study as Sec.~\ref{sec:results:mri:diff-targets} but at four different targets (locations $1$-$4$ in Fig.~\ref{fig:setup}f). We only consider $p=1$ and 2 for the \HP algorithm due to the comparatively worse performance of the $p=3$ case. Furthermore, we we only considered two $I_{tot}^{mul}$ values, namely, $2$ and $8$, to reduce the computation time of the study. Similarly to Sec.~\ref{sec:results:mri:MC}, Fig.~\ref{fig:mri-results}f plots the relative decrease in the off-target volume stimulated by the \HP and LCMV-E montages. Our results again suggest that the \HP montages stimulate less off-target volume than the LCMV-E montages, with the relative decrease being $50\%$ or more in some cases. 

\subsection{Sea of neurons model studies}\label{sec:results:sea-of-neurons}
{Recently, Caldas-Martinez, Goswami et al.~\cite{Caldas-Martinez2024} used a ``sea of neurons'' model (see Fig.~\ref{fig:setup}) combining biophysically-realistic neuron models (taken from Aberra et al.~\cite{aberra2018biophysically}) with spherical head models to understand neural responses under temporal interference stimulation. This model places realistic neuron models at different locations in a region of interest to infer the locations where the neurons are directly stimulated under tES. We use this sea of neurons model to simulate the direct stimulation of excitatory pyramidal neurons in the Layer 5 of cortex under transcranial pulse stimulation, with square pulses of $200\mu\text{s}$. We chose the Layer 5 pyramidal neuron models as they were observed to have the lowest stimulation threshold~\cite{aberra2018biophysically}. We quantify the off-target stimulation using the number of off-target locations where the neuron spiked, denoted as $N_{act}$. Appendix D explains this model in greater detail.} 

We conducted five studies utilizing the sea of neurons model to compare the \HP and LCMV-E algorithms. The studies, described in sections~\ref{sec:results:sea-of-neurons:c3-c4} and~\ref{sec:results:sea-of-neurons:HD-tDCS}, emulate implementing the \HP and LCMV-E algorithms {experimentally in the following sense}. Many experimental tES studies establish an initial input-output curve to determine the injected current amplitudes required to generate neurological/physiological effects. Two standard electrode montages used in tES studies are the C3-C4 montage~\cite{szelenyi2007transcranial} and the $4\times1$ ring montage used in HD-tDCS~\cite{alam2016spatial} (see Appendix D for an illustration). We use these standard electrode montages to construct input-output curves for our sea of neurons model. The orientation and the magnitude of the desired target electric fields are determined from theses input-output curves (see Appendix D). We use the \HP and LCMV-E algorithms to induce the target field deduced from the input-output curves (ensuring target neural activation) while comparing the reduction in the off-target neural activation for different values of $I_{safe}$ and $I_{tot}^{mul}$. In Sec.~\ref{sec:results:sea-of-neurons:opt-dir}, we compare the performance of the \HP and LCMV-E algorithms under the assumption that the optimal direction (i.e., the direction requiring the least electric field amplitude to stimulate the neuron) and the corresponding electric field amplitude is \textit{a priori} known, e.g., through existing studies or neuron models. Finally, Sec.~\ref{sec:results:sea-of-neurons:neuron-type} and Sec.~\ref{sec:results:sea-of-neurons:diff-target} explore the effect of different neuron-type and different targets, respectively, on the performance of \textit{HingePlace}. 

\begin{figure*}[ht]
    \centering
   \includegraphics[width=\textwidth]{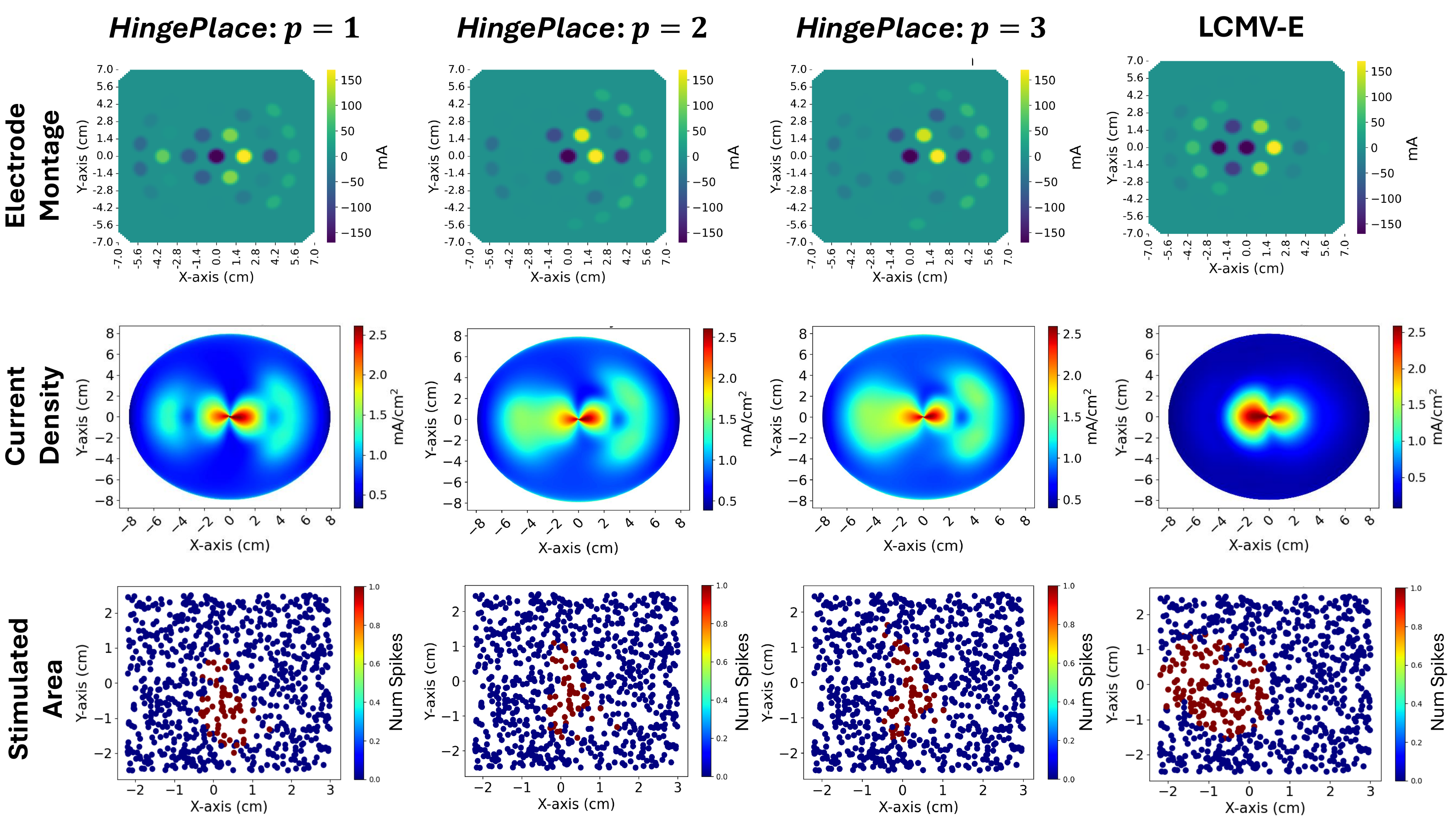}
    \caption{A representative plot of the electrode montage, the current density magnitude at a depth of $1$mm from the cortical surface (representing Layer 5), and the locations where the neurons fired in the Layer 5 of our sea of neurons model (denoted by the red dots) for the \HP and LCMV-E algorithms in Sec.~\ref{sec:results:sea-of-neurons:opt-dir}. The values of $I_{safe}$ and $I_{tot}^{mul}$ were $170$mA and $4$, respectively. We observe that despite LCMV-E producing the least off-target electric field, it produces the largest off-target neural activation, showing the importance of harnessing the thresholding behavior of neural response. Additional representative plots are provided in Appendix E.}
    \label{fig:neuron:visualisation}
\end{figure*}
\subsubsection{C3-C4 electrode montage study}\label{sec:results:sea-of-neurons:c3-c4}
Fig.~\ref{fig:neuron:results}a-c plot the relative decrease in the off-target stimulation between the \HP and LCMV-E montages when the input-output curve was constructed using the C3-C4 electrode montage. We measure the relative decrease as $(N_{act}^{LMCV-E}-N_{act}^{HP})/N_{act}^{LCMV-E}\times 100$, where $N_{act}^{HP}$ and $N_{act}^{LCMV-E}$ are off-target stimulation caused by the \HP and LCMV-E montages, respectively. Our results are consistent with Sec.~\ref{sec:results:mri}, showing that \HP montages cause significantly less off-target stimulation than the LCMV-E montages (as much as $75\%$ in some cases).

\textbf{Different directions require different tolerances}: We show the importance of choosing different tolerances $([E_{tol}^{x}$,$E_{tol}^{y}$,$E_{tol}^{z}])$ instead of a single tolerance ($E_{tol}$) for all directions (as is done in L1L1-norm optimization). Sec.~\ref{sec:algo:equiv-HP-L1L1} shows that all solutions of the L1L1-norm optimization can be obtained by the \HP algorithm for $p=1$ and the same tolerance value $(E_{tol})$ across all directions.\footnote{We replace the $\myVec{I}\preceq I_{safe}\myVec{1}_N$ by the $l_{\infty}$ norm constraint.} Consequently, we analyzed the performance of the L1L1-norm optimization by implementing the \HP algorithm with $p=1$ and the same tolerance ($E_{tol}$) across all directions. Fig.~\ref{fig:neuron:results}a-c plot the relative decrease in off-target stimulation achieved by the L1L1-norm optimization compared to the LCMV-E algorithm.  We observe that the performance of the L1L1-norm optimization suffers considerably due to assuming the same tolerance value across all directions. Hence, it is essential to consider different tolerances across different directions to harness the full potential of the \HP algorithm.
\subsubsection{$4\times1$ ring electrode montage study}\label{sec:results:sea-of-neurons:HD-tDCS}
Fig.~\ref{fig:neuron:results}d-f plot the relative decrease in the off-target neural activation between the \HP and LCMV-E montages, when input-output curve was constructed using the $4\times1$ ring montage. We observe much lower gains in the performance of the \HP algorithm compared to Sec.~\ref{sec:results:sea-of-neurons:HD-tDCS}, suggesting that the baseline pattern used for estimating $E_{des}$ plays an important role in the performance of the \HP algorithm.
\subsubsection{Optimal Direction study}\label{sec:results:sea-of-neurons:opt-dir}
Fig.~\ref{fig:neuron:results}g-i plot the relative decrease in the off-target neural activation between the \HP and LCMV-E montages when the optimal direction and amplitude of the target electric field is known \textit{a priori}. We choose the north pole at a depth $1$mm from the cortical surface (start of Layer 5) as our target point. The optimal direction and magnitude for our neuron model were $[\myminus0.75,\myminus0.41,0.51]$ and $70.27$Vm$^{-1}$ (based on the results in Appendix D). We observe the \HP montages have significantly lower off-target neural activation than the LCMV-E montages (as much as $65\%$ in some cases) for this study. Combining the results of sections~\ref{sec:results:sea-of-neurons:c3-c4},~\ref{sec:results:sea-of-neurons:HD-tDCS}, and~\ref{sec:results:sea-of-neurons:opt-dir}, we observe that choosing the target electric field's direction similar to the preferred direction of stimulation for the underlying neurons helps \HP perform better. Fig.~\ref{fig:neuron:visualisation} provides a visualization of the representative electrode montages, electric field magnitude, and the corresponding neural activation produce by the \HP and LCMV-E algorithms.

\subsubsection{Effect of neuron type and morphology}\label{sec:results:sea-of-neurons:neuron-type}
We also check the robustness of the \HP algorithm to different neuron types. We run the same study as Sec.~\ref{sec:results:sea-of-neurons:c3-c4} but change the neuron type from a layer 5 excitatory pyramidal neuron to a layer 2/3 excitatory pyramidal neuron. Fig.~\ref{fig:neuron:results}j plots the relative decrease between the off-target stimulation caused by the \HP and LCMV-E montages. The \HP montages continue to cause less off-target stimulation than their LCMV-E counterparts, showing that \HP is not sensitive to neuron type and morphology.

\subsubsection{Effect of different targets}\label{sec:results:sea-of-neurons:diff-target}
\begin{figure*}[htbp]
    \centering
    \includegraphics[width=0.9\textwidth]{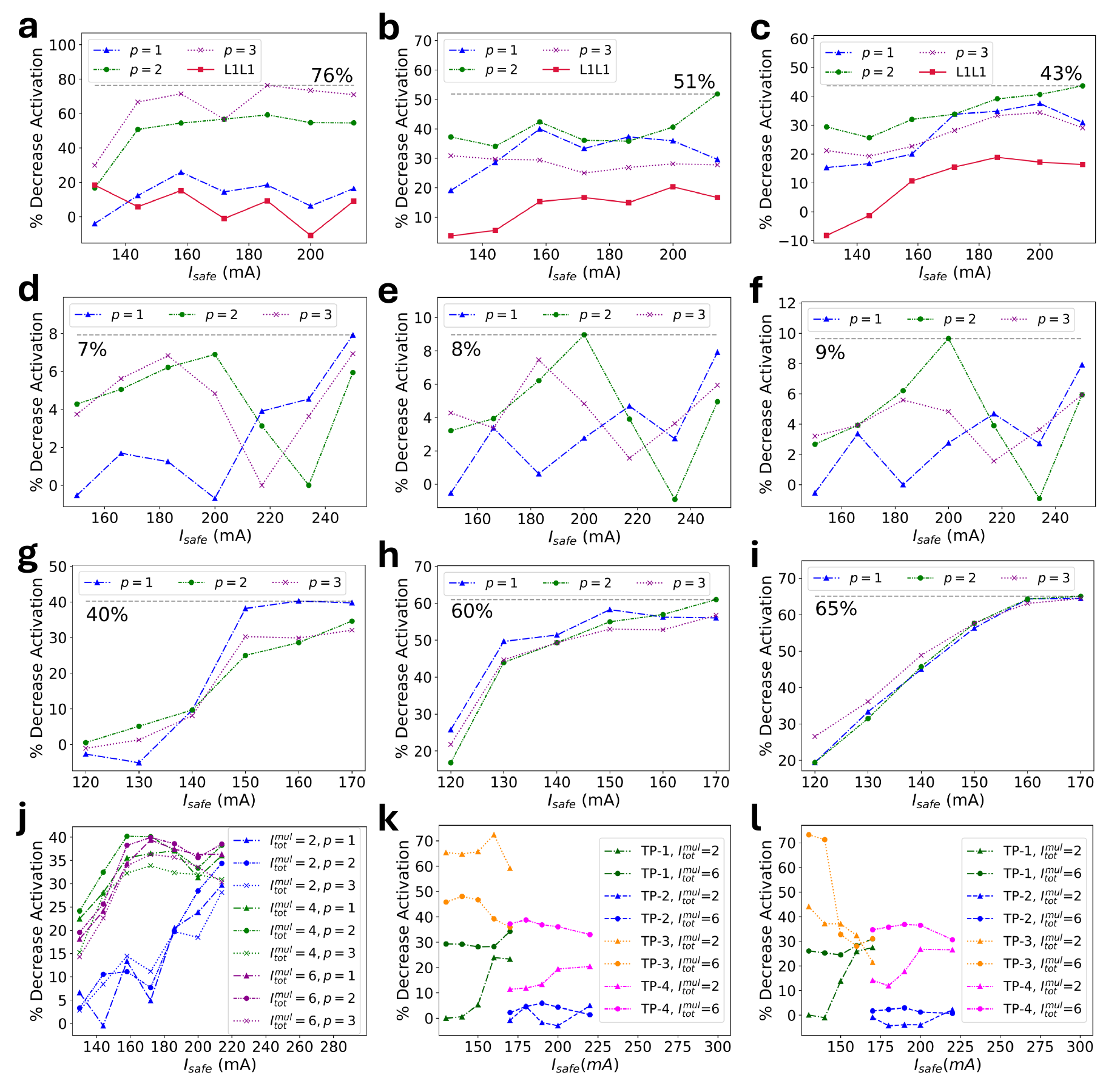}
    \caption{\textbf{a}-\textbf{c}, \textbf{d}-\textbf{f}, and \textbf{g}-\textbf{i} plot the relative decrease in the off-target stimulation between the \HP and LCMV-E montages for the simulation studies described in Sec.~\ref{sec:results:sea-of-neurons:c3-c4}, Sec.~\ref{sec:results:sea-of-neurons:HD-tDCS} and Sec.~\ref{sec:results:sea-of-neurons:opt-dir}, respectively. The groups of \{\textbf{a}, \textbf{d}, \textbf{g}\}, \{\textbf{b}, \textbf{e}, \textbf{h}\}, and \{\textbf{c}, \textbf{f}, \textbf{i}\} correspond to different values of $I_{tot}^{mul}$, namely, $2$, $4$, and $6$, respectively. Similarly, \textbf{j} plots the relative decrease in the off-target stimulation caused by the \HP and LCMV-E montages for Sec.~\ref{sec:results:sea-of-neurons:neuron-type}. \textbf{k} and \textbf{l} also plot the relative decrease in off-target stimulation between the \HP and LCMV-E montages for Sec.~\ref{sec:results:sea-of-neurons:diff-target} corresponding to $p=1$ and $p=2$ cases, respectively. TP refers to target point in \textbf{l} and \textbf{k}}
    \label{fig:neuron:results}
\end{figure*}

This study aimed to test the sensitivity of the \HP algorithm to different targets. We run the same study as Sec.~\ref{sec:results:sea-of-neurons:opt-dir}, but at four different target points. These target points are at the same depth as the target point of Sec.~\ref{sec:results:sea-of-neurons:opt-dir}. Target points 1, 2, 3, and 4 are $+2$cm (along $x$-axis), $+1$cm (along $y$-axis), $-2$cm (along $x$-axis), and $-1$cm (along the $y$-axis) from the target point of Sec.~\ref{sec:results:sea-of-neurons:opt-dir}, respectively. We only consider $p=1,2$ and $I_{tot}^{mul}=2,6$ for this study to reduce computation time. Fig.~\ref{fig:neuron:results}k and l plot the relative decrease between the off-target stimulation caused by the \HP and LCMV-E montages, respectively. Our results again show that \HP continues to outperform the LCMV-E algorithm. 
\section{Discussion and Conclusion}\label{sec:discussion}
We begin this section by briefly summarizing the main contributions of this work:
\begin{enumerate}
    \item We extend the unification result of Fern\'{a}ndez-Corazza et al.~\cite{fernandez2020unification} by theoretically establishing that LCMV-E and CDM approaches are essentially equivalent, thereby unifying the four major classes of electrode placement algorithms: WLS approaches, reciprocity-based approaches, CDM approaches, and LCMV-E approaches. We further show that magnitude maximization approaches proposed in~\cite{sadleir2012target,park2011novel} can be interpreted as a special case of the CDM approach. 
    \item We utilize our unification result to identify that traditional electrode placement algorithms over-penalize the off-target electric field and do not harness the non-linear thresholding behavior of neural response.
    \item We propose the \HP algorithm that uses a symmetrized hinge loss to better harness the non-linear thresholding behavior of neurons and consequently, generates more focal neural responses than the traditional electrode placement algorithm (validated through extensive stimulations).
\end{enumerate}

Traditional electrode placement algorithms focus on designing electrode montages that create a \emph{focal electric field}, which is related but not equivalent to the \emph{focal neural response}. In contrast, the \HP algorithm aims to \emph{directly} create neural responses by harnessing the underlying neural properties. Consequently, \HP suggests a new philosophy for designing electrode montages that aims to directly create focal neural responses instead of focal electric fields. This design philosophy provides an exciting research direction in which neural properties (other than the thresholding behavior) could be incorporated to improve the performance of electrode placement algorithms further. Note that \HP only utilizes the thresholding behavior of neurons to considerably improve the focality of stimulation (see Sec.~\ref{sec:results}), suggesting significant gains can be achieved by appropriately incorporating other neural properties.

Another advantage of the \HP algorithm is that it harnesses the thresholding behavior of neural response while ensuring that the underlying optimization problem remains convex. Therefore, \HP retains all the computational advantages of solving a convex optimization problem. The only extra cost that \HP incurs compared to existing electrode placement algorithms is the need to specify three additional hyperparameters $[\myVec{E}_{tol}^{x+},\myVec{E}_{tol}^{y+},\myVec{E}_{tol}^{z+}]$. These hyperparameters can be specified by using experimental or simulated neuron studies, such as~\cite{aberra2018biophysically}, or can be tuned empirically using a random search (utilizing only $30$ points in our study) in a similar fashion to our studies in Sec.~\ref{sec:results:sea-of-neurons:c3-c4} and Sec.~\ref{sec:results:sea-of-neurons:HD-tDCS} (see Appendix D). 

Although the main focus in this work is on tES applications, the \HP algorithm, similar to other existing electrode placement algorithms, can be utilized for other applications. For example, multi-electrode intracranial electrical stimulation using electrocorticography (ECoG), deep brain stimulation with stereo-EEG electrodes~\cite{guler2018computationally}, or high-frequency alternating electric fields, known as TTFields, to treat malignant tumors~\cite{miranda2014predicting}. Furthermore, our \HP algorithm could be potentially beneficial for optimizing multi-electrode transcranial temporal interference stimulation (tTIS). In tTIS, off-target electric fields tend to be higher than the target field but do not induce neural activation due to waveforms resembling high-frequency sinusoides~\cite{Caldas-Martinez2024}. This information can be easily encoded through the tolerance vectors $\myVec{E}_{tol}^{+}$ and $\myVec{E}_{tol}^{-}$. Another interesting result we obtained in this work was to reframe the magnitude maximization optimization as a bi-convex optimization formulation (see Appendix B), enabling us to utilize the vast literature on bi-convex optimization to solve the non-convex magnitude maximization approaches proposed in~\cite{sadleir2012target,park2011novel}.

\textbf{Limitations}: In this work, we did not consider some uncommon constraints, namely, the $l_2$-norm constraint proposed in~\cite{khan2022individually} or the constraint on the total number of active electrodes proposed in~\cite{saturnino2019accessibility,guler2016optimization}. The $l_2$-norm constraint was proposed in~\cite{khan2022individually} to reduce scalp pain during non-invasive stimulation. We can easily incorporate the $l_2$-norm constraint in our optimization formulation. Since the L1L1-norm optimization proposed in~\cite{prieto2022l1} is equivalent to a special case of \HP algorithm, we can easily use the meta-heuristic method proposed in~\cite{prieto2022l1} to constraint the number of active electrodes for \HP algorithm too. The branch and bound method proposed in~\cite{saturnino2019accessibility} to constrain the number of active electrodes can also be used for the \HP algorithm. We do not explore the role of the network under transcranial stimulation, as constructing a realistic network with biophysically realistic neurons would be too computationally expensive. The role of an underlying network under transcranial stimulation is not well understood and deserves a more targeted study. 

The sea of neurons model utilizes the spherical head model instead of the MRI head model due to the requirement of very high-resolution electric field simulation for modeling extracellular stimulation. Note that spherical head models simulate much higher-resolution electric fields (due to having analytical solutions) than the MRI head models. Combining MRI head models with bio-physical neuron models in a computationally tractable manner is an important yet challenging goal we aim to pursue in subsequent studies. While simulation studies have been the primary tool to validate electrode placement algorithms ~\cite{im2008determination,park2011novel,dmochowski2011optimized,sadleir2012target,dutta2013using,ruffini2014optimization,guler2016optimization,salman2016concurrency,cancelli2016simple,wagner2016optimization,fernandez2016transcranial,dmochowski2017optimal,guler2018computationally,saturnino2019accessibility,khan2022individually,prieto2022l1,wang2023multi}, experimentally validating the \HP algorithm is important. We aim to pursue experimental validation of \HP in subsequent studies. 

We conclude by stating that designing computational techniques that appropriately harness different neural behaviors is an exciting research direction and has the potential to increase the efficacy of non-invasive stimulation significantly by reducing off-target neural activation. 

\section*{Funding Sources}
This material is based upon work supported by the Naval Information Warfare Center (NIWC) Atlantic and the Defense Advanced Research Projects Agency (DARPA) under Contract No. N65236-19-C-8017. Any opinions, findings and conclusions or recommendations expressed in this material are those of the authors and do not necessarily reflect the views of the NIWC Atlantic and DARPA. Chaitanya Goswami was also supported by the Center for Machine Learning and Health Fellowship.

\section*{Declaration of Competing Interests}
Chaitanya Goswami and Pulkit Grover have a patent (US20240278011A1) for the \HP algorithm issued to Carnegie Mellon University. 

\section*{Acknowledgment}
The authors would like to thank Dr. Praveen Venkatesh, Dr. Mats Forssell, and Dr. Vishal Jain for helpful discussions. The authors would also like to thank Amanda Merkley, Yuxin Guo, and Jeehyun Kim for proof-reading some sections of the manuscript.

\section*{Code Availability}
All the related code for reproducing the results provided in this manuscript can be found at \texttt{https://github.com/chaitanyagoswami/HingePlace-Code}.


\begin{table}[ht]
\normalsize
    \centering
    \begin{tabular}{|c|c|}
    \hline
         Notation & Explanation \\
         \hline
         \hline
         Bold-font & Used for denoting vectors and matrices.\\
         \hline
        $\R$ & The set of all real numbers.\\
        \hline
        \multirow{2}{1.5em}{$\R^{d}$} & The standard $d$-dimensional vector spaces constructed\\ 
        &using the $\R$.\\
        \hline
        \multirow{2}{3em}{$\R^{m\times n}$} & The standard vector space of matrices of size $m\times n$ \\
        &whose entries are real numbers.\\
        \hline
        $\R_{+}$ & The set of all non-negative real numbers.\\
        \hline
        \multirow{2}{1.5em}{$\R^{d}_{+}$} & The standard $d$-dimensional vector spaces constructed\\ 
        &using the $\R_+$.\\
        \hline
       \multirow{2}{3em}{$\R^{m\times n}_+$} & The standard vector space of matrices of size $m\times n$ whose\\ 
        &entries are non-negative real numbers.\\
        \hline
        $\N$ & The set of all natural numbers.\\
        \hline
        \multirow{2}{1.5em}{$\N^{d}$} & The standard $d$-dimensional vector spaces constructed\\ 
        &using the $\N$.\\
        \hline
        \multirow{2}{3em}{$\N^{m\times n}$} & The standard vector space of matrices of size $m\times n$\\ 
        &whose entries are natural numbers.\\
        \hline
        $\myVec{1}_d$ & The $d$-dimensional vector having all elements as 1.\\
        \hline
        $\myVec{0}_d$ & The $d$-dimensional vector having all elements as 0.\\
        \hline
        $\myVec{e}_{j}$ & The $j$-th standard vector of $\R^N$ having the $j$-th element\\ 
        &as one and the rest as zero.\\
        \hline
        \multirow{2}{4em}{$\prec$, $\succ$, $\preceq$, $\succeq$} & Denote element-wise inequalities between\\ 
        &vectors and matrices.\\ 
        \hline
        $\|\myVec{x}\|_p$ & The $l_p$-norm of $\myVec{x}$ for $p\geq 1$, defined as $\|\myVec{x}\|_p\coloneqq  \left(\sum_{i=1}^d x_i^p\right)^{\nicefrac{1}{p}}$.\\
        \hline
    \end{tabular}
    \caption{The table containing explanation of the mathematical notations used in this work.}
    \label{tab:math-not}
\end{table}

\newpage
\appendix
\section{}\label{appx:A}
In this appendix, we discuss the mathematical proof of Theorem~\ref{theorem:CDM-LCMV-equiv}. We have divided the section into six parts describing the problem, discussing an informal proof sketch, stating the required background, providing the formal proof, a numerical illustration of the proof, and the accompanying lemmas used in the technical arguments of the proof. 
\subsection{Problem Description}
The goal of this appendix is to show that the CDM optimization formulation stated in~\eqref{eq:cdm:update-not} and the LCMV-E optimization formulation stated in~\eqref{eq:lcmv-e} are essentially equivalent. We re-state the CDM optimization formulation and the LCMV-E optimization formulation here for convenience:
\begin{align}
   \text{\underline{\textbf{CDM}} : } \myVec{I}_1^*=&\argmax_{\myVec{I}\in\R^N}\myVec{A}_f\myVec{I},\nonumber\\
   &\text{s.t. }\left\|\myVec{A}_c\myVec{I}\right\|_2^2\leq \alpha,\|\myVec{I}\|_1\leq I_{tot},\nonumber\\
   &\|\myVec{I}\|_{\infty}\leq I_{safe},\text{ and }\myVec{1}_N^T\myVec{I}=0.\label{eq:appx:a:CDM}\\
    \text{\underline{\textbf{LCMV-E}} : } \myVec{I}_2^*=&\argmin_{\myVec{I}\in\R^N}\|\myVec{A}_c\myVec{I}\|_2^2,\nonumber\\
   &\text{s.t. }\myVec{A}_f\myVec{I}=E_{des},\|\myVec{I}\|_1\leq I_{tot},\nonumber\\
   &\|\myVec{I}\|_{\infty}\leq I_{safe},\text{ and }\myVec{1}_N^T\myVec{I}=0.\label{eq:appx:a:CDM-alt}
\end{align}
We provide a mathematical proof accompanied by a numerical illustration showing that the CDM optimization problem stated in~\eqref{eq:appx:a:CDM} and the LCMV-E optimization problem (LCMV-E) stated in~\eqref{eq:appx:a:CDM-alt} are essentially equivalent.
\subsection{Informal Description of the Proof}
To prove the equivalence of CDM and LCMV-E optimization problems, we will show that for each non-zero value of $\alpha$, we can always find a corresponding value of $E_{des}$ such that the optimum obtained by solving the LCMV-E optimization problem, denoted as $\myVec{I}^*_{2}$, is a solution to the CDM optimization problem. Furthermore, we show that if the value of $\alpha$ is below a certain threshold (denoted as $\alpha_{MAX}$), $\myVec{I}_2^*$ is a unique solution of the CDM optimization problem. For $\alpha$ values greater than $\alpha_{MAX}$, the CDM optimization problem may not have a unique solution. In this case, $\myVec{I}_2^*$ is the solution of the CDM optimization with the least off-target electric field energy (represented by the term $\|\myVec{A}_{c}\myVec{I}\|_2^2$ for the electrode montage $\myVec{I}$). Note that other solutions of the CDM optimization problem for this case have larger values of off-target electric field energy without providing higher intensity at the target. Hence, $\myVec{I}_2^*$ is the most ``relevant'' solution of the CDM optimization problem for this case.

The proof structure is divided into two cases: $\alpha<\alpha_{MAX}$ and $\alpha\geq\alpha_{MAX}$. For $\alpha<\alpha_{MAX}$, we use the Karush–Kuhn–Tucker (K.K.T.) conditions~\cite{boyd2004convex} to show that $\myVec{I}_1^*$ solves the K.K.T. conditions of the LCMV-E optimization for appropriately chosen $E_{des}$. Utilizing the fact that LCMV-E has a unique solution, we argue that $\myVec{I}_1^*$ (also being a solution of LCMV-E optimization problem) must be unique. Consequently, both LCMV-E and CDM must have the same unique optimums. For $\alpha\geq\alpha_{MAX}$, we utilize the fact that the solution of the CDM optimization problem with the least off-target electric field energy is also the solution of a simpler directional intensity maximization form. Then, we show that $\myVec{I}_2^*$ (for appropriately chosen $E_{des}$) is equal to this least off-target electroc field energy solution of the simpler directional intensity maximization problem, concluding our proof.
\subsection{Background} 
The main argument of our proof requires some background before we can precisely state it. Hence, we begin our proof by covering the required background.
\subsubsection{Directional Intensity Maximization Framework}
Our proof makes key use of a directional intensity maximization problem and its solution $I_{ME}^*$ having the least off-target electric energy. We re-state the directional intensity maximization problem in~\eqref{eq:appx:a:CDM:max-intensity} (earlier stated in~\eqref{eq:max-intenstity}) and define $I_{ME}^*$ using~\eqref{eq:appx:a:I_ME:defn} (earlier stated in).
\begin{align}
    \myVec{I}^*_3 =& \argmin_{\myVec{I}\in\R^N}-\myVec{A}_f\myVec{I},\text{ s.t. }\|\myVec{I}\|_{\infty}\leq I_{safe},\|\myVec{I}\|_{1}\leq I_{tot},\text{ and }\myVec{1}_N^T\myVec{I} = 0.\label{eq:appx:a:CDM:max-intensity}\\
    \myVec{I}_{ME}^* =& \argmin_{\myVec{I}_3^*\in\mathcal{I}^*}\|\myVec{A}_c\myVec{I}_3^*\|_2^2,\label{eq:appx:a:I_ME:defn}
\end{align}
where $\mathcal{I}^*$ is the set of all solutions to~\eqref{eq:appx:a:CDM:max-intensity}. Furthermore, we re-state the definition of $\alpha_{MAX}$ for convenience as follows:
\begin{align}
    \alpha_{MAX} = \|\myVec{A}_c\myVec{I}_{ME}^*\|_2^2.\label{eq:appx:alpha_max-defn}
\end{align}
\subsubsection{$\textit{l}_1$ and $\textit{l}_{\infty}$ constraint alternative formulation}
 The $l_1$ and $l_{\infty}$ constraints of~\eqref{eq:appx:a:CDM} and~\eqref{eq:appx:a:CDM-alt} can be alternatively formulated in terms of affine inequalities which makes certain parts of proof easier to analyze. Hence, we briefly discuss this alternative formulation of $l_1$ and $l_{\infty}$ constraint.
 
\noindent\textbf{$\mathbf{L_1}$ constraint alternative formulation}: We note that $l_1$ constraint on a $N$ dimensional variable can be represented using $2^{N}$ linear inequalities (see Appendix of ~\cite{fernandez2020unification}). We provide an example for representing $l_1$ constraint for a 3-dimensional vector below:, 
\begin{align}
    &\|\myVec{I}\|_1\leq 2I_{tot},\nonumber\\
    \Rightarrow &\sum_{j=1}^3|i_j|\leq 2I_{tot},\nonumber\\
    \Rightarrow & \begin{bmatrix} 1&1&1\\
    1&1&-1\\
    1&-1&1\\
    1&-1&-1\\
    -1&1&1\\
    -1&1&-1\\
    -1&-1&1\\
    -1&-1&-1\\\end{bmatrix}\ \begin{bmatrix}i_1\\i_2\\i_3\end{bmatrix}\preceq \begin{bmatrix}2I_{tot}\\2I_{tot}\\2I_{tot}\end{bmatrix},\nonumber\\
    \Rightarrow &\myVec{G}\myVec{I} \preceq 2I_{tot}\1_3.
\end{align}
A similar $\myVec{G}$ can be constructed for the $N$-dimensional case. We denote the $i$-th row of $\myVec{G}$ as $\myVec{g}_i^T$.

\noindent\textbf{$\mathbf{L_{\infty}}$ constraint alternative formulation}: The $l_{\infty}$ constraint on a $N$-dimensional variable can be represented using $2N$ inequalities:
\begin{align}
    &\|\myVec{I}\|_{\infty}\leq I_{safe}\Rightarrow \myVec{I}\preceq I_{safe}\myVec{1}_N \text{ and } \myVec{I}\succeq -I_{safe}\myVec{1}_N.
\end{align}
\subsubsection{Assumptions and Properties used in the proof}
We state four key properties and assumptions that are required for the technical arguments stated later in the proof.
\begin{enumerate}[label={\bfseries (P\arabic*)}]
    \item The CDM optimization problem stated in~\eqref{eq:appx:a:CDM}, the directional-intensity optimization problem stated in~\eqref{eq:appx:a:CDM:max-intensity} and the LCMV-E optimization problem stated in~\eqref{eq:appx:a:CDM-alt} are convex optimization problems. We can easily verify this claim by noticing that the objectives of these optimization problems are convex functions and the corresponding constraint sets are convex sets.
    \item We will assume that $\myVec{A}_c$ is a full-rank matrix. This is a reasonable assumption as typically $\myVec{A}_{c}$ is $C\times N$ matrix where $C\gg N$ (typically $N\sim 100$ whereas the number of off-target voxels are typically greater than $10000$). Hence, for most realistic head models, $\myVec{A}_c$ turns out to be a full-rank matrix. Under the assumption $\myVec{A}_c$ is a full-rank matrix, we can show that the objective of~\eqref{eq:appx:a:CDM-alt} is strictly convex and consequently, the optimizing $\myVec{I}^*_2$ for LCMV-E is unique~\cite{boyd2004convex}.
    \item We also assume that $I_{safe}$, $\alpha$, and $I_{tot}$ are strictly greater than zero. Note that if either $I_{safe}=0$ or $I_{tot}=0$, then the constraint sets of CDM and LCMV-E reduce to a singleton set consisting of the vector $\myVec{0}_N$. Similarly, under the assumption of full-rank $\myVec{A}_c$, $\alpha=0\Rightarrow \myVec{A}_c\myVec{I}=\myVec{0}_{|\mathcal{C}|}\Rightarrow \myVec{I}=\myVec{0}_N$, which again implies the constraint set for CDM is a singleton set consisting of the vector $\myVec{0}_N$. Consequently, if $\alpha=0$, $I_{safe}=0$, or $I_{tot}=0$, then the solution of the CDM optimization problem is $\myVec{0}_N$, which is not a practical/useful case for analysis.
    \item Under the assumption of $\alpha>0$, $I_{safe}>0$, and $I_{tot}>0$, the CDM optimization problem and the directional intensity maximization problem satisfy Slater's condition~\cite{boyd2004convex} as $\myVec{0}_N$ is an interior point of both CDM's and directional intensity maximization's constraint set. We can easily verify the validity of $\myVec{0}_N$ as an interior point since $\|\myVec{0}_N\|_1<I_{tot}$, $\|\myVec{0}_N\|_{\infty}<I_{safe}$, $\myVec{1}_N^T\myVec{0}_N=0$, and $\|\myVec{A}_c\myVec{0}_N\|_2^2<\alpha$. Hence, strong duality holds for~\eqref{eq:appx:a:CDM} and~\eqref{eq:appx:a:CDM:max-intensity} and we can use the Lagrangian dual and K.K.T.conditions to find the optimum of~\eqref{eq:appx:a:CDM} and~\eqref{eq:appx:a:CDM:max-intensity}.
\end{enumerate}

\subsection{Proof}
Our proof is divided into two cases depending on the value chosen for $\alpha$, namely, $\alpha<\alpha_{MAX}$ and $\alpha\geq\alpha_{MAX}$. Before proceeding to the main arguments of the proof, we briefly state the canonical form, the Lagrangian, the dual formulation and the corresponding K.K.T. conditions of~\eqref{eq:appx:a:CDM},~\eqref{eq:appx:a:CDM-alt}, and~\eqref{eq:appx:a:CDM:max-intensity} to ease the Lagrangian analysis we extensively use in this proof.

\subsubsection*{CDM Lagrangian analysis preliminaries}
We specify the canonical form, the Lagrangian, and dual formulation of~\eqref{eq:appx:a:CDM} in~\eqref{eq:appx:a:CDM:canon},~\eqref{eq:appx:a:CDM:lagrangian}, and~\eqref{eq:1}, respectively. We also appropriately replace the $l_1$-norm and $l_{\infty}$-norm constraints with linear inequalities to ease our analysis.\\
\textbf{Canonical Form of~\eqref{eq:appx:a:CDM}}:
\begin{align}
      \myVec{I}_1^{*} =& \argmin_{\myVec{I}\in\R^N} -\myVec{A}_f\myVec{I},\text{ s.t. } \|\myVec{A}_c\myVec{I}\|_2^2-\alpha\leq 0, \myVec{1}_N^T\myVec{I} = 0,\myVec{G}\myVec{I}-2I_{tot}\1_{2^N}\preceq\myVec{0}_{2^N},\nonumber\\ &\myVec{I}-I_{safe}\1_N\preceq\myVec{0}_N,\text{ and}-\myVec{I}-I_{safe}\1_N\preceq\myVec{0}_N.\label{eq:appx:a:CDM:canon}
\end{align}
\textbf{Lagrangian of~\eqref{eq:appx:a:CDM}:}
\begin{align}
    \mathcal{L}^{(1)}(\myVec{I}) =& -\myVec{A}_f\myVec{I}+\lambda^{(1)}\left(\|\myVec{A}_c\myVec{I}\|_2^2-\alpha\right)+\mu^{(1)}(\1_N^T\myVec{I})+\sum_{j=1}^{2^N} \delta_j^{(1)}(\myVec{g}_j^T\myVec{I}-2I_{tot})\nonumber\\&+\sum_{j=1}^{N} \nu_j^{(1)}(\myVec{e}_j^T\myVec{I}-I_{safe})+\sum_{j=1}^{N}\kappa_j^{(1)}(-\myVec{e}_j^T\myVec{I}-I_{safe}).\label{eq:appx:a:CDM:lagrangian}
\end{align}
\textbf{Dual of~\eqref{eq:appx:a:CDM}:}
\begin{align}
    \left\{\begin{array}{c}
     \myVec{I}_1^*,\lambda^{(1)*},\\
         \{\delta_j^{(1)*}\}_{j=1}^{2^N},\\ 
         \mu^{(1)*}\\
         \{\nu_j^{(1)*}\}_{j=1}^N,\\\{\kappa_j^{(1)*}\}_{j=1}^N 
\end{array}\right\}\myeq&\argmax_{\substack{\lambda^{(1)}\geq0,\mu^{(1)}\in\mathbb{R},\\\{\delta_j^{(1)}\}_{j=1}^{2^N}\geq0,\\ \{\nu_j^{(1)}\}_{j=1}^N\geq0\\\{\kappa_j^{(1)}\}_{j=1}^N\geq0}}\argmin_{\myVec{I}\in\mathbb{R}^N}\mathcal{L}^{(1)}(\myVec{I}).\label{eq:1}
\end{align}
By \textbf{(P4)}, strong duality holds for~\eqref{eq:appx:a:CDM}. Hence, all solutions of~\eqref{eq:1} satisfy their respective K.K.T. conditions~\cite{boyd2004convex} specified in~\eqref{1}-\eqref{5}.\\

\noindent\textbf{K.K.T. conditions of~\eqref{eq:appx:a:CDM}:}
\begin{align}
    &\left.\frac{\partial \mathcal{L}^{(1)}(\myVec{I})}{\partial \myVec{I}}\right|_{\myVec{I}=\myVec{I}_1^*} = -\myVec{A}_f^T+2\lambda^{(1)*}\myVec{A}_f^T\myVec{A}_f\myVec{I}_1^* +\mu^{(1)*}\1_N+\sum_{j=1}^{2^N} \delta_j^{(1)*}\myVec{g}_j+\nonumber\\&\sum_{j=1}^{N} (\nu_j^{(1)*}-\kappa_j^{(1)*})\myVec{e}_j = 0.\label{1}\\
    &\lambda^{(1)*}(\|\myVec{A}_c\myVec{I}_1^*\|_2^2-\alpha) = 0.\label{eq:appx:a:CDM-kk1:l2}\\
    &\1_N^T\myVec{I}_1^* = 0.\label{2}\\
    &\delta^{(1)*}(\myVec{g}_j^T\myVec{I}_1^*-2I_{tot}) = 0\ \forall\ j\in\{1,\hdots,2^N\}.\label{3}\\
    &\nu^{(1)*}_j(\myVec{e}_j^T\myVec{I}_1^*-I_{safe}) = 0\ \forall\ j\in\{1,\hdots,N\}.\label{4}\\
    &\kappa^{(1)*}_j(-\myVec{e}_j^T\myVec{I}_1^*-I_{safe}) = 0\ \forall\ j\in\{1,\hdots,N\}.\label{5}
\end{align}

\subsubsection*{LCMV-E Lagrangian analysis preliminaries}
We specify the canonical form, the Lagrangian, and dual formulation of~\eqref{eq:appx:a:CDM-alt} in~\eqref{eq:appx:a:CDM-alt:canon},~\eqref{eq:2}, and~\eqref{eq:3}, respectively. We also appropriately replace the $l_1$-norm and $l_{\infty}$-norm constraints with linear inequalities to ease our analysis.\\
\textbf{Canonical Form of~\eqref{eq:appx:a:CDM-alt}}:
\begin{align}
\myVec{I}_2^{*} =& \argmin_{\myVec{I}\in\R^N}  \|\myVec{A}_c\myVec{I}\|_2^2,\text{ s.t. }\myVec{A}_f\myVec{I}-E_{des}= 0,\ \ \myVec{1}_N^T\myVec{I} = 0,\nonumber\\
&\myVec{G}\myVec{I}-2I_{tot}\1_{2^N}\preceq\myVec{0}_{2^N},\ \  \myVec{I}-I_{safe}\1_N\preceq\myVec{0}_N, \nonumber\\
&\text{ and }-\myVec{I}-I_{safe}\1_N\preceq\myVec{0}_N.\label{eq:appx:a:CDM-alt:canon}
\end{align}
\textbf{Lagrangian of~\eqref{eq:appx:a:CDM-alt}:}
\begin{align}
    \mathcal{L}^{(2)}(\myVec{I}) =& \|\myVec{A}_c\myVec{I}\|_2^2+\beta^{(2)}\left(\myVec{A}_f\myVec{I}-E_{des}\right)+\mu^{(2)}(\myVec{1}_N^T\myVec{I})+\sum_{j=1}^{2^N} \delta_j^{(2)}(\myVec{g}_j^T\myVec{I}-2I_{tot})\nonumber\\&+\sum_{j=1}^{N} \nu_j^{(2)}(\myVec{e}_j^T\myVec{I}-I_{safe})+\sum_{j=1}^{N}\kappa_j^{(2)}(-\myVec{e}_j^T\myVec{I}-I_{safe}).\label{eq:2}
\end{align}
\textbf{Dual of~\eqref{eq:appx:a:CDM-alt}:}
\begin{align}
    \left\{\begin{array}{c}
     \myVec{I}_2^*,\beta^{(2)*},\mu^{(2)*}\\
         \{\delta_j^{(2)*}\}_{j=1}^{2^N},\\ 
         \{\nu_j^{(2)*}\}_{j=1}^N,\\\{\kappa_j^{(2)*}\}_{j=1}^N 
\end{array}\right\}\myeq\argmax_{\substack{\beta^2,\mu^{(1)}\in\mathbb{R},\\\{\delta_j^{(2)}\}_{j=1}^{2^N}\geq0,\\ \{\nu_j^{(2)},\kappa_j^{(2)}\}\geq0}}\argmin_{\myVec{I}\in\mathbb{R}^N}\mathcal{L}^{(2)}(\myVec{I}).\label{eq:3}
\end{align}
Since all the inequality constraints inf~\eqref{eq:appx:a:CDM-alt:canon} are affine,~\eqref{eq:appx:a:CDM-alt:canon} satisfies relaxed Slater's condition, as long as, its constraint set is not empty. We will show later on in the proof that our choice of $E_{des}$ guarantees that the constraint set of LCMV-E is not empty. Therefore, the solution of~\eqref{eq:3} must satisfy its K.K.T. conditions specified in~\eqref{eq:4}-\eqref{eq:5}.\\

\noindent\textbf{K.K.T. conditions of~\eqref{eq:appx:a:CDM-alt}:}
\begin{align}
    &\left.\frac{\partial \mathcal{L}^{(2)}(\myVec{I})}{\partial \myVec{I}}\right|_{\myVec{I}=\myVec{I}_2^*} = \beta^{(2)*}\myVec{A}_f^T+2\myVec{A}_f^T\myVec{A}_f\myVec{I}_2^{*} +\mu^{(2)*}\myVec{1}_N+\sum_{j=1}^{2^N} \delta_j^{(2)*}\myVec{g}_j+\nonumber\\&\sum_{j=1}^{N} (\nu_j^{(2)*}-\kappa_j^{(2)*})\myVec{e}_j = 0.\label{eq:4}\\
    &\myVec{A}_f\myVec{I}_2^*=E_{des}.\\
    &\myVec{1}_N^T\myVec{I}_2^* = 0.\\
    &\delta^{(2)*}(\myVec{g}_j^T\myVec{I}_2^*-2I_{tot}) = 0\ \forall\ j\in\{1,\hdots,2^N\}\\
    &\nu^{(2)*}_j(\myVec{e}_j^T\myVec{I}_2^*-I_{safe}) = 0\ \forall\ j\in\{1,\hdots,N\}\\
    &\kappa^{(2)*}_j(-\myVec{e}_j^T\myVec{I}_2^*-I_{safe}) = 0\ \forall\ j\in\{1,\hdots,N\}\label{eq:5}
\end{align}
\subsubsection*{Directional Intensity Maximization Lagrangian Preliminaries} We specify the canonical form, the Lagrangian, and dual formulation of~\eqref{eq:appx:a:CDM:max-intensity} in~\eqref{eq:appx:a:CDM-max-intensity:canon},~\eqref{eq:appx:a:CDM-max-intensity:lagrangian}, and~\eqref{eq:appx:a:CDM-max-intensity:dual}, respectively:\\
\textbf{Canonical Form of~\eqref{eq:appx:a:CDM:max-intensity}}:
\begin{align}
      \myVec{I}_3^{*} =& \argmin_{\myVec{I}\in\R^N}  -\myVec{A}_f\myVec{I},\text{ s.t. } \myVec{1}_N^T\myVec{I} = 0,\myVec{G}\myVec{I}-2I_{tot}\1_{2^N}\preceq\myVec{0}_{2^N},\myVec{I}-I_{safe}\1_N\preceq\myVec{0}_N,\nonumber\\&\text{ and }-\myVec{I}-I_{safe}\1_N\preceq\myVec{0}_N.\label{eq:appx:a:CDM-max-intensity:canon}
\end{align}
\textbf{Lagrangian of~\eqref{eq:appx:a:CDM-max-intensity:canon}:}
\begin{align}
   \mathcal{L}^{(3)}(\myVec{I}) =& -\myVec{A}_f\myVec{I}+\mu^{(3)}(\1_N^T\myVec{I})+\sum_{j=1}^{2^N} \delta_j^{(3)}(\myVec{g}_j^T\myVec{I}-2I_{tot})+\sum_{j=1}^{N} \nu_j^{(3)}(\myVec{e}_j^T\myVec{I}\myminus I_{safe})\nonumber\\&+\sum_{j=1}^{N}\kappa_j^{(3)}(\myminus\myVec{e}_j^T\myVec{I}\myminus I_{safe}).\label{eq:appx:a:CDM-max-intensity:lagrangian}
\end{align}
\textbf{Dual of~\eqref{eq:appx:a:CDM-max-intensity:canon}:}
\begin{align}
    \left\{\begin{array}{c}
     \myVec{I}_3^*,\\
         \{\delta_j^{(3)*}\}_{j=1}^{2^N},\\ 
         \mu^{(3)*}\\
         \{\nu_j^{(3)*}\}_{j=1}^N,\\\{\kappa_j^{(3)*}\}_{j=1}^N 
\end{array}\right\}\myeq&\argmax_{\substack{\mu^{(3)}\in\mathbb{R},\\\{\delta_j^{(3)}\}_{j=1}^{2^N}\geq0,\\ \{\nu_j^{(3)},\kappa_j^{(3)}\}_{j=1}^N\geq0}}\argmin_{\myVec{I}\in\mathbb{R}^N}\mathcal{L}^{(3)}(\myVec{I}).\label{eq:appx:a:CDM-max-intensity:dual}
\end{align}
By \textbf{(P4)}, strong duality holds for~\eqref{eq:appx:a:CDM:max-intensity}. Hence, all solutions of~\eqref{eq:appx:a:CDM-max-intensity:dual} satisfy the K.K.T. conditions specified in~\eqref{eq:kkt:CDM-max:1}-\eqref{eq:kkt:CDM-max:5}.\\

\noindent\textbf{K.K.T. conditions of~\eqref{eq:appx:a:CDM-max-intensity:canon}:}
\begin{align}
    &\left.\frac{\partial \mathcal{L}^{(3)}(\myVec{I})}{\partial \myVec{I}}\right|_{\myVec{I}=\myVec{I}_3^*} = -\myVec{A}_f^T \myplus\mu^{(3)*}\1_N\myplus\sum_{j=1}^{2^N} \delta_j^{(3)*}\myVec{g}_j\myplus\sum_{j=1}^{N} (\nu_j^{(3)*}\myminus\kappa_j^{(3)*})\myVec{e}_j = 0.\label{eq:kkt:CDM-max:1}\\
    &\1_N^T\myVec{I}_3^* = 0.\label{eq:kkt:CDM-max:2}\\
    &\delta^{(3)*}(\myVec{g}_j^T\myVec{I}_3^*-2I_{tot}) = 0\ \forall\ j\in\{1,\hdots,2^N\}.\label{eq:kkt:CDM-max:3}\\
    &\nu^{(3)*}_j(\myVec{e}_j^T\myVec{I}_3^*-I_{safe}) = 0\ \forall\ j\in\{1,\hdots,N\}.\label{eq:kkt:CDM-max:4}\\
    &\kappa^{(3)*}_j(-\myVec{e}_j^T\myVec{I}_3^*-I_{safe}) = 0\ \forall\ j\in\{1,\hdots,N\}.\label{eq:kkt:CDM-max:5}
\end{align}
\subsubsection*{\textbf{Case 1}: $\alpha<\alpha_{MAX}$}
For showing the equivalence between LCMV-E and CDM in this case, we first need to choose an appropriate value of $E_{des}$ such that the solutions of both LCMV-E and CDM optimization problems are same. We choose $E_{des}= \myVec{A}_f\myVec{I}_1^*$, i.e., the optimal directional current intensity we found by solving~\eqref{eq:appx:a:CDM}. Note that $\|\myVec{I}_1^*\|_{\infty}\leq I_{safe}$, $\|\myVec{I}_1^*\|_1\leq I_{tot}$, and $\myVec{1}_N^T\myVec{I}_1^*=0$ due to $\myVec{I}_1^*$ being a solution of~\eqref{eq:appx:a:CDM}. Furthermore, we can see that $\myVec{I}_{1}^*$ trivially satisfies the constraint $\myVec{A}_f\myVec{I}_1^*=E_{des}$ as $E_{des}=\myVec{A}_f\myVec{I}_1^*$. Hence, we conclude that the constraint set of~\eqref{eq:appx:a:CDM-alt} is not empty as $\myVec{I}_1^*$ must lie in the constraint set of~\eqref{eq:appx:a:CDM-alt}. Hence,~\eqref{eq:appx:a:CDM-alt} satisfies the relaxed Slater's condition and the solution of~\eqref{eq:3} must satisfy its K.K.T. conditions stated in~\eqref{eq:4}-\eqref{eq:5}. 

We prove the equivalence between CDM and LCMV-E by showing that all solutions $(\myVec{I}_1^*)$ of the CDM optimization problem (which necessarily satisfy their K.K.T. conditions in~\eqref{1}-\eqref{5}) also are a solution of the K.K.T. system of LCMV-E stated in~\eqref{eq:4}-\eqref{eq:5} (for $E_{des}=\myVec{A}_f\myVec{I}_1^*$). Hence, all solutions of the CDM optimization problem must also be the solution of the LCMV-E optimization problem with appropriately chosen $E_{des}$. Note that by lemma~\ref{prop:2}, we know that all solutions of the CDM's K.K.T. system will have $\lambda^{(1)*}\neq 0$ for this case as $\alpha<\alpha_{MAX}$. Consider the following substitution in the equations \eqref{eq:4}-\eqref{eq:5}:
\begin{align}
 &\myVec{I}_2^*=\myVec{I}_1^*,\label{11}\\
 &\beta^{(2)*}=-\frac{1}{\lambda^{(1)*}}, \mu^{(2)*}=\frac{\mu^{(1)*}}{\lambda^{(1)*}}\label{12} \\
 &\delta^{(2)*}_j = \frac{\delta^{(1)*}}{\lambda^{(1)*}}\ \forall\ j\in\{1,\hdots,2^N\},\label{13}\\
 &\nu^{(2)*}_j = \frac{\nu^{(1)*}}{\lambda^{(1)*}}, \kappa^{(2)*}_j = \frac{\kappa^{(1)*}}{\lambda^{(1)*}}\ \forall\ j\in\{1,\hdots,N\}.\label{14}
\end{align}
After making the substitutions, we observe: 
\begin{align}
    \left.\frac{\partial \mathcal{L}^{(2)}(\myVec{I})}{\partial \myVec{I}}\right|_{\myVec{I}=\myVec{I}_1^*} =&\frac{1}{\lambda^{(1)*}}\left(-\myVec{A}_f^T+2\lambda^{(1)*}\myVec{A}_f^T\myVec{A}_f\myVec{I}_1^* +\mu^{(1)*}\myVec{1}_N\myplus\sum_{j=1}^{2^N} \delta_j^{(1)*}\myVec{g}_j\myplus\right.\nonumber\\&\left.\sum_{j=1}^{N} (\nu_j^{(1)*}\myminus\kappa_j^{(1)*})\myVec{e}_j \right)\overset{(a)}{=}0,\label{eq:sol:1}
\end{align}
where $(a)$ is due to eq. ~\eqref{1}.
\begin{align}
    &\myVec{A}_f\myVec{I}_1^* = E_{des} = \myVec{A}_f\myVec{I}_1^*.\\
&\underbrace{\myVec{1}_N^T\myVec{I}_1^*}_{=0\text{ due to eq. ~\eqref{2}}} = 0.\\
    &\frac{1}{\lambda^{(1)*}}\left(\underbrace{\delta^{(1)*}(\myVec{g}_j^T\myVec{I}_1^*-2I_{tot})}_{=0 \text{ due to eq. ~\eqref{3}}}\right) = 0\ \forall\ j\in\{1,\hdots,2^N\}.\\
    &\frac{1}{\lambda^{(1)*}}\left(\underbrace{\nu^{(1)*}_j(\myVec{e}_j^T\myVec{I}_1^*-I_{safe})}_{=0\text{ due to eq. ~\eqref{4}}}\right) = 0\ \forall\ j\in\{1,\hdots,N\}.
\end{align}
\begin{align}
    &\frac{1}{\lambda^{(1)*}}\left(\underbrace{\kappa^{(1)*}_j(\myminus\myVec{e}_j^T\myVec{I}_1^*\myminus I_{safe})}_{=0\text{ due to eq. ~\eqref{5}}}\right) \myeq 0\ \forall\ j\in\{1,\hdots,N\}.\label{eq:sol:2}
\end{align}
From~\eqref{eq:sol:1}-\eqref{eq:sol:2}, we can conclude that the values of $\myVec{I}_2^*$, $\beta^{(2)*}$, $\mu^{(2)*}$, $\{\delta_j^{(2)*}\}_{j=1}^{2^N}$, $\{\nu_j^{(2)*}\}_{j=1}^N$, and $\{\kappa_j^{(2)*}\}_{j=1}^N$ given by \eqref{11}-\eqref{14} are the solution of~\eqref{eq:3} as they satisfy the K.K.T. conditions of~\eqref{eq:appx:a:CDM-alt:canon}. Hence, the optimal $\myVec{I}^*$ for both~\eqref{eq:appx:a:CDM} and \eqref{eq:appx:a:CDM-alt} is the same by virtue of~\eqref{11}. 

Additionally, the above analysis also shows that the $\myVec{I}_1^*$ is a unique solution of~\eqref{eq:appx:a:CDM} for $\alpha<\alpha_{MAX}$. We can verify the uniqueness of $\myVec{I}_1^*$ through a contradiction argument. Suppose that $\myVec{I}_1^*$ is not a unique solution of~\eqref{eq:appx:a:CDM}. Consequently, we can find multiple solutions for~\eqref{eq:appx:a:CDM-alt} by using the above analysis, which provides a contradiction as the solution of~\eqref{eq:appx:a:CDM-alt} must be unique by \textbf{(P2)}. Therefore, if $\alpha<\alpha_{MAX}$, then there always exists a value $E_{des}$ such that the solutions of~\eqref{eq:appx:a:CDM} and~\eqref{eq:appx:a:CDM-alt} are same and unique. Hence, the optimization problems of CDM and LCMV-E are equivalent.

Note that using a similar analysis, we can also show that for any $E_{des}<\myVec{A}_{f}\myVec{I}_{ME}^*$, the solution $\myVec{I}_2^*$ is a unique solution of the CDM optimization with $\alpha=\|\myVec{A}_c\myVec{I}_2^*\|_2^2$. Note that for $E_{des}<\myVec{A}_{f}\myVec{I}_{ME}^*$ the corresponding $\alpha=\|\myVec{A}_c\myVec{I}_2^*\|_2^2<\alpha_{MAX}$ thereby ensuring that $\lambda^{(1)*}\neq0$. Furthermore, Observe that $\myVec{I}_1^*=\myVec{I}_2^*$, $\lambda^{(1)*}\myeq\nicefrac{-1}{\beta^{(2)}}$, $\mu^{(1)*}\myeq\nicefrac{-\mu^{(1)*}}{\beta^{(2)*}}$, $\kappa^{(1)*}\myeq\nicefrac{-\kappa^{(1)*}}{\beta^{(2)*}}$,$\nu^{(1)*}$ $\myeq$ $\nicefrac{-\nu^{(1)*}}{\beta^{(2)*}}$, and $\delta^{(1)*}\myeq\nicefrac{-\delta^{(1)*}}{\beta^{(2)*}}$ solves the K.K.T. system of CDM for $\alpha=\|\myVec{A}_c\myVec{I}_2^*\|$. Consequently, $\myVec{I}_2^*$ is the unique solution of the CDM optimization for $\alpha=\|\myVec{A}_c\myVec{I}_2^*\|$. Hence, the uniqueness of the solutions in this case allows us to transform CDM into LCMV-E by choosing $E_{des}=\myVec{A}_f\myVec{I}_1^*$ or transform LCMV-E into a CDM optimization problem with $\alpha=\|\myVec{A}_c\myVec{I}\|_2^2$.

\subsubsection*{\textbf{Case 2}: $\alpha\geq \alpha_{MAX}$}
For this case, we first show that the solution of~\eqref{eq:appx:a:CDM:max-intensity} having the least off-target electric field energy ($\myVec{I}_{ME}^*$) is also a solution of the CDM optimization problem. Then, for an appropriately chosen $E_{des}$, we show that $\myVec{I}_{ME}^*$ is equal to $\myVec{I}_2^*$ of LCMV-E formulation, thereby showing that $\myVec{I}_2^*$ is the solution of~\eqref{eq:appx:a:CDM} with least off-target electric field energy.

We argue that $\myVec{I}_{ME}^*$ is always a solution of~\eqref{eq:appx:a:CDM} if $\alpha\geq\alpha_{MAX}$. Note that the objectives of~\eqref{eq:appx:a:CDM} and~\eqref{eq:appx:a:CDM:max-intensity} are same and the constraint set $\mathcal{C}_{CDM}$ of~\eqref{eq:appx:a:CDM} is always a subset of the constraint set $\mathcal{C}$ of~\eqref{eq:appx:a:CDM:max-intensity}. Hence, we have:
\begin{align}
    \myVec{A}_f\myVec{I}_{ME}^*\geq     \myVec{A}_f\myVec{I}_1^*\ \forall\ \alpha>0.\label{eq:alpha-max:1}
\end{align}
For $\alpha>\alpha_{MAX}$, we can easily verify that $\myVec{I}_{ME}^{*}\in\mathcal{C}_{CDM}$ as $\|\myVec{A}_c\myVec{I}_{ME}^*\|_2^2=\alpha_{MAX}\leq \alpha$, and the rest of the constraints are satisfied due to $\myVec{I}_{ME}^{*}$ being a solution of~\eqref{eq:appx:a:CDM:max-intensity}. Hence, we can conclude that $\myVec{I}_{ME}^{*}$ is also a solution of~\eqref{eq:appx:a:CDM} as a consequence of~\eqref{eq:alpha-max:1}. 

\noindent\underline{\textit{Showing that $\myVec{I}_{ME}^*$ is the solution of~\eqref{eq:appx:a:CDM-alt}}}: For showing that $\myVec{I}_{ME}^*$ is the solution of~\eqref{eq:appx:a:CDM-alt}, we first need to choose an appropriate value of $E_{des}$. In this case, we choose the value of $E_{des}=\myVec{A}_f\myVec{I}_{ME}^*$. We now analyze the constraint set of~\eqref{eq:appx:a:CDM-alt} under this particular choice of $E_{des}$. Let us denote the constraint set of~\eqref{eq:appx:a:CDM-alt} as $\mathcal{C}_{LCMV}$. Then, $\mathcal{C}_{LCMV}$ is a collection of all $\myVec{I}\in\R^N$ such that:
\begin{align}
    \|\myVec{I}\|_{\infty}\mylesseq I_{safe},\|\myVec{I}\|_{1}\mylesseq I_{tot},\myVec{1}_N^T\myVec{I}\myeq 0,\text{ and }\myVec{A}_f\myVec{I}\myeq E_{des}.
\end{align}
Similarly, the constraint set $\mathcal{C}$ of~\eqref{eq:appx:a:CDM:max-intensity} is a a collection of all $\myVec{I}\in\R^N$ such that:
\begin{align}
    \|\myVec{I}\|_{\infty}\mylesseq I_{safe},\|\myVec{I}\|_{1}\mylesseq I_{tot},\text{ and }\myVec{1}_N^T\myVec{I}\myeq 0.
\end{align}
Note that $\mathcal{C}_{LCMV}$ is a strict subset of $\mathcal{C}$, where $\mathcal{C}_{LCMV}$ only contains the elements of $\mathcal{C}$ which also satisfy the equation $\myVec{A}_f\myVec{I}=E_{des}$. For the particular choice of $E_{des}=\myVec{A}_f\myVec{I}_{ME}^*$, we can deduce that the elements in $\mathcal{C}$ which satisfy the equation $\myVec{A}_f\myVec{I}=\myVec{A}_f\myVec{I}_{ME}^*$ are the solutions of~\eqref{eq:appx:a:CDM:max-intensity}, as for any other $\myVec{I}\in\mathcal{C}$ (which is not a solution of~\eqref{eq:appx:a:CDM:max-intensity}), we know that $\myVec{A}_f\myVec{I}<E_{des}=\myVec{A}_f\myVec{I}_{ME}^*$. Hence, the constraint set $\mathcal{C}_{LCMV}$ only contains elements which are the solution of~\eqref{eq:appx:a:CDM:max-intensity} under the choice $E_{des}=\myVec{A}_f\myVec{I}_{ME}^*$. Hence,~\eqref{eq:appx:a:CDM-alt} reduces to: 
\begin{align}
    \myVec{I}_2^* = \argmin_{\myVec{I}\in\mathcal{I}^*}\|\myVec{A}_{c}\myVec{I}\|_2^2.\label{eq:appx:a:LCMV:reduced}
\end{align}
Observing~\eqref{eq:appx:a:I_ME:defn} and~\eqref{eq:appx:a:LCMV:reduced}, we can conclude that $\myVec{I}_2^*=\myVec{I}_{ME}^*$. Using the fact that $\myVec{I}_{ME}^*$ is the solution of~\eqref{eq:appx:a:CDM}, we can further conclude that $\myVec{I}_2^*$ must also be a solution of~\eqref{eq:appx:a:CDM}. Consequently, $\myVec{I}_2^*$ is the solution of~\eqref{eq:appx:a:CDM} with the least value of $\|\myVec{A}_c\myVec{I}\|_2^2$ due to the definition of $\myVec{I}_{ME}^*$. 

\begin{conclusion}
    Combining the results of case 1 and 2, we can show that for any value of $\alpha<\myVec{A_f}\myVec{I}_{ME}^*$, we can always find a value of $E_{des}$, such that the solution of CDM and LCMV-E are same and unique. For $\alpha\geq \myVec{A}_f\myVec{I}_{ME}^*$, CDM optimization essentially reduces to constrained maximum intensity optimization (also noted earlier in~\cite{fernandez2020unification}), in which case, the goal of CDM reduces to maximizing intensity regardless of off-target stimulation. For this case, we can equivalently choose $E_{des}=\myVec{A}_f\myVec{I}_{ME}^*$, and show that the solution of LCMV-E is the solution of CDM having the least value of $\|\myVec{A}_c\myVec{I}\|_2^2$. Therefore, ifr~\eqref{eq:appx:a:CDM:max-intensity} has a unique solution, then LCMV-E and CDM are exactly equivalent optimization formulations. If~\eqref{eq:appx:a:CDM:max-intensity} has multiple solutions, then LCMV-E and CDM are exactly equivalent for $\alpha<\myVec{A_f}\myVec{I}_{ME}^*$, and approximately equivalent for $\alpha>\myVec{A}_f\myVec{I}_{ME}^*$, where LCMV-E picks the most-relevant solution of CDM (choosing a solution of CDM having a larger value of $\|\myVec{A}_c\myVec{I}\|_2^2$ offers no additional advantage as the intensity at the target region remains the same). 
\end{conclusion}
\subsection{Numerical Simulations}
\begin{figure*}[ht]
    \centering
    \includegraphics[width=\textwidth]{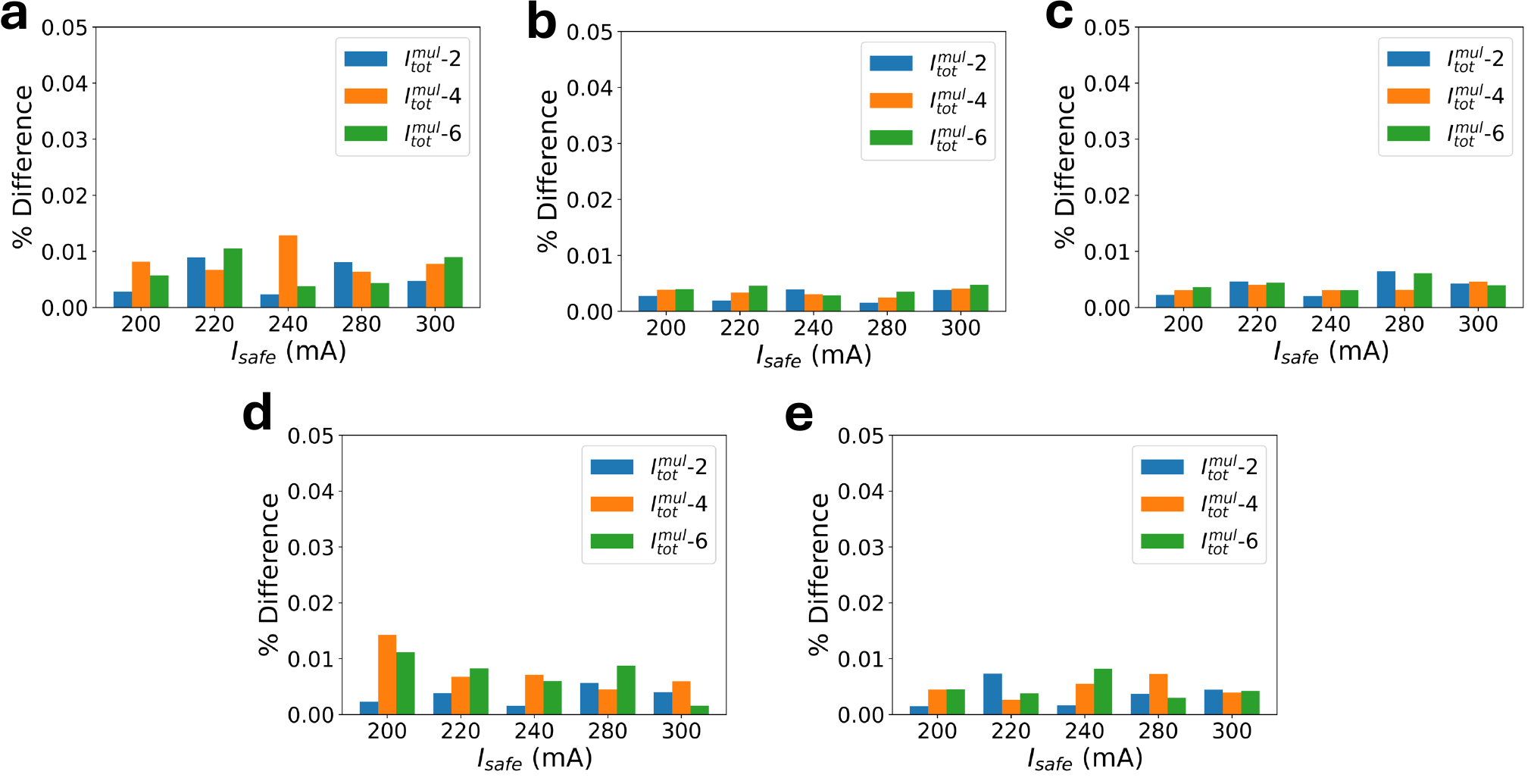}
    \caption{\textbf{a}-\textbf{e} show the difference between the outputs of the LCMV-E and CDM approaches for the target locations $1$, $2$, $3$, $4$, and $5$, respectively discussed in the Numerical Simulations section of Appendix A. The difference between the outputs of LCMV-E ($\myVec{I}_{LCMV-E}^*$) and CDM ($\myVec{I}_{CDM}^*$) was measured as $\nicefrac{\|\myVec{I}^*_{LCMV-E}-\myVec{I}^*_{CDM}\|_1}{\|\myVec{I}^*_{CDM}\|_1}\times 100$. At each target, we calculated the difference between LCMV-E and CDM outputs at five different values of $I_{safe}$ and at three different values of $I_{tot}^{mul}$ resulting in $15$ different combinations of $I_{safe}$ and $I_{tot}$. Each bar plot corresponds to the median difference between $\myVec{I}_{LCMV-E}^*$ and $\myVec{I}_{CDM}^*$ across $11$ different values of $E_{des}$.}
    \label{fig:lcmv-e-and-dcm-equiv}
\end{figure*}

We also provide a numerical illustration of Theorem~\ref{theorem:CDM-LCMV-equiv}. For this study, we used the spherical head model used in our sea of neurons models. Appendix D contains the details of the spherical head model. The results of Theorem~\ref{theorem:CDM-LCMV-equiv} (discussed in the Proof section above) provide us with a recipe to transform the LCMV-E optimization problem to the CDM optimization problem by choosing $\alpha=\|\myVec{A}_c\myVec{I}_{LCMV-E}^*\|_2^2$ where $\myVec{I}_{LCMV-E}^*$ is the solution of the LCMV-E optimization problem. Hence, if we solve the LCMV-E and CDM optimization problems with the above transformation, the solution of both problems should be same. Therefore, we numerically verify our theorem by the following procedure: 
\begin{enumerate}
    \item We first solve the LCMV-E optimization problem for a chosen set of hyperparameters $E_{des}$, $I_{tot}$, $I_{safe}$, $\myVec{A}_f$, and $\myVec{A}_c$ to design $\myVec{I}^*_{LCMV-E}$.
    \item Using $\myVec{I}^*_{LCMV-E}$, we specify the value $\alpha=\|\myVec{A}_c\myVec{I}_{LCMV-E}^*\|_2^2$ in the CDM optimization problem and obtain the corresponding solution $\myVec{I}_{CDM}^*$. 
    \item We calculate the difference between $\myVec{I}_{CDM}^*$ and $\myVec{I}_{LCMV-E}^*$ empirically by $\nicefrac{\|\myVec{I}_{CDM}^*-\myVec{I}_{LCMV-E}^*\|_1}{\|\myVec{I}_{CDM}^*\|_1}\times 100$.
\end{enumerate}
We quantified the difference between $\myVec{I}_{LCMV-E}^*$ and $\myVec{I}_{CDM}^*$ for five different target locations. The target region was chosen as a disc of $1$cm radius centered at the target location. The corresponding off-target region was constructed as a hollow disc of inner-radius $1.1$cm and outer radius $7$cm. The corresponding spherical coordinates for the five target locations are as follows: 
\begin{itemize}
    \item Target location 1: $[7.7,0,0]$ (north pole at the depth of $1.5$cm)
    \item Target location 2: $[7.7,\nicefrac{2}{7.7},0]$ ($+2$cm along the $x$-axis from the north pole)
    \item Target location 3: $[7.7,\nicefrac{2}{7.7},\nicefrac{\pi}{2}]$ ($+2$cm along the $y$-axis from the north pole)
    \item Target location 4: $[7.7,\nicefrac{2}{7.7},\pi]$ ($-2$cm along the $x$-axis from the north pole)
    \item Target location 5: $[7.7,\nicefrac{2}{7.7},\nicefrac{3\pi}{2}]$ ($-2$cm along the $y$-axis from the north pole). 
\end{itemize}
For each location, we tested five different values of $I_{safe}$, namely, $200$mA, $220$mA, $240$mA, $280$mA, and $300$mA. The $I_{tot}$ was parameterized in a similar manner as done in Sec.~\ref{sec:results}, with $I_{tot}=I_{tot}^{mul}I_{safe}$. For each value of $I_{safe}$, three different values of $I_{tot}^{mul}$ were tested, namely, $2$, $4$, and $6$ (similar to our studies in Sec.~\ref{sec:results}). For each target location, $I_{safe}$, and $I_{tot}^{mul}$, $11$ different values of $E_{des}$ were tested. The values of $E_{des}$ and the corresponding direction were chosen by calculating the electric field of a $2$ electrode pattern at the target location (used in Sec.~\ref{sec:results:sea-of-neurons:c3-c4}) at $11$ equally spaced levels of injected currents between $80$mA and $120$mA to represent realistic electric field values that can be induced by traditional electrode patterns (the values $80$mA and $120$mA were arbitrarily chosen). Fig.~\ref{fig:lcmv-e-and-dcm-equiv} shows the aggregate differences between $\myVec{I}_{CDM}^*$ and $\myVec{I}_{LCMV-E}^*$. Fig.~\ref{fig:lcmv-e-and-dcm-equiv}(a)-(e) represent the five different target locations, with (a), (b), (c), (d), and (e) representing target locations $1$, $2$, $3$, $4$, and $5$, respectively. Each bar plot shows the median difference across the $11$ different values of $E_{des}$ for each $I_{safe}$ and $I_{tot}^{mul}$ value. Fig.~\ref{fig:lcmv-e-and-dcm-equiv} shows that across all the parameters tested, the corresponding $\myVec{I}_{CDM}^*$ and $\myVec{I}_{LCMV-E}^*$ are same (except for small differences due to numerical noise) complementing our theoretical results that predict $\myVec{I}_{CDM}^*=\myVec{I}_{LCMV-E}^*$.
\subsection{Lemmas}
\begin{lemma}\label{prop:1}
    $\myVec{I}_{ME}^{*}$ is a solution of ~\eqref{eq:appx:a:CDM} if the K.K.T. system of~\eqref{eq:appx:a:CDM} stated in~\eqref{1}-\eqref{5} has a solution with $\lambda^{(1)*}=0$.
\end{lemma}
\begin{proof}
To show that $\myVec{I}_{ME}^{*}$ is a solution of the CDM optimization problem whenever a solution of CDM's K.K.T. system has $\lambda^{(1)*}=0$, we will prove the following two claims.
\begin{claim}
    All solutions of~\eqref{eq:appx:a:CDM} for which $\lambda^{(1)*}=0$ must also be the solution of~\eqref{eq:appx:a:CDM:max-intensity}.
\end{claim} 
\begin{subproof} To prove the above claim, we show that the solutions of the K.K.T. system of~\eqref{eq:appx:a:CDM} with $\lambda^{(1)*}=0$ also solve the K.K.T. system of~\eqref{eq:appx:a:CDM:max-intensity}. Consequently, all solutions of~\eqref{eq:appx:a:CDM} for which $\lambda^{(1)*}=0$ must also be the solution of~\eqref{eq:appx:a:CDM:max-intensity}. We first make the following substitutions in \eqref{eq:kkt:CDM-max:1}-\eqref{eq:kkt:CDM-max:5}:
\begin{align}
 &\myVec{I}_3^*=\myVec{I}_1^*,\label{eq:subs:1}\\
 &\delta^{(3)*}_j =\delta^{(1)*}\ \forall\ j\in\{1,\hdots,2^N\},\label{eq:subs:2}\\
 &\nu^{(3)*}_j = \nu^{(1)*}, \kappa^{(3)*}_j = \kappa^{(1)*}\ \forall\ j\in\{1,\hdots,N\}.\label{eq:subs:3}
\end{align}
After making the above substitutions, we observe: 
\begin{align}
    &\left.\frac{\partial \mathcal{L}^{(3)}(\myVec{I})}{\partial \myVec{I}}\right|_{\myVec{I}=\myVec{I}_1^*} = -\myVec{A}_f^T+\mu^{(1)*}\1_N+\sum_{j=1}^{2^N} \delta_j^{(1)*}\myVec{g}_j+\nonumber\\&+\sum_{j=1}^{N} (\nu_j^{(1)*}-\kappa_j^{(1)*})\myVec{e}_j.
\intertext{Adding $2\lambda^{(1)*}\myVec{A}_f^T\myVec{A}_f\myVec{I}_1^*$ to both sides of the above equation:}
&\left.\frac{\partial \mathcal{L}^{(3)}(\myVec{I})}{\partial \myVec{I}}\right|_{\myVec{I}=\myVec{I}_1^*}+2\lambda^{(1)*}\myVec{A}_f^T\myVec{A}_f\myVec{I}_1^* = -\myVec{A}_f^T+\mu^{(1)*}\1_N+2\lambda^{(1)*}\myVec{A}_f^T\myVec{A}_f\myVec{I}_1^*\myplus\nonumber\\&\sum_{j=1}^{2^N} \delta_j^{(1)*}\myVec{g}_j\myplus\sum_{j=1}^{N} (\nu_j^{(1)*}\myminus\kappa_j^{(1)*})\myVec{e}_j.
\intertext{Substituting $2\lambda^{(1)*}\myVec{A}_f^T\myVec{A}_f\myVec{I}_1^*=\myVec{0}_N$ (due to $\lambda^{(1)*}=0$) in the L.H.S of the above equation:}
&\left.\frac{\partial \mathcal{L}^{(3)}(\myVec{I})}{\partial \myVec{I}}\right|_{\myVec{I}=\myVec{I}_1^*}= -\myVec{A}_f^T+\mu^{(1)*}\1_N+2\lambda^{(1)*}\myVec{A}_f^T\myVec{A}_f\myVec{I}_1^*+\sum_{j=1}^{2^N} \delta_j^{(1)*}\myVec{g}_j\myplus\nonumber\\&\sum_{j=1}^{N} (\nu_j^{(1)*}\myminus\kappa_j^{(1)*})\myVec{e}_j
\overset{(a)}{=}\myVec{0}_N,
\end{align}
where $(a)$ follows from~\eqref{1}. Similarly, we have
\begin{align}
    &\underbrace{\1_N^T\myVec{I}_1^*}_{=0\text{ due to~\eqref{2}}} = 0.\\
    &\underbrace{\delta^{(3)*}(\myVec{g}_j^T\myVec{I}_1^*-2I_{tot})}_{=0\text{ due to~\eqref{3}}} = 0\ \forall\ j\in\{1,\hdots,2^N\}.\\
&\underbrace{\nu^{(3)*}_j(\myVec{e}_j^T\myVec{I}_1^*-I_{safe})}_{=0\text{ due to~\eqref{4}}} = 0\ \forall\ j\in\{1,\hdots,N\}.\\
    &\underbrace{\kappa^{(3)*}_j(-\myVec{e}_j^T\myVec{I}_1^*-I_{safe})}_{=0\text{ due to~\eqref{5}}} = 0\ \forall\ j\in\{1,\hdots,N\}.
\end{align}
Hence, $\{\myVec{I}_1^*$, $\mu^{(1)*}$, $\{\delta_j^{(1)*}\}_{j=1}^{2^N}$, $\{\nu_j^{(1)*},\kappa_j^{(1)*}\}_{j=1}^N\}$ solves the K.K.T. system of \eqref{eq:appx:a:CDM:max-intensity}, i.e.,~\eqref{eq:kkt:CDM-max:1}-\eqref{eq:kkt:CDM-max:5}. Consequently, $\{\myVec{I}_1^*$, $\mu^{(1)*}$, $\{\delta_j^{(1)*}\}_{j=1}^{2^N}$, $\{\nu_j^{(1)*},\kappa_j^{(1)*}\}_{j=1}^N\}$ must be the solution of the Lagrangian dual of~\eqref{eq:appx:a:CDM:max-intensity} stated in~\eqref{eq:appx:a:CDM-max-intensity:dual}. Hence, all the solutions of CDM optimization problem which have $\lambda^{(1)*}=0$ also solve~\eqref{eq:appx:a:CDM:max-intensity} by virtue of~\eqref{eq:subs:1}.
\end{subproof}
\begin{claim}
    The solutions of~\eqref{eq:appx:a:CDM:max-intensity} which are not the solutions of~\eqref{eq:appx:a:CDM} must not satisfy the constraint $\|\myVec{A}_{c}\myVec{I}\|_2^2\leq \alpha$ in~\eqref{eq:appx:a:CDM}.
\end{claim} 
\begin{subproof}
\noindent Note that all solutions of~\eqref{eq:appx:a:CDM:max-intensity} have the same value of the objective
($\myVec{A}_f\myVec{I}$) of~\eqref{eq:appx:a:CDM} due to being a solution of~\eqref{eq:appx:a:CDM:max-intensity}. Furthermore, all solutions of~\eqref{eq:appx:a:CDM:max-intensity} are guaranteed to satisfy all constraints of~\eqref{eq:appx:a:CDM} except the constraint $\|\myVec{A}_{c}\myVec{I}\|_2^2\leq \alpha$ again due to being a solution of~\eqref{eq:appx:a:CDM:max-intensity}. Hence, the only constraint that a solution of~\eqref{eq:appx:a:CDM:max-intensity} can violate in~\eqref{eq:appx:a:CDM} is the $\|\myVec{A}_{c}\myVec{I}\|_2^2\leq \alpha$ constraint. 
\end{subproof}
Combining claims 1 and 2, we observe that $\myVec{I}_{ME}^*$ must always be a solution of~\eqref{eq:appx:a:CDM}. Note that, if $\myVec{I}_{ME}^*$ is not a solution of~\eqref{eq:appx:a:CDM}, then it must be the case that $\|\myVec{A}_c\myVec{I}_{ME}^*\|_2^2>\alpha$. Consequently, all other solutions of~\eqref{eq:appx:a:CDM:max-intensity} must also violate the constraint $\|\myVec{A}_{c}\myVec{I}\|_2^2\leq \alpha$ due to the definition of $\myVec{I}_{ME}^*$ (see~\eqref{eq:appx:a:I_ME:defn}) implying that no solution of~\eqref{eq:appx:a:CDM:max-intensity} is a solution of~\eqref{eq:appx:a:CDM}. This is a contradiction as under the assumption $\lambda^{(1)*}=0$ there must be at least one solution of~\eqref{eq:appx:a:CDM:max-intensity} that should also be a solution of~\eqref{eq:appx:a:CDM}. Hence, we have that $\myVec{I}_{ME}^*$ is always a solution of~\eqref{eq:appx:a:CDM} if there exists a solution to the K.K.T. system of~\eqref{eq:appx:a:CDM} with the corresponding $\lambda^{(1)*}=0$.
\end{proof}
\begin{lemma}\label{prop:2}
For $\alpha<\alpha_{MAX}$, all solutions of the CDM's K.K.T. system stated in~\eqref{1}-\eqref{5} have their corresponding $\lambda^{(1)*}\neq 0$.
\end{lemma} 
\begin{proof}
From lemma~\ref{prop:1}, we know that $\lambda^{(1)*}=0$ implies that $\myVec{I}_{ME}^*$ is a solution of~\eqref{eq:appx:a:CDM}. For $\alpha<\alpha_{MAX}=\|\myVec{A}_c\myVec{I}_{ME}^*\|_2^2$, it is easy to verify that $\myVec{I}_{ME}^*$ is not in the constraint set of~\eqref{eq:appx:a:CDM} and consequently, cannot be a solution~\eqref{eq:appx:a:CDM}. Therefore, it must be the case that $\lambda^{(1)*}\neq 0$. 
\end{proof}
\newpage
\section{}\label{appx:B}
\noindent\textbf{Goal}: To show that the electrode placement algorithms proposed in~\cite{sadleir2012target,park2011novel} over-penalize the off-target electric field.\\
\textbf{Argument}: We argue that the magnitude maximization algorithm proposed in~\cite{sadleir2012target} and~\cite{park2011novel} also force the off-target electric field to be as close to zero as possible by highlighting a connection between CDM approaches and magnitude maximization approaches. To show this connection, we use the optimization formulation proposed in~\cite{sadleir2012target} (it was argued in~\cite{gomez2024perspectives} that the algorithm in~\cite{park2011novel} is the same as in~\cite{sadleir2012target}). Furthermore, we ignore the constraint (5) of equation (7) proposed in Sadlier et al.~\cite{sadleir2012target} as this constraint has largely not been used in subsequent studies such as~\cite{fernandez2016transcranial,fernandez2020unification,prieto2022l1}. 

Let us now write out the magnitude maximization electrode placement algorithm proposed in Sadlier et al.~\cite{sadleir2012target}:
\begin{align}
&\argmax_{\myVec{I}\in\R^N}\sum_{i=1}^{|\mathcal{F}|}\left\|\myVec{t}^f_i
\myVec{I}\right\|_2,\text{s.t. }\myVec{1}_N^T\myVec{I}\myeq0,\|\myVec{A}_c\myVec{I}\|_2^2\leq\alpha,\|\myVec{I}\|_{\infty}\leq I_{safe},\text{ and }\nonumber\\&\|\myVec{I}\|_1\leq2I_{tot},\text{ where }\myVec{t}^f_i = \begin{bmatrix}
    (\myVec{T}_f^x)_i\\
    (\myVec{T}_f^y)_i\\
    (\myVec{T}_f^z)_i
\end{bmatrix},\label{eq:mag-max}
\end{align}
$(\myVec{T}_f^x)_i$,$(\myVec{T}_f^y)_i$, and $(\myVec{T}_f^z)_i$ are the $i$-th rows of $\myVec{T}_f^x$, $\myVec{T}_f^y$, and $\myVec{T}_f^z$, respectively, corresponding to the $i$-th voxel in the target region $\mathcal{F}$. $\myVec{A}_c$ is as defined in Sec.~\ref{sec:prob-description:sys-model} with $\myVec{\Gamma}_C$ chosen to be the identity matrix. Let $\myVec{I}_{mag}^*$ be a solution of~\eqref{eq:mag-max}, then we define the vector $\myVec{D}^*$ as follows:
\begin{align}
&\myVec{D^*} = \begin{bmatrix}
    \myVec{D}_x^*\\
    \myVec{D}_y^*\\
    \myVec{D}_z^*
\end{bmatrix},\text{where }\myVec{D}_j^*=\begin{bmatrix}
    \frac{E_{1}^j}{\sqrt{(E_1^x)^2\myplus(E_1^y)^2\myplus(E_1^z)^2}}\\
    \vdots\\
    \frac{E_{|\mathcal{F}|}^j}{\sqrt{(E_{|\mathcal{F}|}^x)^2\myplus(E_{|\mathcal{F}|}^y)^2\myplus(E_{|\mathcal{F}|}^z)^2}}
\end{bmatrix}\nonumber\\&\text{ and } 
\myVec{T}_f^j\myVec{I}_{mag}^* = \begin{bmatrix}
        E_1^{j}\\
        \vdots\\
        E_{|\mathcal{F}|}^j
    \end{bmatrix}\text{ for }j=x,y,\text{ and }z.
\end{align}
Now, consider the following optimization problem:
\begin{align}
&\argmax_{\myVec{I}\in\R^N}(\myVec{D}^*)^T\myVec{T}_f\myVec{I},\text{s.t. }\myVec{1}_N^T\myVec{I}\myeq0,\|\myVec{A}_c\myVec{I}\|_2^2\leq\alpha,\|\myVec{I}\|_{\infty}\leq I_{safe},\text{ and }\|\myVec{I}\|_1\leq2I_{tot}.\label{eq:cdm-opt}  
\end{align}
The optimization formulation in~\eqref{eq:cdm-opt} is the same as the optimization formulation of CDM (discussed in Sec.~\ref{sec:equiv-LCMV-EDM}) with $\myVec{\Gamma}_F$ chosen to be an identity matrix. A subtle difference between CDM approach and~\eqref{eq:cdm-opt} is that the vector $\myVec{D}^*$ in~\eqref{eq:cdm-opt} does not exactly represent the directions as it did in the CDM approach. CDM does not explicitly define the structure of $\myVec{D}$ and works for any real vector $\myVec{D}$. Hence, the exact physical meaning of $\myVec{D}$ does not affect the mathematical properties of the CDM optimization problem, and all the equivalence results derived for CDM approaches would continue to hold for~\eqref{eq:cdm-opt}. 

Before proceeding with our proof, we restate~\eqref{eq:cdm-opt} and~\eqref{eq:mag-max} for brevity's sake by utilizing the fact that the constraint set $\mathcal{C}$ of~\eqref{eq:cdm-opt} and~\eqref{eq:mag-max} is same:
\begin{align}
\argmax_{\myVec{I}\in\mathcal{C}}\sum_{i=1}^{|\mathcal{F}|}\left\|\myVec{t}^f_i\myVec{I}\right\|_2,\label{eq:mag-max:short}\\
\argmax_{\myVec{I}\in\mathcal{C}}(\myVec{D}^*)^T\myVec{T}_f\myVec{I}.\label{eq:cdm-opt:short}
\end{align}
We now claim that the $\myVec{I}^*_{mag}$ is also a solution of~\eqref{eq:cdm-opt:short}. We prove our claim by using the method of contradiction. Let us assume that the solution of~\eqref{eq:cdm-opt}, denoted as $\myVec{I}^*_{CDM}$, is not a solution of~\eqref{eq:mag-max:short}. Then, we have:
\begin{align}
 &(\myVec{D}^*)^T\myVec{T}_f\myVec{I}_{CDM}^* = \sum_{i=1}^{|\mathcal{F}|}\frac{(\myVec{t}^f_i\myVec{I}_{mag}^*)^T(\myVec{t}^f_i\myVec{I}_{CDM}^*)}{\|\myVec{t}^f_i\myVec{I}_{mag}^*\|_2}.\label{eq:appx:mag:5}
 \intertext{Using the fact that absolute value of a number is always equal or bigger than the number, we have for all $i\in\{1,\hdots,|\mathcal{F}|\}$:}
&\frac{(\myVec{t}^f_i\myVec{I}_{mag}^*)^T(\myVec{t}^f_i\myVec{I}_{CDM}^*)}{\|\myVec{t}^f_i\myVec{I}_{mag}^*\|_2}\mylesseq \frac{\left|(\myVec{t}^f_i\myVec{I}_{mag}^*)^T(\myVec{t}^f_i\myVec{I}_{CDM}^*)\right|}{\|\myVec{t}^f_i\myVec{I}_{mag}^*\|_2}.\label{eq:appx:mag:4}
\intertext{Using Cauchy-Schwarz inequality on R.H.S of the above equation, we obtain:}
&\frac{\left|(\myVec{t}^f_i\myVec{I}_{mag}^*)^T(\myVec{t}^f_i\myVec{I}_{CDM}^*)\right|}{\|\myVec{t}^f_i\myVec{I}_{mag}^*\|_2}\leq \frac{\|\myVec{t}^f_i\myVec{I}_{mag}^*\|_2\|\myVec{t}^f_i\myVec{I}_{CDM}^*\|_2}{\|\myVec{t}^f_i\myVec{I}_{mag}^*\|_2}.\\
&\Rightarrow \frac{\left|(\myVec{t}^f_i\myVec{I}_{mag}^*)^T(\myVec{t}^f_i\myVec{I}_{CDM}^*)\right|}{\|\myVec{t}^f_i\myVec{I}_{mag}^*\|_2}\leq \left\|\myVec{t}^f_i\myVec{I}_{CDM}^*\right\|_2.
\label{eq:appx:mag:1}
\intertext{Since $\myVec{I}^*_{mag}$ is a solution of~\eqref{eq:mag-max:short} and $\myVec{I}^*_{CDM}\in\mathcal{C}$ is not a solution of~\eqref{eq:mag-max:short}, we know:}
&\sum_{i=1}^{|\mathcal{F}|}\left\|\myVec{t}^f_i\myVec{I}_{mag}^*\right\|_2>\sum_{i=1}^{|\mathcal{F}|}\left\|\myVec{t}^f_i\myVec{I}_{CDM}^*\right\|_2.\label{eq:appx:mag:2}
\intertext{Combining~\eqref{eq:appx:mag:1} and~\eqref{eq:appx:mag:2}, we obtain:}
&\sum_{i=1}^{|\mathcal{F}|}\frac{\left|(\myVec{t}^f_i\myVec{I}_{mag}^*)^T(\myVec{t}^f_i\myVec{I}_{CDM}^*)\right|}{\|\myVec{t}^f_i\myVec{I}_{mag}^*\|_2}<\sum_{i=1}^{|\mathcal{F}|}\|\myVec{t}_i^{f}\myVec{I}_{mag}^*\|_2.\label{eq:appx:mag:3}
\intertext{Combining~\eqref{eq:appx:mag:3} and~\eqref{eq:appx:mag:4}, we further obtain:}
&\sum_{i=1}^{|\mathcal{F}|}\frac{(\myVec{t}^f_i\myVec{I}_{mag}^*)^T(\myVec{t}^f_i\myVec{I}_{CDM}^*)}{\|\myVec{t}^f_i\myVec{I}_{mag}^*\|_2}<\sum_{i=1}^{|\mathcal{F}|}\left\|\myVec{t}^f_i\myVec{I}_{mag}^*\right\|_2.\label{eq:appx:mag:6}
\intertext{Using~\eqref{eq:appx:mag:5} in~\eqref{eq:appx:mag:6}, we have:}
&(\myVec{D}^*)^T\myVec{T}_f\myVec{I}_{CDM}^*<\sum_{i=1}^{|\mathcal{F}|}\left\|\myVec{t}^f_i\myVec{I}_{mag}^*\right\|_2.
 \intertext{Using the fact that $\sum_{i=1}^{|\mathcal{F}|}\|\myVec{t}^f_i\myVec{I}_{mag}^*\|_2=(\myVec{D}^*)^T\myVec{T}_f\myVec{I}_{mag}^*$ in the above equation, we have:}
 &(\myVec{D}^*)^T\myVec{T}_f\myVec{I}_{CDM}^*<(\myVec{D}^*)^T\myVec{T}_f\myVec{I}_{mag}^*.\label{eq:contardiction}
\end{align}
Hence,~\eqref{eq:contardiction} provides a contradiction as $\myVec{I}^*_{mag}\in\mathcal{C}$ and has a larger value of the objective in~\eqref{eq:cdm-opt:short} compared to $\myVec{I}_{CDM}^*$ (which is not possible as $\myVec{I}_{CDM}^*$ is a solution of~\eqref{eq:cdm-opt:short}). Therefore, $\myVec{I}^*_{mag}$ is also a solution of~\eqref{eq:cdm-opt:short}. 

From the above analysis we can conclude that a solution of magnitude maximization problem of~\cite{park2011novel,sadleir2012target} must also be a solution of some CDM-approach with an appropriately chosen $\myVec{D}$. Since, $\myVec{I}^*_{mag}$ is also a solution of the CDM approach, it forces the  off-target electric field magnitude as close to zero as possible.

As a side note, one can extend the above analysis to show that~\eqref{eq:mag-max:short} is equivalent to the following bi-convex optimization:
\begin{align}
\myVec{I}^*,\{\myVec{d}_i\}_{i=1}^{|\mathcal{F}|}=\argmax_{\substack{{\left\{\|\myVec{d}_i\|_2\leq1\right\}_{i=1}^{|\mathcal{F}|}},\\{\myVec{I}\in\mathcal{C}}}}\sum_{i=1}^{|\mathcal{F}|}\myVec{d}^T_i\myVec{t}^f_i\myVec{I}, \label{eq:bi-convex}
\end{align}
where $\myVec{d}_i\in\R^3\ \forall\ i\in\{1,\hdots,|\mathcal{F}|\}$. A brief proof is described below. Using Cauchy-Schwarz inequality, we have 
\begin{align}
\sum_{i=1}^{|\mathcal{F}|}\myVec{d}_i^T\myVec{t}^f_i\myVec{I}\leq\sum_{i=1}^{|\mathcal{F}|}\|\myVec{d}_i\|_2\|\myVec{t}^f_i\myVec{I}\|_2,
\intertext{Using $\|\myVec{d}_i\|_2\leq1$ in the above equation, we have:}
\sum_{i=1}^{|\mathcal{F}|}\myVec{d}_i^T\myVec{t}^f_i\myVec{I}\leq\sum_{i=1}^{|\mathcal{F}|}\left\|\myVec{t}^f_i\myVec{I}\right\|_2.\label{eq:appx:mag:11}
\end{align}
Since $\myVec{I}_{mag}^*$ is a solution to~\eqref{eq:mag-max:short}, we have:
\begin{align}
\sum_{i=1}^{|\mathcal{F}|}\left\|\myVec{t}^f_i\myVec{I}\right\|_2\leq \sum_{i=1}^{|\mathcal{F}|}\left\|\myVec{t}^f_i\myVec{I}_{mag}^*\right\|_2\ \forall\ \myVec{I}\in\mathcal{C}.\label{eq:appx:mag:10} 
\end{align}
Combining~\eqref{eq:appx:mag:11} and~\eqref{eq:appx:mag:10}, we have:
\begin{align}
\sum_{i=1}^{|\mathcal{F}|}\myVec{d}_i^T\myVec{t}^f_i\myVec{I}\leq   \sum_{i=1}^{|\mathcal{F}|}\left\|\myVec{t}^f_i\myVec{I}_{mag}^*\right\|_2\ \forall\ \myVec{I}\in\mathcal{C}.\label{eq:appx:upper-bound}
\end{align}
Hence,~\eqref{eq:appx:upper-bound} provides an upper-bound for~\eqref{eq:bi-convex} which is achievable at $\myVec{I}^*=\myVec{I}_{mag}^*$ and $\myVec{d}_i^*=\nicefrac{1}{\|\myVec{t}^f_i\myVec{I}^*_{mag}\|_2}\myVec{t}^f_i\myVec{I}_{mag}^*$. Hence, $\myVec{I}_{mag}^*$ is also a solution of~\eqref{eq:bi-convex}. This can be helpful as a large body of work on bi-convex optimization~\cite{gorski2007biconvex} can be leveraged to solve the magnitude maximization problem for designing electrode placement.
\newpage
\section{}\label{appx:C}
In this appendix, we discuss the mathematical proof showing that the solutions obtained by solving L1L1-norm optimization proposed in Prieto et al.~\cite{prieto2022l1} can be equivalently obtained by solving the \HP optimization with $p=1$ and the $l_{\infty}$ constraint appropriately modified. We have divided the section into five parts describing the problem, discussing an informal proof sketch, the formal proof, the numerical illustration, and the accompanying lemmas used to support the technical arguments of the proof section. 
\subsection{Problem Description}
 The goal of this appendix is to show that the solutions obtained by L1L1-norm optimization formulation proposed in Prieto et al.~\cite{prieto2022l1} can be equivalently obtained by solving a special case of the \HP optimization formulation (see~\eqref{eq:algo:HP-main-opt}) for $p=1$ and $\myVec{E}_{tol}^{+}=\myVec{E}_{tol}^{-}=E_{tol}\myVec{1}_{|\mathcal{C}|},E_{tol}>0$ with an appropriately modified $l_{\infty}$ constraint. For this proof, we will assume that $\myVec{\Gamma}_C$ and $\myVec{\Gamma}_F$ to be identity matrices to match the assumptions of Prieto et al.~\cite{prieto2022l1}. We re-write the special case of the modified \HP optimization formulation and the L1L1-norm optimization formulation here for convenience:
\begin{align}
   \text{\underline{\textbf{L1L1:}} } \myVec{I}_1^* \myeq& \argmin_{\myVec{I}\in\R^N} \left\|\begin{array}{c}
          \myVec{T}_{f}\myVec{I}-\myVec{E}_{des} \\
          \Psi_{\epsilon}\left[\nu^{-1}\myVec{T}_{c}\myVec{I}\right]
    \end{array}\right\|_1\myplus \alpha\zeta\|\myVec{I}\|_1,
    \text{ s.t. }\myVec{I}\preceq I_{safe}\myVec{1}_N,\|\myVec{I}\|_1\mylesseq 2I_{tot},\nonumber\\ &\text{and } \myVec{1}_N^T\myVec{I}\myeq0,\label{eq:l1l1norm:opt}
\end{align}
where $\zeta=\|\myVec{T}_f\|_1$,\footnote{Note that $\|\myVec{T}_f\|_1$ is the $l_1$ matrix-norm of the matrix $\myVec{T}_f$.} $\nu=\|\myVec{E}_{des}\|_{\infty}$, and the function $\Psi_{\epsilon}\left[\cdot\right]$ is defined as follows:
\begin{align}
    \Psi_{\epsilon}\left[\myVec{w}\right] = \max\left\{\epsilon\myVec{1}_d,\left |\myVec{w}\right| \right\},\label{eq:l1l1norm:psi-func-defn}
\end{align}
with $\max\{\cdot,\cdot\}$ and $|\cdot|$ applied element-wise between the vectors $\epsilon\myVec{1}_{d}$ and $\myVec{w}\in\R^{d}$. 
\begin{align}
   \text{\textbf{\underline{Modified-\HP:}} } \myVec{I}_2^*=&\argmin_{\myVec{I}\in\R^N}\mathcal{L}_{hinge}(\myVec{I}),\text{ s.t. }\myVec{T}_f\myVec{I}=\widetilde{\myVec{E}_{des}},\nonumber\\&\myVec{I}\preceq I_{safe}\myVec{1}_N,\|\myVec{I}\|_1\mylesseq\Tilde{I}_{tot},\text{and }\myVec{1}_N^T\myVec{I}\myeq0,\label{eq:appx:b:hp}\
\end{align}
where $\mathcal{L}_{hinge}(\myVec{I})$ is defined as follows:
\begin{align}
   \mathcal{L}_{hinge}(\myVec{I})=&\|\max\{\myVec{0}_{|\mathcal{C}|},\myVec{T}_c\myVec{I}-\myVec{E}_{tol}^+\}\|_1+\|\max\{\myVec{0}_{|\mathcal{C}|},-\myVec{T}_c\myVec{I}-\myVec{E}_{tol}^{-}\}\|_1.
\end{align}
We provide a mathematical proof accompanied by a numerical illustration showing that the solutions of the L1L1-norm optimization (stated in~\eqref{eq:l1l1norm:opt}) can be equivalently obtained by solving a special case of \HP optimization formulation (stated in~\eqref{eq:appx:b:hp}) for $p=1$ and $\myVec{E}_{tol}^{+}=\myVec{E}_{tol}^{-}=\epsilon\nu\myVec{1}_{|\mathcal{C}|}$ with appropriately modified $l_{\infty}$ constraint.
\subsection{Informal Description of the Proof}
To show that the solutions~\eqref{eq:l1l1norm:opt} can be equivalently obtained by solving~\eqref{eq:appx:b:hp}, we will show that for each value of $\myVec{E}_{des}$, $\alpha$, and $\epsilon$ chosen in~\eqref{eq:l1l1norm:opt}, there exists a corresponding value of $\widetilde{\myVec{E}_{des}}$, $\Tilde{I}_{tot}$, $\myVec{E}_{tol}^+$ and $\myVec{E}_{tol}^-$ such that the optimum obtained by solving~\eqref{eq:l1l1norm:opt}, denoted as $\myVec{I}^*_{1}$, is also a solution of~\eqref{eq:appx:b:hp}, denoted as $\myVec{I}^*_{2}$. To do so, we will make use of the K.K.T. conditions to show that the $\myVec{I}_1^*$ also solves the K.K.T. conditions of~\eqref{eq:appx:b:hp}.

\subsection{Proof}
We start our proof by simplifying~\eqref{eq:l1l1norm:opt} as follows:
\begin{align}
    \myVec{I}^*_1 = &\argmin_{\myVec{I}\in\R^N} \left\|
          \myVec{T}_{f}\myVec{I}-\myVec{E}_{des} \right\|_1\myplus\left\|\Psi_{\epsilon}\left[\nu^{-1}\myVec{T}_{c}\myVec{I}\right]\right\|_1\myplus \alpha\zeta\|\myVec{I}\|_1,\text{ s.t. }\myVec{I}\preceq I_{safe}\myVec{1}_N,\nonumber\\&\|\myVec{I}\|_1\leq 2I_{tot}, \text{ and } \myVec{1}_N^T\myVec{I}=0,\label{eq:l1l1norm:opt:simplify:1}
\end{align}
where~\eqref{eq:l1l1norm:opt:simplify:1} follows from the fact $l_1$-norm of a vector is equal to the sum of the $l_1$-norm of its disjoint vector components provided that each element of the vector is contained in one of its vector components. Furthermore, substituting the form of $\Psi_{\epsilon}\left[\nu^{-1}\myVec{T}_c\myVec{I}\right]$ from~\eqref{eq:l1l1norm:psi-func-defn} in~\eqref{eq:l1l1norm:opt:simplify:1}, we obtain:
\begin{align}
    \myVec{I}^*_1 = &\argmin_{\myVec{I}\in\R^N} \left\|
          \myVec{T}_{f}\myVec{I}-\myVec{E}_{des} \right\|_1\myplus\left\|\max\left\{\epsilon\myVec{1}_{|\mathcal{C}|},|\nu^{-1}\myVec{T}_{c}\myVec{I}|\right\}\right\|_1 +\alpha\zeta\|\myVec{I}\|_1,\nonumber\\
    &\text{ s.t. }\myVec{I}\preceq I_{safe}\myVec{1}_N,\|\myVec{I}\|_1\leq 2I_{tot}, \text{ and } \myVec{1}_N^T\myVec{I}=0,\label{eq:l1l1norm:opt:simplify:2}
\end{align}
Using the result of lemma~\ref{lemma:l1l1norm:hp-equiv}, we can simplify the term $\left\|\max\left\{\epsilon\myVec{1}_{|\mathcal{C}|},|\nu^{-1}\myVec{T}_{c}\myVec{I}|\right\}\right\|_1$ in~\eqref{eq:l1l1norm:opt:simplify:2} to obtain the following optimization form:
\begin{align}
    \myVec{I}^*_1 \myeq &\argmin_{\myVec{I}\in\R^N} \left\|\myVec{T}_{f}\myVec{I}\myminus\myVec{E}_{des} \right\|_1\myplus\left\|\max\left\{\myVec{0}_{|\mathcal{C}|},\frac{1}{\nu}\myVec{T}_{c}\myVec{I}\myminus\epsilon\myVec{1}_{|\mathcal{C}|}\right\}\right\|_1 +\epsilon C+\alpha\zeta\|\myVec{I}\|_1+\nonumber\\
    &\left\|\max\left\{\myVec{0}_{|\mathcal{C}|},-\frac{1}{\nu}\myVec{T}_{c}\myVec{I}\myminus\epsilon\myVec{1}_{|\mathcal{C}|}\right\}\right\|_1,\text{ s.t. }\myVec{I}\preceq I_{safe}\myVec{1}_N,\|\myVec{I}\|_1\leq 2I_{tot},\nonumber\\&\text{ and } \myVec{1}_N^T\myVec{I}=0,\label{eq:l1l1norm:opt:simplify:3}
\end{align}
Since $\epsilon C$ is just a constant in the loss of~\eqref{eq:l1l1norm:opt:simplify:3}, it does not affect the outcome of the optimization problem. Consequently, the optimization problem of~\eqref{eq:l1l1norm:opt:simplify:3} is equivalent to the following optimization problem described in~\eqref{eq:l1l1norm:opt:simplify:4}:
\begin{align}
    \myVec{I}^*_1 \myeq&\argmin_{\myVec{I}\in\R^N} \left\|\myVec{T}_{f}\myVec{I}\myminus\myVec{E}_{des} \right\|_1\myplus\left\|\max\left\{\myVec{0}_{|\mathcal{C}|},\frac{1}{\nu}\myVec{T}_{c}\myVec{I}\myminus\epsilon\myVec{1}_{|\mathcal{C}|}\right\}\right\|_1 +\alpha\zeta\|\myVec{I}\|_1+\nonumber\\
    &\left\|\max\left\{\myVec{0}_{|\mathcal{C}|},-\frac{1}{\nu}\myVec{T}_{c}\myVec{I}\myminus\epsilon\myVec{1}_{|\mathcal{C}|}\right\}\right\|_1,\text{ s.t. }\myVec{I}\preceq I_{safe}\myVec{1}_N,\|\myVec{I}\|_1\leq 2I_{tot},\nonumber\\ &\text{ and } \myVec{1}_N^T\myVec{I}=0.\label{eq:l1l1norm:opt:simplify:4}
\end{align}
We now multiply the objective of~\eqref{eq:l1l1norm:opt:simplify:4} with $\nu$. Note that multiplying the objective in an optimization problem by a positive constant does not affect the point of optimum. Hence, we can equivalently write~\eqref{eq:l1l1norm:opt:simplify:4} as:
\begin{align}
    \myVec{I}^*_1 \myeq&\argmin_{\myVec{I}\in\R^N} \nu\left\|\myVec{T}_{f}\myVec{I}\myminus\myVec{E}_{des} \right\|_1\mkern0.2mu{+}\mkern0.2mu\left\|\max\left\{\myVec{0}_{|\mathcal{C}|},\myVec{T}_{c}\myVec{I}\myminus\nu\epsilon\myVec{1}_{|\mathcal{C}|}\right\}\right\|_1 +\alpha\nu\zeta\|\myVec{I}\|_1+\nonumber\\
    &\left\|\max\left\{\myVec{0}_{|\mathcal{C}|},-\myVec{T}_{c}\myVec{I}\myminus\nu\epsilon\myVec{1}_{|\mathcal{C}|}\right\}\right\|_1,\text{ s.t. }\myVec{I}\preceq I_{safe}\myVec{1}_N,\|\myVec{I}\|_1\leq 2I_{tot},\nonumber\\ 
    &\text{ and } \myVec{1}_N^T\myVec{I}=0.\label{eq:l1l1norm:opt:simplify:5}
\end{align}
Note that the term ``$\|\max\{\myVec{0}_{|\mathcal{C}|},\myVec{T}_c\myVec{I}\myminus\nu\epsilon\myVec{1}_{|\mathcal{C}|}\}\|_1$ $\myplus$ $\|\max\{\myVec{0}_{|\mathcal{C}|},-\myVec{T}_c\myVec{I}\myminus\nu\epsilon\myVec{1}_{|\mathcal{C}|}\}\|_1$'' is exactly the hinge loss for $p=1$ and $\myVec{E}_{tol}^+=\myVec{E}_{tol}^{-}=\nu\epsilon\myVec{I}$. Hence, we can alternatively write out the above optimization problem as:
\begin{align}
\myVec{I}_1^*=&\argmin_{\myVec{I}\in\R^N}\mathcal{L}_{hinge}(\myVec{I})\myplus\nu\|\myVec{T}_f\myVec{I}\myminus\myVec{E}_{des}\|_1+\alpha\zeta\nu\|\myVec{I}\|_1,\nonumber\text{ s.t. }\myVec{I}\preceq I_{safe}\myVec{1}_N,\nonumber\\&\|\myVec{I}\|_1\leq 2I_{tot}, \text{ and } \myVec{1}_N^T\myVec{I}=0\label{eq:l1l1norm:opt:simplify:6}.
\end{align}
For brevity's sake, we introduce an extra notation: 
\begin{align}
 \mathbb{L}(\myVec{I}) =& \mathcal{L}_{hinge}(\myVec{I})+\nu\left\|\myVec{T}_{f}\myVec{I}\myminus\myVec{E}_{des} \right\|_1. 
\end{align}
Hence, we rewrite~\eqref{eq:l1l1norm:opt:simplify:6} as 
\begin{align}
    \myVec{I}^*_1 = &\argmin_{\myVec{I}\in\R^N} \mathbb{L}(\myVec{I})+\alpha\zeta\nu\|\myVec{I}\|_1,\text{ s.t. }\myVec{I}\preceq I_{safe}\myVec{1}_N,\|\myVec{I}\|_1\leq 2I_{tot}, \text{ and } \myVec{1}_N^T\myVec{I}=0,\label{eq:l1l1norm:opt:simplify:7}
\end{align}

\noindent\textbf{Part 1: Showing the equivalence of $\mathbf{L_1}$ regularizer and constraints}\\
For showing equivalence of~\eqref{eq:appx:b:hp} and~\eqref{eq:l1l1norm:opt}, we will first show that~\eqref{eq:l1l1norm:opt} is equivalent to the optimization problem stated below:  
\begin{align}
    \myVec{I}^*_3 = &\argmin_{\myVec{I}\in\R^N} \mathbb{L}(\myVec{I}),\text{ s.t. },\myVec{1}_N^T\myVec{I}=0, \myVec{I}\myminus I_{safe}\myVec{1}_N\preceq\myVec{0}_{N},\text{ and } \|\myVec{I}\|_1\myminus 2\Tilde{I}_{tot}\leq 0,\label{eq:l1l1norm:cannon-opt:no-l1norm}
\end{align}
where $\Tilde{I}_{tot}$ represents a new constraint on total current and is not necessarily equal to $I_{tot}$. 

We employ the same proof technique used earlier in Appendix A for showing the equivalence of~\eqref{eq:l1l1norm:cannon-opt:no-l1norm} and~\eqref{eq:l1l1norm:opt}. We begin by writing out the canonical form of the simplified version of L1L1-norm optimization stated in~\eqref{eq:l1l1norm:opt:simplify:7}: 
\begin{align}
    \myVec{I}^*_1 = &\argmin_{\myVec{I}\in\R^N} \mathbb{L}(\myVec{I})\myplus \alpha\zeta\nu\|\myVec{I}\|_1,\text{ s.t. },\myVec{1}_N^T\myVec{I}=0, \myVec{I}\myminus I_{safe}\myVec{1}_N\preceq\myVec{0}_{N},\nonumber\\&\text{ and } \|\myVec{I}\|_1\myminus 2I_{tot}\leq 0.\label{eq:l1l1norm:cannon-opt}
\end{align}
We now use the canonical form of~\eqref{eq:l1l1norm:opt} (stated in~\eqref{eq:l1l1norm:cannon-opt}) to write out its Lagrangian:
\begin{align}
    \mathcal{L}^{(1)}(\myVec{I})\myeq& \mathbb{L}(\myVec{I})\myplus \alpha\zeta\nu\|\myVec{I}\|_1+\mu^{(1)}\left(\myVec{1}_N^T\myVec{I}\right)+\delta^{(1)}(\|\myVec{I}\|_1-2I_{tot})+\nonumber\\&\sum_{j=1}^N\nu_j^{(1)}(\myVec{e}_j^T\myVec{I}-I_{safe}).
\end{align}
Let $\myVec{I}_1^*$, $\mu^{(1)*}$, $\delta^{(1)*}$, and $\{\nu_j^{(1)*}\}_{j=1}^d$ denote the optimal values obtained by solving the Lagrangian dual of~\eqref{eq:l1l1norm:cannon-opt}, i.e.,
\begin{align}
\left\{\begin{array}{c}
     \myVec{I}_1^*,\\
              \mu^{(1)*}\\
         \delta^{(1)*},\\ 
        \{\nu_j^{(1)*}\}_{j=1}^N,
\end{array}\right\}
\myeq \argmax_{\substack{\mu^{(1)}\in\mathbb{R},\\\delta^{(1)}\geq0,\\ \{\nu_j^{(1)}\}_{j=1}^N\geq0}}\argmin_{\myVec{I}\in\mathbb{R}^N}\mathcal{L}^{(1)}(\myVec{I}).\label{eq:l1l1norm:dual}
\end{align}
 Note that~\eqref{eq:l1l1norm:cannon-opt} is a convex optimization problem which only has affine inequality constraints (note that from Appendix A, we know that the $l_1$ constraint can be equivalently written as an affine inequality constraint). Since, $\myVec{0}_N$ always lies in the constraint set of~\eqref{eq:l1l1norm:cannon-opt}, the optimization problem stated in~\eqref{eq:l1l1norm:cannon-opt} satisfies the relaxed Slater's condition and consequently has strong duality. Consequently, the solution of~\eqref{eq:l1l1norm:dual} must satisfy its K.K.T. conditions. 
 
 We introduce some extra notations before describing the K.K.T. conditions of~\eqref{eq:l1l1norm:cannon-opt}, as the terms $\|\myVec{I}\|_1$  and $\mathbb{L}(\myVec{I})$ in the loss of~\eqref{eq:l1l1norm:cannon-opt} are not differentiable and require us to formulate the K.K.T. conditions in terms of sub-gradients and sub-differentials. We refer the reader~\cite{boyd2004convex,hiriart1996convex,bertsekas2009convex} for a discussion on sub-gradients and sub-differentials. For non-differential functions, the traditional stationarity condition of $\nabla f(\myVec{x}^*)=0$ changes to $\partial f(\myVec{x}^*)\ni \myVec{0}$, where $\partial f(\myVec{x}^*)$ (known as the sub-differential) is the set of all possible sub-gradients of $f(\cdot)$ at the point $\myVec{x}^*$~\cite{bertsekas2009convex,hiriart1996convex,boyd2004convex}. For convenience, we state some properties of sub-differentials taken from~\cite{boyd2004convex,hiriart1996convex,bertsekas2009convex}:
 \begin{enumerate}[label={\bfseries (R\arabic*)}]
     \item Let $\partial f$ be a sub-differential of $f$, then $\partial \alpha f=\alpha\partial f$ for some $\alpha\geq 0$. Note that the operation $\alpha\partial f=\{\alpha \myVec{l}| \myVec{l}\in\partial f\} $.
     \item $\partial \sum_{j=1}^K f_i=\sum_{j=1}^K\partial f_i$, where the addition on the R.H.S of the equation is a ``Minkowski sum''.
     \item Let $h(\myVec{x})=f(\myVec{T}\myVec{x}+\myVec{b})$, then $\partial h(\myVec{x})= \myVec{T}^T\partial f(\myVec{T}\myVec{x}+\myVec{b})$, where the operation $\myVec{T}^T\partial f(\myVec{T}\myVec{x}+\myVec{b})=\{\myVec{T}^T\myVec{l}|\myVec{l}\in\partial f(\myVec{T}\myVec{x}+\myVec{b})\}$.
 \end{enumerate}
We now write out the K.K.T. conditions for~\eqref{eq:l1l1norm:cannon-opt}:
\begin{align}
&\partial \mathcal{L}^{(1)}(\myVec{I}_1^*)=\partial \mathbb{L}(\myVec{I}_1^*)\myplus (\alpha\zeta\nu+\delta^{(1)*})\partial \|\myVec{I}_1^*\|_1+\mu^{(1)*}\myVec{1}_N+\sum_{j=1}^N\nu_j^{(1)*}\myVec{e}_j\ni\myVec{0}_N,\label{eq:l1l1norm:kkt:1}\\
    &\1_N^T\myVec{I}_1^* = 0,\label{eq:l1l1norm:kkt:2}\\
    &\delta^{(1)*}(\|\myVec{I}_1^*\|_1-2I_{tot}) = 0,\label{eq:l1l1norm:kkt:3}\\
    &\nu^{(1)*}_j(\myVec{e}_j^T\myVec{I}_1^*-I_{safe}) = 0,\label{eq:l1l1norm:kkt:4}\ \forall\ j\in\{1,\hdots,N\}. 
\end{align}
Similarly, we write the Largrangian of~\eqref{eq:l1l1norm:cannon-opt:no-l1norm}:
\begin{align}
    \mathcal{L}^{(3)}(\myVec{I})\myeq& \mathbb{L}(\myVec{I})+\mu^{(3)}\left(\myVec{1}_N^T\myVec{I}\right)+\delta^{(3)}(\|\myVec{I}\|_1-2I_{tot})+\sum_{j=1}^N\nu_j^{(3)}(\myVec{e}_j^T\myVec{I}-I_{safe}).\label{eq:l1l1norm:dual:nol1norm}
\end{align}
Furthermore, let $\myVec{I}_3^*$, $\mu^{(3)*}$, $\delta^{(3)*}$, and $\{\nu_j^{(3)*}\}_{j=1}^d$ denote the optimal values obtained by solving the Lagrangian dual of~\eqref{eq:l1l1norm:cannon-opt:no-l1norm}, i.e.,
\begin{align}
\left\{\begin{array}{c}
     \myVec{I}_3^*,\\
              \mu^{(3)*}\\
         \delta^{(3)*},\\ 
        \{\nu_j^{(3)*}\}_{j=1}^N,
\end{array}\right\}
    = \argmax_{\substack{\mu^{(3)}\in\mathbb{R},\\\delta_j^{(3)}\geq0,\\ \{\nu_j^{(3)}\}_{j=1}^N\geq0}}\argmin_{\myVec{I}\in\mathbb{R}^N}\mathcal{L}^{(3)}(\myVec{I}).\label{eq:l1l1norm:no-l1norm:dual}
\end{align}
The optimization problem stated in~\eqref{eq:l1l1norm:cannon-opt:no-l1norm} also satisfies the relaxed Slater's condition, as~\eqref{eq:l1l1norm:cannon-opt:no-l1norm} is a convex optimization problem with only affine inequalities constraints and $\myVec{0}_N$ always lies in its constraint set. Consequently, strong duality holds for~\eqref{eq:l1l1norm:cannon-opt:no-l1norm} and $\myVec{I}_3^*$, $\mu^{(3)*}$, $\delta^{(3)*}$, and $\{\nu_j^{(3)*}\}_{j=1}^d$ must satisfy their respective K.K.T conditions. Hence,
\begin{align}
&\partial\mathcal{L}^{(3)}(\myVec{I}_3^*)=\partial\mathbb{L}(\myVec{I}_3^*)\myplus\mu^{(3)*}\myVec{1}_N+\delta^{(3)*}\partial\|\myVec{I}_3^*\|_1+\sum_{j=1}^N\nu_j^{(3)*}\myVec{e}_j\ni\myVec{0}_N,\label{eq:l1l1norm:kkt:l1no-norm:1}\\
    &\myVec{1}_N^T\myVec{I}_3^* = 0,\label{eq:l1l1norm:kkt:l1no-norm:2}\\
    &\delta^{(3)*}(\|\myVec{I}_3^*\|_1-2\Tilde{I}_{tot}) = 0,\label{eq:l1l1norm:kkt:l1no-norm:3}\\
    &\nu^{(3)*}_j(\myVec{e}_j^T\myVec{I}_3^*-I_{safe}) = 0\ \forall\ j\in\{1,\hdots,N\}. \label{eq:l1l1norm:kkt:l1no-norm:4}
\end{align}
We now choose the value of $\Tilde{I}_{tot}$ as  $\nicefrac{\|\myVec{I}_1^*\|_1}{2}$, i.e., $\Tilde{I}_{tot}=\nicefrac{\|\myVec{I}_1^*\|_1}{2}$. Substituting $\myVec{I}_3^*=\myVec{I}_1^*$, $\delta^{(3)*}=\delta^{(1)*}+\alpha\zeta\nu$, $\nu_j^{(3)*}=\nu_j^{(1)*}\ \forall\ j\in\{1,\hdots,N\}$ in~\eqref{eq:l1l1norm:kkt:l1no-norm:1}-\eqref{eq:l1l1norm:kkt:l1no-norm:4}, we see that: 
\begin{align}
  \partial \mathcal{L}^{(3)}(\myVec{I}_1^*)=&\ \partial\mathbb{L}(\myVec{I}_1^*)+\mu^{(1)*}\myVec{1}_N+(\delta^{(1)*}+\alpha\zeta\nu)\partial\|\myVec{I}_1^*\|_1+\sum_{j=1}^N\nu^{(1)*}_j\myVec{e}_j.
  \intertext{Now note that due to~\eqref{eq:l1l1norm:kkt:1}, we know that $\partial \mathcal{L}^{(3)}(\myVec{I}_1^*)=\partial \mathcal{L}^{(1)}(\myVec{I}_1^*)$ and consequently, must contain the element $\myVec{0}_N$. Hence,}
\mathcal{L}^{(3)}(\myVec{I}_1^*)\ni&\ \myVec{0}_N.\label{eq:l1l1norm:kkt:l1no-norm:equiv:1}
\end{align}
Similarly,
\begin{flalign}
    &\underbrace{\myVec{1}_N^T\myVec{I}_1^*}_{=0 \text{ due to~\eqref{eq:l1l1norm:kkt:2}}} = 0,&\label{eq:l1l1norm:kkt:l1no-norm:equiv:2}\\
    &\underbrace{\delta^{(1)*}(\|\myVec{I}_1^*\|_1-2\Tilde{I}_{tot})}_{=0\text{ due to }\Tilde{I}_{tot}=\nicefrac{\|\myVec{I}_1^*\|_1}{2}} = 0,&\label{eq:l1l1norm:kkt:equiv:l1no-norm:3}\\
&\underbrace{\nu^{(1)*}_j(\myVec{e}_j^T\myVec{I}_1^*-I_{safe})}_{=0\text{ due to~\eqref{eq:l1l1norm:kkt:4}]}} = 0\ \forall\ j\in\{1,\hdots,N\}.& \label{eq:l1l1norm:kkt:l1no-norm:equiv:4}
\end{flalign}
Hence, the solution set $\{\myVec{I}_1^*,\mu^{(1)*},\delta^{(1)*}+\alpha\zeta\nu,\{\nu^{(1)*}_j\}_{j=1}^N\}$ also solves the K.K.T system of the optimization problem in~\eqref{eq:l1l1norm:cannon-opt:no-l1norm}, specified in~\eqref{eq:l1l1norm:kkt:l1no-norm:1}-\eqref{eq:l1l1norm:kkt:l1no-norm:4}. Consequently, $\{\myVec{I}_1^*,\mu^{(1)*},\delta^{(1)*}+\alpha\zeta\nu,\{\nu^{(1)*}_j\}_{j=1}^N\}$ solves the Lagrangian dual of~\eqref{eq:l1l1norm:cannon-opt:no-l1norm} specified in~\eqref{eq:l1l1norm:dual:nol1norm}. Hence, $\myVec{I}_1^*$ is also the solution to the optimization problems specified in~\eqref{eq:l1l1norm:cannon-opt:no-l1norm}. 

Similarly, we can also show the other direction, namely, that $\myVec{I}_3^*$ is a solution to the~\eqref{eq:l1l1norm:cannon-opt}. Let us assume that $\Tilde{I}_{tot}$ is chosen small enough that $\|\myVec{I}_3^*\|_1-2\Tilde{I}_{tot}=0$, i.e., the $l_1$ constraint is achieved at the boundary which implies $\delta^{(3)*}\neq 0$. Then, either choosing $I_{tot}>\nicefrac{\|\myVec{I}_3^*\|_1}{2}$, $\alpha=\nicefrac{\delta^{(3)*}}{\nu\zeta}$, $\myVec{I}_1^*=\myVec{I}_3^*$, $\mu^{(1)*}=\mu^{(3)*}$, $\nu^{(1)*}=\nu^{(3)*}$, and $\delta^{(1)*}=0$ (due to complementary slackness as $\|\myVec{I}_3^*\|_1<I_{tot}$), or $I_{tot}=\nicefrac{\|\myVec{I}_3^*\|_1}{2}$, $\alpha<\nicefrac{\delta^{(3)*}}{\nu\zeta}$, $\myVec{I}_1^*=\myVec{I}_3^*$, $\mu^{(1)*}=\mu^{(3)*}$, $\nu^{(1)*}=\nu^{(3)*}$, and $\delta^{(1)*}=\delta^{(3)*}-\alpha\zeta\nu$ would solve the K.K.T. system of~\eqref{eq:l1l1norm:cannon-opt} specified in~\eqref{eq:l1l1norm:kkt:1}-\eqref{eq:l1l1norm:kkt:4}. Hence, $\myVec{I}_3^*$ is also a solution to~\eqref{eq:l1l1norm:cannon-opt}.

Consequently, the solutions of~\eqref{eq:l1l1norm:cannon-opt} and~\eqref{eq:l1l1norm:cannon-opt:no-l1norm} are the same. Therefore,~\eqref{eq:l1l1norm:cannon-opt}  can be equivalently written as~\eqref{eq:l1l1norm:cannon-opt:no-l1norm} where the effect of $l_1$-regularizer in~\eqref{eq:l1l1norm:cannon-opt} can be alternatively achieved by appropriately reducing the value of total current constraint, i.e., reducing the value of $I_{tot}$ to $\Tilde{I}_{tot}$.
\begin{conclusion}\label{conc:1}
    From the above analysis, we can conclude that the optimization problem proposed in~\eqref{eq:l1l1norm:opt} can be equivalently represented using~\eqref{eq:l1l1norm:cannon-opt:no-l1norm}.
\end{conclusion}

\noindent\textbf{Part 2: Showing that the solutions of~\eqref{eq:l1l1norm:cannon-opt:no-l1norm} can be equivalently obtained by solving~\eqref{eq:appx:b:hp}}

We now show that solutions of~\eqref{eq:l1l1norm:cannon-opt:no-l1norm} can be equivalently obtained by solving~\eqref{eq:appx:b:hp} to conclude that solutions of~\eqref{eq:l1l1norm:opt} can be equivalently obtained by solving~\eqref{eq:appx:b:hp}.
We re-write the optimization of~\eqref{eq:l1l1norm:cannon-opt:no-l1norm} (expanding the term $\mathbb{L}(\myVec{I})$) to make the connections between~\eqref{eq:appx:b:hp} and~\eqref{eq:l1l1norm:cannon-opt} more apparent. 
\begin{align}
\myVec{I}_3^*=&\argmin_{\myVec{I}\in\R^N}\mathcal{L}_{hinge}(\myVec{I})\myplus\nu\|\myVec{T}_f\myVec{I}\myminus\myVec{E}_{des}\|_1,\text{ s.t. }\myVec{1}_N^T\myVec{I}=0,\myVec{I}-I_{safe}\myVec{1}_N\preceq\myVec{0}_N,\nonumber\\&\text{ and }\|\myVec{I}\|_1-2\Tilde{I}_{tot}\leq 0.\label{eq:l1l1norm:canon-opt:wls-hinge}
\end{align}
We will now show that~\eqref{eq:l1l1norm:canon-opt:wls-hinge} is equivalent to~\eqref{eq:appx:b:hp}. We begin by writing the canonical form of~\eqref{eq:appx:b:hp}:
\begin{align}
\myVec{I}_2^*=&\argmin_{\myVec{I}\in\R^N}\mathcal{L}_{hinge}(\myVec{I})\text{ s.t. }\myVec{1}_N^T\myVec{I}=0, \myVec{T}_f\myVec{I}\myminus\widetilde{\myVec{E}_{des}}\myeq \myVec{0}_{|\mathcal{F}|}, \myVec{I}-I_{safe}\myVec{1}_N\preceq\myVec{0}_N,\nonumber\\&\text{ and }\|\myVec{I}\|_1-2\Tilde{I}_{tot}\leq 0,\label{eq:l1l1norm:canon-opt:hinge}
\end{align}
where we choose the value of $\widetilde{\myVec{E}_{des}}=\myVec{T}_f\myVec{I}_3^*$. Now note that~\eqref{eq:l1l1norm:canon-opt:hinge} is a convex optimization problem with only affine inequalities. Furthermore, by the particular choice of $\widetilde{\myVec{E}_{des}}=\myVec{T}_f\myVec{I}_3^*$, we know that $\myVec{I}_3^*$ lies in the constraint set of~\eqref{eq:l1l1norm:canon-opt:hinge}. Consequently,~\eqref{eq:l1l1norm:canon-opt:hinge} satisfies the relaxed Slater's condition and strong duality holds for~\eqref{eq:l1l1norm:cannon-opt}. We now describe the corresponding Lagrangian dual and the K.K.T. conditions for~\eqref{eq:l1l1norm:canon-opt:hinge}, which we will utilize to show the equivalence between~\eqref{eq:l1l1norm:canon-opt:hinge} and~\eqref{eq:l1l1norm:canon-opt:wls-hinge}.  

\noindent\textbf{Lagrangian dual of~\eqref{eq:l1l1norm:canon-opt:hinge}}:
\begin{align}
    \mathcal{L}^{(2)}(\myVec{I})\myeq& \mathcal{L}_{hinge}(\myVec{I})+\boldsymbol{\beta}^T(\myVec{T}_f\myVec{I}-\myVec{E}_{des})+\mu^{(2)}\left(\myVec{1}_N^T\myVec{I}\right)+\delta^{(2)}(\|\myVec{I}\|_1-2I_{tot})+\nonumber\\&\sum_{j=1}^N\nu_j^{(2)}(\myVec{e}_j^T\myVec{I}-I_{safe}),\label{eq:dual:2}
\end{align}
where $\boldsymbol{\beta}=\begin{bmatrix}
    \beta_1&\hdots&\beta_{F}
\end{bmatrix}\in\R^d$.\\

\noindent\textbf{K.K.T. of~\eqref{eq:l1l1norm:canon-opt:hinge}}:
\begin{align}
  &\partial \mathcal{L}^{(2)}(\myVec{I}_2^*)=\partial\mathcal{L}_{hinge}(\myVec{I}_2^*)\myplus\myVec{T}_f^T\boldsymbol{\beta}^{*} \myplus\delta^{(2)*}\|\partial\myVec{I}_2^*\|_1+\mu^{(2)*}\myVec{1}_N+\sum_{j=1}^N\nu_j^{(2)*}\myVec{e}_j\ni\myVec{0}_N,\label{eq:kkt-2:1}\\
    &\1_N^T\myVec{I}_2^* = 0,\label{eq:kkt-2:2}\\
    &\delta^{(2)*}(\|\myVec{I}_2^*\|_1-2\Tilde{I}_{tot}) = 0,\label{eq:kkt-2:3}\\
    &\nu^{(2)*}_j(\myVec{e}_j^T\myVec{I}_2^*-I_{safe}) = 0\ \forall\ j\in\{1,\hdots,N\}, \label{eq:kkt-2:4}\\
    &\myVec{T}_f\myVec{I}_2^*-\widetilde{\myVec{E}_{des}}=\myVec{0}_{|\mathcal{F}|}.\label{eq:kkt-2:5}
\end{align}
From lemma~\ref{lemma:2}, we also can simplify~\eqref{eq:l1l1norm:kkt:l1no-norm:1} as follows:
\begin{align}
    &\partial \mathcal{L}^{(3)}(\myVec{I}_3^*)=\ \partial\mathcal{L}_{hinge}(\myVec{I}_3^*)+\myVec{T}_f^T(\nu\partial\|\myVec{T}_f\myVec{I}_3^*-\myVec{E}_{des}\|_1)+\delta^{(3)*}\partial\|\myVec{I}_3^*\|_1+\nonumber\\&\sum_{j=1}^N\nu^{(3)*}_j\myVec{e}_j+\mu^{(3)*}\myVec{1}_N\ni\myVec{0}_N.
\end{align}
Let $\myVec{T}\in\partial\|\myVec{T}_f\myVec{I}_3^*-\myVec{E}_{des}\|$, $\myVec{b}\in\partial\|\myVec{I}_3^*\|_1$, and $\myVec{c}\in\partial\mathcal{L}_{hinge}(\myVec{I}_3^*)$ such that:
\begin{align}
    \myVec{c}\myplus\myVec{T}_f^T(\nu\myVec{T})\myplus \delta^{(3)*}\myVec{b}\myplus\sum_{j=1}^N\nu^{(3)*}_j\myVec{e}_j\myplus\mu^{(3)*}\myVec{1}_N=\myVec{0}_N.\label{eq:final:?}
\end{align}
Note that by~\eqref{eq:l1l1norm:kkt:l1no-norm:1}, we are guaranteed the existence of at least one such $\myVec{T}$, $\myVec{b}$, and $\myVec{c}$. Then, we make the following substitutions in K.K.T. conditions of~\eqref{eq:l1l1norm:canon-opt:hinge}, i.e., in~\eqref{eq:kkt-2:1}-~\eqref{eq:kkt-2:1}:
\begin{align}
&\myVec{I}_2^*=\myVec{I}_3^*,\\
&\boldsymbol{\beta}^* = \nu\myVec{b},\\
&\delta^{(2)*}=\delta^{(3)*},\\
&\nu_j^{(2)*}=\nu_j^{(3)*}\ \forall\ j\in\{1,\hdots,N\},\\
&\mu^{(2)*}=\mu^{(3)*}.
\end{align}
After making the above substitutions, we can make the following conclusions. First, note that
\begin{align}
    \partial \mathcal{L}^{(2)}(\myVec{I}_3^*) =& \partial\mathcal{L}_{hinge}(\myVec{I}_3^*)\myplus\myVec{T}_f^T(\nu\myVec{b})\myplus\delta^{(3)*}\partial\|\myVec{I}_3^*\|_1\myplus\mu^{(3)*}\myVec{1}_N+\sum_{j=1}^N\nu^{(3)*}_j\myVec{e}_j.
\end{align}
We know that $\myVec{T}\in\partial\mathcal{L}_{hinge}(\myVec{I}_3^*)$ and $\myVec{c}\in\partial\|\myVec{I}_3^*\|_1$ by their definition in~\eqref{eq:final:?}. Hence, we can conclude that:
\begin{align}
      \partial \mathcal{L}^{(2)}(\myVec{I}_3^*)\ni \myVec{c}\myplus\myVec{T}_f^T&(\nu\myVec{T})\myplus \delta^{(3)*}\myVec{b}\myplus\sum_{j=1}^N\nu^{(3)*}_j\myVec{e}_j\myplus\mu^{(3)*}\myVec{1}_N.\label{eq:L}
      \intertext{Combining~\eqref{eq:L} and~\eqref{eq:final:?}, we can conclude that:}
    \partial \mathcal{L}^{(2)}(\myVec{I}_3^*)\ni \myVec{0}_N.\label{eq:kkt-equiv:1}
\end{align}
Similarly,
\begin{align}
    &\underbrace{\1_N^T\myVec{I}_3^*}_{=0\text{ due to~\eqref{eq:l1l1norm:kkt:l1no-norm:2}}} = 0,\label{eq:kkt-equiv:2}\\
    &\underbrace{\delta^{(3)*}(\|\myVec{I}_3^*\|_1-2\Tilde{I}_{tot})}_{=0\text{ due to~\eqref{eq:l1l1norm:kkt:l1no-norm:3}}} = 0,\label{eq:kkt-equiv:3}\\
&\underbrace{\nu^{(3)*}_j(\myVec{e}_j^T\myVec{I}_3^*-I_{safe})}_{=0\text{ due to~\eqref{eq:l1l1norm:kkt:l1no-norm:4}}} = 0\ \forall\ j\in\{1,\hdots,N\}, \label{eq:kkt-equiv:4}\\
    &\underbrace{\myVec{T}_f\myVec{I}_3^*-\widetilde{\myVec{E}_{des}}}_{=\myVec{0}_{|\mathcal{F}|}\text{ due to }\widetilde{\myVec{E}_{des}}=\myVec{T}_f\myVec{I}_3^*}=\myVec{0}_{|\mathcal{F}|}.\label{eq:kkt-equiv:5}
\end{align}
Hence, the solution set $\{\myVec{I}_3^*,\nu\myVec{b},\delta^{(3)*},\mu^{(3)*},\{\nu_j^{(3)*}\}_{j=1}^{N}\}$ also solves the K.K.T. system of~\eqref{eq:l1l1norm:canon-opt:hinge} specified in~\eqref{eq:kkt-2:1}-\eqref{eq:kkt-2:5}. Consequently, we know that $\myVec{I}_3^*$ must also be a solution of~\eqref{eq:l1l1norm:canon-opt:hinge}. Hence, all solutions obtained using~\eqref{eq:l1l1norm:cannon-opt:no-l1norm} can also be equivalently obtained by solving~\eqref{eq:l1l1norm:canon-opt:hinge}.

\begin{conclusion}\label{conc:2}
    From the above analysis, we can conclude that the solutions of the optimization problem proposed in~\eqref{eq:l1l1norm:cannon-opt:no-l1norm} can be equivalently obtained by solving~\eqref{eq:appx:b:hp}.
\end{conclusion}

\noindent\textbf{Combining the results of parts 1 and 2}

Combining conclusions~\ref{conc:1} and~\ref{conc:2}, we can conclude that the solutions of the optimization problem in~\eqref{eq:l1l1norm:opt} can be equivalently obtained by solving~\eqref{eq:l1l1norm:canon-opt:hinge} for appropriately chosen values of $\widetilde{E_{des}}$, $\Tilde{I}_{tot}$, $\myVec{E}_{tol}^+$, and $\myVec{E}_{tol}^+$. For the sake of convenience, we write out the optimization problems of~\eqref{eq:l1l1norm:opt} and~\eqref{eq:l1l1norm:canon-opt:hinge} in their non-canonical forms: 
\begin{align*}
\intertext{Optimization problem proposed in~\cite{prieto2022l1}}
    \myVec{I}^* = &\argmin_{\myVec{I}\in\R^N} \left\|\begin{array}{c}
          \myVec{T}_{f}\myVec{I}-\myVec{E}_{des} \\
          \Psi_{\epsilon}\left[\nu^{-1}\myVec{T}_{c}\myVec{I}\right]
    \end{array}\right\|_1\myplus \alpha\zeta\|\myVec{I}\|_1,\text{ s.t. }\myVec{I}\preceq I_{safe}\myVec{1}_N,\|\myVec{I}\|_1\leq 2I_{tot},\nonumber\\ &\text{ and } \myVec{1}_N^T\myVec{I}=0.\\
    \intertext{Non-canonical form of~\eqref{eq:l1l1norm:canon-opt:hinge}}
        \myVec{I}^*=&\argmin_{\myVec{I}\in\R^N}
            \|\max\{\myVec{0}_{|\mathcal{C}|},\myVec{T}_c\myVec{I}\myminus\epsilon\nu\myVec{1}_N\}\|_1+\|\max\{\myVec{0}_{|\mathcal{C}|},\myminus\myVec{T}_c\myVec{I}\myminus\epsilon\nu\myVec{1}_N\}\|_1
        ,\nonumber\\
    &\text{ s.t. }\myVec{1}_N^T\myVec{I}\myeq0, \myVec{T}_f\myVec{I}\myeq \widetilde{\myVec{E}_{des}},\myVec{I}\preceq I_{safe}\myVec{1}_N,\text{and }\|\myVec{I}\|_1\mylesseq 2\Tilde{I}_{tot}. 
\end{align*}
Now note that the only difference between the non-canonical form of~\eqref{eq:l1l1norm:canon-opt:hinge} and our \HP formulation described in~\eqref{eq:algo:HP-main-opt} (for $p=1$) is in the constraint $\|\myVec{I}\|_{\infty}\leq I_{safe}$ (in~\eqref{eq:algo:HP-main-opt}) and the one-sided version of this $l_{\infty}$ constraint in~\eqref{eq:l1l1norm:canon-opt:hinge}, i.e., $\myVec{I}\preceq I_{safe}\myVec{1}_N$. Since, the $l_{\infty}$ constraint is used to ensure that maximum electrode per electrode is less than the specified limit, which is not ensured by the constraint $\myVec{I}\preceq I_{safe}\myVec{1}_N$ (it only ensures that the positive current is less than $I_{safe}$), we believe it is more appropriate to use the $l_{\infty}$ constraint. Hence, swapping out the one-sided constraint $\myVec{I}\preceq I_{safe}\myVec{1}_N$ in the optimization problem of Prieto et al.~\cite{prieto2022l1} with $l_{\infty}$ constraint, we find that the solutiona obtained by L1L1-norm optimization can be equivalently obtained by solving a special case of our \HP formulation with $p=1$ and $\myVec{E}_{tol}^+=\myVec{E}_{tol}^{-}=E_{tol}\myVec{1}_{|\mathcal{C}|}$.  

\subsection{Numerical Simulations}
\begin{figure*}[ht]
    \centering
\includegraphics[width=\textwidth]{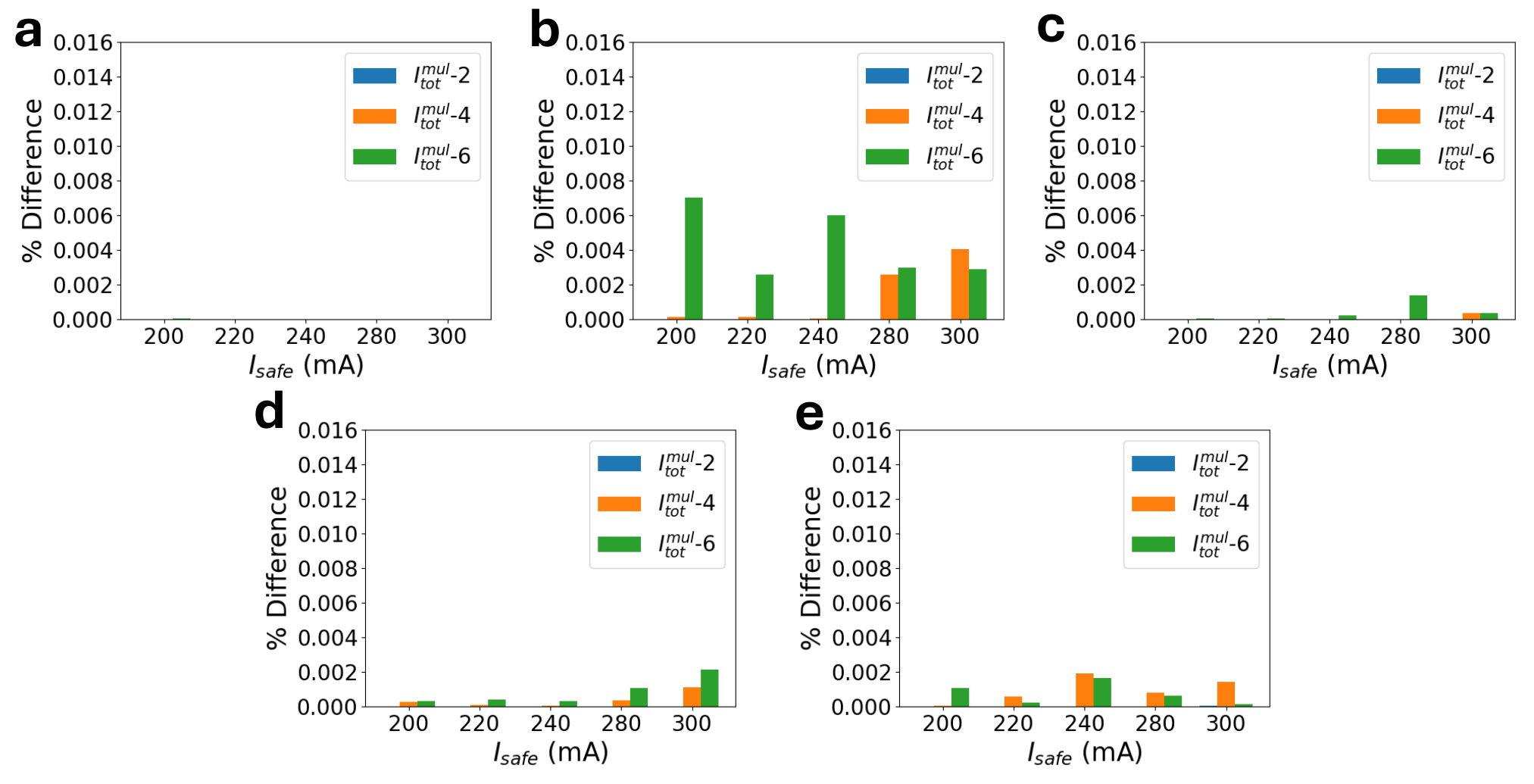}
    \caption{\textbf{a}-\textbf{e} show the difference between the outputs of the L1L1-norm and \HP approaches for targets locations $1$, $2$, $3$, $4$, and $5$, respectively discussed in the Numerical Simulations section in Appendix C. The difference between the outputs of L1L1-norm ($\myVec{I}_{L1L1}^*$) and \HP ($\myVec{I}_{HP}^*$) was measured as $\nicefrac{\|\myVec{I}^*_{L1L1}-\myVec{I}^*_{HP}\|_1}{\|\myVec{I}^*_{HP}\|_1}\times 100$. At each target, we calculated the difference between L1L1 and \HP outputs at five different values of $I_{safe}$ and at three different values of $I_{tot}^{mul}$ resulting in $15$ different combinations of $I_{safe}$ and $I_{tot}$. Each bar plot corresponds to the median difference between $\myVec{I}_{L1L1}^*$ and $\myVec{I}_{HP}^*$ across $25$ different values of the pair $(\alpha,\epsilon))$.}
    \label{fig:l1l1-and-hp-equiv}
\end{figure*}

We also provide a numerical illustration of our theoretical results discussed in the previous Proof section. For this study, we used the spherical head model used in our sea of neurons models. Appendix D contains the details of the spherical head model. The target region was chosen as a disc of $1$cm centered at the target location. The corresponding off-target region was constructed as a hollow disc of inner-radius $1.1$cm and outer radius $7$cm. We use the results of the proof section above to transform the L1L1-norm optimization problem into the \HP optimization problem by choosing $\widetilde{\myVec{E}_{des}}=\myVec{A}_f\myVec{I}^*_{L1L1}$, $I_{tot}=\|\myVec{I}^*_{L1L1}\|_1/2$, and $\myVec{E}_{tol}^+=\myVec{E}_{tol}^{-}=\nu\epsilon\myVec{1}_{|\mathcal{C}|}$, where $\myVec{I}_{L1L1}^*$ is the solution of the L1L1-norm optimization problem. We perform the following study to numerically verify our theoretical results
\begin{enumerate}
    \item We first solve the L1L1-norm optimization problem and obtain the solution $\myVec{I}^*_{L1L1}$.
    \item Using $\myVec{I}^*_{L1L1}$, we specify the values of the values of $\myVec{E}_{des}=\myVec{A}_f\myVec{I}^*_{L1L1}$, $I_{tot}=\|\myVec{I}^*_{L1L1}\|_1/2$, and $\myVec{E}_{tol}^+=\myVec{E}_{tol}^{-}=\nu\epsilon\myVec{1}_{|\mathcal{C}|}$ in the modified \HP optimization problem (stated in~\eqref{eq:appx:b:hp}), and obtain the corresponding solution $\myVec{I}_{HP}^*$. 
    \item We calculate the difference between $\myVec{I}_{L1L1}^*$ and $\myVec{I}_{HP}^*$ empirically by, $$\frac{\|\myVec{I}_{L1L1}^*-\myVec{I}_{HP}^*\|_1}{\|\myVec{I}_{HP}^*\|_1}\times 100$$.
\end{enumerate}
We quantified the difference between $\myVec{I}_{L1L1}^*$ and $\myVec{I}_{HP}^*$ for five different target locations. The corresponding spherical coordinates for the five targets are as follows: 
\begin{itemize}
    \item Target location 1: $[7.7,0,0]$ (north pole at the depth of $1.5$cm)
    \item Target location 2: $[7.7,\nicefrac{2}{7.7},0]$ ($+2$cm along the $x$-axis from the north pole)
    \item Target location 3: $[7.7,\nicefrac{2}{7.7},\nicefrac{\pi}{2}]$ ($+2$cm along the $y$-axis from the north pole)
    \item Target location 4: $[7.7,\nicefrac{2}{7.7},\pi]$ ($-2$cm along the $x$-axis from the north pole)
    \item Target location 5: $[7.7,\nicefrac{2}{7.7},\nicefrac{3\pi}{2}]$ ($-2$cm along the $y$-axis from the north pole). 
\end{itemize}
For each location, we tested five different values of $I_{safe}$, namely, $200$mA, $220$mA, $240$mA, $280$mA, and $300$mA. The $I_{tot}$ was parameterized in a similar manner as done in Sec.~\ref{sec:results}, with $I_{tot}=I_{tot}^{mul}I_{safe}$. For each value of $I_{safe}$, three different values of $I_{tot}^{mul}$ were tested, namely, $2$, $4$, and $6$ (similar to our studies in Sec.~\ref{sec:results}). For each target location, $I_{safe}$, and $I_{tot}^{mul}$, $25$ different values of the pair $(\alpha,\epsilon)$ were tested. Five different values of $\epsilon$ were chosen on a equally spaced log-scale between the values $1$ and $0.01$. Similarly, five values of $\alpha$ were chosen on a equally spaced log-scale between the values $0.1$ and $10^{-5}$. All $25$ possible pairs of $(\alpha,\epsilon)$ were tested. Fig.~\ref{fig:l1l1-and-hp-equiv} shows the aggregate differences between $\myVec{I}_{L1L1}^*$ and $\myVec{I}_{HP}^*$. To ensure uniqueness of the solution in L1L1-norm optimization problem and modified \HP optimization problem, we added a small amount of $l_2$ norm to the objectives of both optimization. Specifically, we added $10^{-9}\|\myVec{I}\|_2$ and $\nu10^{-9}\|\myVec{I}\|_2$ to the objectives of L1L1-norm and modified \HP algorithms, respectively. 

Fig.~\ref{fig:l1l1-and-hp-equiv}(a)-(e) represent the five different target locations, with (a), (b), (c), (d), and (e) representing target locations $1$, $2$, $3$, $4$, and $5$, respectively. Each bar plot shows the median difference across the $25$ different values of $\alpha$ and $\epsilon$ for each $I_{safe}$ and $I_{tot}^{mul}$ value. Fig.~\ref{fig:l1l1-and-hp-equiv} shows that across all the parameters tested, the corresponding $\myVec{I}_{L1L1}^*$ and $\myVec{I}_{HP}^*$ are same (except for small differences due to numerical noise) complementing our theoretical results that predict $\myVec{I}_{HP}^*=\myVec{I}_{L1L1}^*$.
\subsection{Lemmas}
\begin{lemma}\label{lemma:l1l1norm:hp-equiv}
    Let $\myVec{w}\in\R^d$ and $\epsilon>0$. Then,
    \begin{align}
       \|\Psi_{\epsilon}\left[\myVec{w}\right]\|_1 =\left\|\max\{\epsilon\myVec{1}_{d},|\myVec{w}|\}\right\|_1
       =&\left\|\max\{\myVec{0}_d,\myVec{w}\myminus\epsilon\myVec{1}_d\}\right\|_1+\epsilon d+\nonumber\\ &\left\|\max\{\myVec{0}_d,-\myVec{w}-\epsilon\myVec{1}_d\}\right\|_1.\nonumber
    \end{align}
\end{lemma}
\begin{proof}
    Let $\myVec{w}=\begin{bmatrix}
        w_1&\hdots&w_d
    \end{bmatrix}^T$. Then, expanding the vector $\max\{\epsilon\myVec{1}_d,\myVec{w}\}$ in terms of its element, we obtain:
    \begin{align}
        \max\{\epsilon\myVec{1}_d,|\myVec{w}|\}=\begin{bmatrix}
            \max\{\epsilon,|w_1|\}\\
            \vdots\\
            \max\{\epsilon,|w_d|\}\label{eq:l1l1norm:hp-equiv:1}
        \end{bmatrix}.
    \end{align}
    We now use the fact that $\max\{\epsilon,|w_i|\}=\max\{0,|w_i|-\epsilon\}+\epsilon\ \forall\ i\in\{1,\hdots,d\}$ in~\eqref{eq:l1l1norm:hp-equiv:1} to show:
    \begin{align}
        \max\{\epsilon\myVec{1}_d,|\myVec{w}|\}&=\begin{bmatrix}
            \max\{0,|w_1|-\epsilon\}+\epsilon\\
            \vdots\\
            \max\{0,|w_d|-\epsilon\}+\epsilon
        \end{bmatrix}\label{eq:l1l1norm:hp-equiv:2},\\
        &=\begin{bmatrix}
            \max\{0,|w_1|-\epsilon\}\\
            \vdots\\
            \max\{0,|w_d|-\epsilon\}
        \end{bmatrix}+\begin{bmatrix}
            \epsilon\\
            \vdots\\
            \epsilon
        \end{bmatrix},\label{eq:l1l1norm:hp-equiv:3}\\
        &=\max\{\myVec{0}_d,|\myVec{w}|-\epsilon\myVec{1}_d\}+\epsilon\myVec{1}_d.\label{eq:l1l1norm:hp-equiv:4}
    \end{align}
    Now, we derive another fact regarding $\max\{0,|\kappa|-\epsilon\}$, namely that $\max\{0,|\kappa|-\epsilon\}=\max\{0,\kappa-\epsilon\}+\max\{0,-\kappa-\epsilon\}$. First, let us consider the case $\kappa\geq 0$:
    \begin{align}
        \max\{0,|\kappa|-\epsilon\}&\overset{(a)}{=}\max\{0,\kappa-\epsilon\},\nonumber\\
        &\overset{(b)}{=}\max\{0,\kappa-\epsilon\}+\max\{0,-\kappa-\epsilon\},\label{eq:l1l1norm:hp-equiv:5}
    \end{align}
    where $(a)$ is due to the assumption that $\kappa\geq 0$ and $(b)$ stems from the fact that $\max\{0,-\kappa-\epsilon\}=0$ due to the fact that $-(\kappa+\epsilon)\leq 0$ if we assume $\kappa\geq 0$ and we know that $\epsilon>0$ by the assumption in the lemma statement. Now, we consider the case $\kappa\leq 0$, then we have:  
        \begin{align}
        \max\{0,|\kappa|-\epsilon\}&\overset{(c)}{=}\max\{0,-\kappa-\epsilon\},\nonumber\\
        &\overset{(d)}{=}\max\{0,\kappa-\epsilon\}+\max\{0,-\kappa-\epsilon\},\label{eq:l1l1norm:hp-equiv:6}
    \end{align}
    where $(c)$ is due to the assumption that $\kappa\leq 0$ and $(b)$ stems from the fact that $\max\{0,\kappa-\epsilon\}=0$ due to the fact that $\kappa-\epsilon)\leq 0$ if we assume $\kappa\leq 0$ and we know that $\epsilon>0$ by the assumption in the lemma statement. Combining~\eqref{eq:l1l1norm:hp-equiv:5} and~\eqref{eq:l1l1norm:hp-equiv:6}, we obtain:
    \begin{align}
         \max\{0,|\kappa|-\epsilon\}=\max\{0,\kappa-\epsilon\}+\max\{0,-\kappa-\epsilon\}.\label{eq:l1l1norm:hp-equiv:7}
    \end{align}
    Substituting the result of~\eqref{eq:l1l1norm:hp-equiv:7} in~\eqref{eq:l1l1norm:hp-equiv:4}, we can simplify $\max\{\epsilon\myVec{1}_d, |\myVec{w}|\}$ as:
    \begin{align}
        \max\{\epsilon\myVec{1}_d,|\myVec{w}|\}=&\max\{\myVec{0}_d,|\myVec{w}|-\epsilon\myVec{1}_d\}+\epsilon\myVec{1}_d,\nonumber\\
        =&\begin{bmatrix}
            \max\{0,|w_1|-\epsilon\}\\
            \vdots\\
            \max\{0,|w_d|-\epsilon\}
        \end{bmatrix}+\epsilon\myVec{1}_d,\nonumber\\
        =&\begin{bmatrix}
            \max\{0,w_1-\epsilon\}+\max\{0,\myminus w_1\myminus \epsilon\}\\
            \vdots\\
            \max\{0,w_d-\epsilon\}+\max\{0,\myminus w_d\myminus\epsilon\}
        \end{bmatrix}+\epsilon\myVec{1}_d,\nonumber\\
        =&\begin{bmatrix}
            \max\{0,w_1-\epsilon\}\\
            \vdots\\
            \max\{0,w_d.-\epsilon\}
        \end{bmatrix}+\epsilon\myVec{1}_d+\begin{bmatrix}
           \max\{0,\myminus w_1\myminus \epsilon\}\\
            \vdots\\
           \max\{0,\myminus w_d\myminus\epsilon\}
        \end{bmatrix},\nonumber\\
        =&\max\{\myVec{0}_d,\myVec{w}-\epsilon\myVec{1}_d\}+\epsilon\myVec{1}_d+\max\{\myVec{0}_d,-\myVec{w}-\epsilon\myVec{1}_d\}.\label{eq:l1l1norm:hp-equiv:main}
    \end{align}
    Hence, from~\eqref{eq:l1l1norm:hp-equiv:main}, we can conclude:
    \begin{align}
        \|\max\{\epsilon\myVec{1}_d,|\myVec{w}|\}\|_1=&\|\max\{\myVec{0}_d,\myVec{w}-\epsilon\myVec{1}_d\}+\epsilon\myVec{1}_d+\max\{\myVec{0}_d,-\myVec{w}-\epsilon\myVec{1}_d\}\|_1.
    \end{align}
    We now use property~\ref{prop:appx-b:l1norm-equiv} in the above equation as the vectors $\epsilon\myVec{1}_d$, $\max\{\myVec{0}_d,\myVec{w}-\epsilon\myVec{1}_d\}$, and $\max\{\myVec{0}_d,-\myVec{w}-\epsilon\myVec{1}_d\}$ only have non-negative elements to show the lemma statement and conclude the proof:
    \begin{align}
        \|\max\{\epsilon\myVec{1}_d,|\myVec{w}|\}\|_1\myeq&\|\max\{\myVec{0}_d,\myVec{w}-\epsilon\myVec{1}_d\}\|_1\myplus\|\epsilon\myVec{1}_d\|_1\myplus\|\max\{\myVec{0}_d,-\myVec{w}-\epsilon\myVec{1}_d\}\|_1,\nonumber\\
        =&\|\max\{\myVec{0}_d,\myVec{w}-\epsilon\myVec{1}_d\}\|_1+\epsilon d+\|\max\{\myVec{0}_d,-\myVec{w}-\epsilon\myVec{1}_d\}\|_1.
    \end{align}    
\end{proof}
\begin{lemma}\label{lemma:2}
    Let $\mathbb{L}(\myVec{I})$ be as defined in~\eqref{eq:l1l1norm:opt:simplify:7}, then
    \begin{align}
        \partial \mathbb{L}(\myVec{I})=\partial\mathcal{L}(\myVec{I})+\nu\myVec{T}_f^T\partial\|\myVec{T}_f\myVec{I}-\myVec{E}_{des}\|_1^1.
    \end{align}
\end{lemma}

\begin{proof}
    From~\eqref{eq:l1l1norm:opt:simplify:7}, we know that:
    \begin{align}
        \mathbb{L}(\myVec{I})=\mathcal{L}_{hinge}(\myVec{I})+\nu\|\myVec{T}_f^T\myVec{I}-\myVec{E}_{des}\|_1.
        \intertext{Taking sub-differential on both sides, we obtain:}
        \partial\mathbb{L}(\myVec{I})=\partial\mathcal{L}_{hinge}(\myVec{I})+\partial\nu\|\myVec{T}_f^T\myVec{I}-\myVec{E}_{des}\|_1.
        \intertext{where the above inequality comes from property \textbf{(R2)}. Using property \textbf{(R1)} and \textbf{(R3)} in the above equation yields:}
        \partial\mathbb{L}(\myVec{I})=\partial\mathcal{L}_{hinge}(\myVec{I})+\nu\myVec{T}_f^T\partial\|\myVec{A}_f^T\myVec{I}-\myVec{E}_{des}\|_1.
    \end{align}
\end{proof}
\begin{property}\label{prop:appx-b:l1norm-equiv}
    If $d$-dimensional real vectors $\myVec{x},\myVec{y},\text{ and }\myVec{z}$ only have non-negative elements then:
    \begin{align}       \|\myVec{x}+\myVec{y}+\myVec{z}\|_1=\|\myVec{x}\|_1+\|\myVec{y}\|_1+\|\myVec{z}\|_1.\label{eq:l1l1norm:hp-equiv:norm-equiv} 
    \end{align}
\end{property}
\begin{proof}
    The proof of~\eqref{eq:l1l1norm:hp-equiv:norm-equiv} can be easily obtained by expanding the L.H.S and R.H.S of the above equation:
    \begin{align}
        \sum_{i=1}^d|x_i+y_i+z_i| =& \sum_{i=1}^d|x_i|+\sum_{i=1}^d|y_i|+\sum_{i=1}^d|z_i|,\nonumber\\
        \overset{(e)}{\Rightarrow} \sum_{i=1}^dx_i+y_i+z_i =& \sum_{i=1}^dx_i+\sum_{i=1}^dy_i+\sum_{i=1}^dz_i,\\
        &=\sum_{i=1}^d x_i+y_i+z_i,
    \end{align}
    where the implication $(e)$ follows from the fact that $x_i,y_i,z_i\geq 0\ \forall\ i\in\{1,\hdots,d\}$ by the assumption in the property statement.
\end{proof}
\newpage

\section{}\label{appx:d}

\subsection{Realistic MRI head model}
We compared the \HP and LCMV-E algorithms in a realistic MRI head model simulated using a standard software ROAST~\cite{huang2019realistic}, which was previously used in numerous studies (e.g.,~\cite{Hermann2020,farahani2024transcranial}) to model tES. We used the default MNI152 head MRI with a $1$mm resolution provided by the ROAST software for constructing the head model. The electrode locations were a subset of all possible electrode locations under the 10-05 EEG electrode placement scheme~\cite{oostenveld2001five}. We only considered the electrode locations within $8$cm of the Cz electrode locations, amounting to $69$ different electrode locations. The ground electrode used for constructing the forward matrix was placed at the ``AF1'' location. Each electrode was modeled as a disc electrode with a height of $2$mm and a radius of $6$mm (the default values in ROAST). All other parameters were chosen to be their default values provided in the software. A visualization of the head model with the corresponding electrodes is provided in Fig.~\ref{fig:setup}a-b. 

\textbf{Quantifying neural activation}: \HP primarily relies on the thresholding behavior of neurons to create focal neural responses. Traditional figures of merits used in previous works, such as full-width half-max radius~\cite{dmochowski2011optimized} and integral focality~\cite{fernandez2020unification}, measure the focality of \emph{electric fields} and do not account for the thresholding behavior of neurons. Consequently, these traditional figures of merit are not appropriate for quantifying the performance of the \HP algorithm. To come up with a more appropriate approximation of neural activation, we make two simplifying assumptions: (i) neurons are primarily activated by electric fields oriented along the direction of their axon; (ii) excitatory pyramidal cortical neurons' axons are primarily oriented along the ``radial-in'' direction (i.e., perpendicular to the cortical surface). Assumption (i) is based on the modeling work done in~\cite{rattay1986analysis}, and assumption (ii) is a well-known fact about neuron physiology~\cite{kandel2000principles}. The overall effect of assumptions (i) and (ii) is that only the electric field oriented along the radial-in direction matters quantifying neural activation. 

Under assumptions (i) and (ii), we formulate an intuitive and simplistic metric, denoted as $\mathcal{V}_{\mathrm{Th}}$, to approximate the neural activation. Nominally, $\mathcal{V}_{\mathrm{Th}}$ corresponds to the volume of the off-target brain regions where the field intensity along the radial direction is greater than $80\%$ of the desired electric field intensity $(E_{des})$. We earlier used this metric in~\cite{goswami2021hingeplace} to quantify neural activation. The threshold of 80\% is borrowed from Forssell \& Jain et al.~\cite{forssellner2021}, where authors experimentally calculated the stimulation area when exposed to transcranial fields using local field potentials (LFPs) in slices. They found no neural response after the field decayed to $\sim 80\%$ of its maximum values.

\textbf{Constructing $\myVec{T}_f$ and $\myVec{T}_c$}: First, we note that under assumptions (i) and (ii), only the electric field along the radial-in direction estimates neural activation. Hence, we simplify the forward matrix $\myVec{T}\in\R^{3M\times N}$ to a new forward matrix $\Tilde{\myVec{T}}\in\R^{M\times N}$ such that the elements of the resultant $\Tilde{\myVec{E}_{I}}=\Tilde{\myVec{T}}\myVec{I}$ represent the electric field along the radial-in direction at each voxel in the head model. We employ the ``radial-in'' option in the ROAST software to construct $\Tilde{\myVec{T}}$. Furthermore, we consider $\myVec{\Gamma}_{C}$ and $\myVec{\Gamma}_F$ to be identity matrices for simplicity.  

We consider five different target locations for our studies (shown in Fig.~\ref{fig:setup}c) having the following MNI coordinates: Target MC at $[-48,-8,50]$ (left motor cortex),  Target $1$ at $[-48, -21, 50]$ (left somatosensory cortex), Target $2$ at $[-22,-20,77]$, Target $3$ at $[5,8,61]$, and Target $4$ at $[48, -8, 50]$ (right motor cortex). 
The target forward matrix $\myVec{T}_f$ was constructed by considering all voxels within $2$mm, $2$mm, $3$mm, $2$mm, and $2$mm from the MNI coordinates for targets MC, $1$, $2$, $3$, and $4$, respectively, except for the study described in Sec.~\ref{sec:results:mri:larger-focus}. In Sec.~\ref{sec:results:mri:larger-focus}, $\myVec{T}_f$ was constructed by considering all voxels within $20$mm from target MC's MNI coordinates. The corresponding $\myVec{A}_f\in\R^{1\times N}$ was constructed by averaging all the rows of $\myVec{T}_f$, except for Sec.~\ref{sec:results:mri:multi-site}, where $\myVec{A}_f\in\R^{2\times N}$ was constructed by separately averaging the left and right motor target regions. The off-target matrix $\myVec{T}_c$ was constructed using the following voxel sub-sampling procedure:
\begin{enumerate}
    \item Add all voxels lying within $40$mm of the target location into the off-target region $\mathcal{C}$ excluding the target region voxels.
    \item Randomly choose $25$\% of the rest of the voxels and add them to $\mathcal{C}$. Construct the matrix $\myVec{T}_c$ by only considering the voxels that lie on $\mathcal{C}$.   
\end{enumerate}
The above sub-sampling procedure utilizes the smoothness of transcranial fields to significantly reduce the size of $\myVec{T}_c$, thereby allowing faster computation. Appendix E compares electrical fields generated using the sub-sampled off-target matrix with the electrical fields generated using $\myVec{T}_c$ containing all brain voxels (except the target voxels). We observed no significant difference in the resultant electric fields, but the sub-sampled matrix required $\sim4\times$ less computation time to design the electrode montage.
\subsection{Sea of neurons model: Spherical head models with realistic biophysical neuron models}
While the metric $\mathcal{V}_{\mathrm{Th}}$ provides a good first-order approximation of neural response, the interactions of transcranial fields and cortical neurons tend to be more complex due to their morphology and membrane dynamics. In order to more accurately model the neural response, we implemented a ``sea of neurons" model. The sea of neurons model combines spherical head models (also frequently used for modeling transcranial electric fields~\cite{forssell2021effect}) with realistic neuron models taken from the Blue Brain Project~\cite{markram2015reconstruction} to gauge the neural response to transcranial fields more accurately. Essentially, the sea of neurons model quantifies the response of a cortical neuron at different locations under a transcranial electrical field. We used this model earlier to understand the mechanisms of transcranial temporal interference stimulation (tTIS)~\cite{Caldas-Martinez2024}. Fig.~\ref{fig:setup}d-f provides a schematic of our sea of neurons model. 

The sea of neurons model consists of two main components: (i) a $4$-layer spherical head model to simulate transcranial electric fields and (ii) realistic bio-physical neuron models that simulate the neural response to the induced electric fields. Briefly, we first use the $4$-sphere head model to simulate the electric field and voltage induced in the brain's cortical layers. Using the simulated voltage, we simulate the extracellular stimulation of bio-physically realistic neuron models placed at different spatial locations within a region of interest in these cortical layers. We use these neuron models' resultant membrane potential (due to the extracellular stimulation) to determine where the neuron models fired an action potential in the region of interest. Consequently, we estimated the stimulated region by constructing a map of the region where the neuron model spiked due to the transcranial electric field.

\textbf{Four-sphere head model}: We used the $4$-sphere head model previously used in~\cite{forssell2021effect} to model transcranial electric fields. We kept the same parameters used in~\cite{forssell2021effect}, namely: the overall radius of the head is $9.2$cm, with the thickness of the scalp, skull, and CSF chosen as $6$mm, $5$mm, and $1$mm, respectively. The scalp, skull, CSF, and brain conductivities were chosen to be $0.33$Sm$^-1$, $0.006$Sm$^{-1}$, $1.79$Sm$^{-1}$, and $0.3$Sm$^{-1}$, respectively. The voltage and electric field are calculated using the analytical expressions provided in~\cite{forssell2021effect}. The electrode grid consisted of $21$ electrodes, each having a $5$mm radius and an inter-electrode spacing of $20$mm. The electrodes are arranged in a circular patch centered at the north pole (see Fig.~\ref{fig:setup}d and e). We discuss our reasoning behind using the spherical head model instead of the MRI-based model in Sec.~\ref{sec:discussion}. Briefly, we chose the spherical head model due to the need for high-resolution electric fields (on the order of $50\mu\text{m}$) to accurately model extracellular stimulation of the neuron models (spherical head models provide arbitrary precision). 

\textbf{Neuron Models}: We use a Layer 5 Pyramidal somatosensory neuron model and a Layer 2/3 Pyramidal somatosensory neuron model with realistic morphology to estimate the activation region. The neuron models were originally published in~\cite{markram2015reconstruction}. The neuron models were modified according to the algorithm presented in~\cite{aberra2018biophysically} for simulating extracellular stimulation and enlarged to match human neurons. The neuron models were simulated using the \texttt{python} environment of NEURON~\cite{carnevale2006neuron,hines2009neuron}. These models were shown to match experimental results~\cite{aberra2018biophysically}. The neuron models were stimulated using a $200\mu\text{s}$ monophasic square pulse. Similar pulses have been widely used to stimulate the cortical regions of brain, primarily the motor cortex~\cite{merton1980stimulation}.

\textbf{Constructing $\myVec{T}_f$ and $\myVec{T}_c$}: Different target locations were chosen for different simulation studies, detailed later in their respective sections. For this study, we only choose one target point as our target region, i.e., $\myVec{T}_f\in\R^{1\times N}$. The corresponding off-target region was constructed as a hollow ring lying on the same shell with an inner radius of $0.5$cm and an outer radius of $7$cm with points sampled with a uniform spacing of $1$mm. This off-target region was then used to construct the matrix $\myVec{T}_c$. We also used this procedure in our earlier conference work~\cite{goswami2021hingeplace}.

\textbf{Quantifying Neural activation}: We count the number of off-target locations where the neuron fired, denoted as $N_{act}$, to infer the focality of the neural response produced by a transcranial field. Across the sea of neurons model studies, the target region was assumed to be a single point and the target electric field (i.e., $E_{des}$ and $\myVec{D}$) was chosen appropriately to guarantee neural activation at the target point (see Sec.~\ref{sec:results:sea-of-neurons}). Hence, we are able to calculate the off-target stimulation by subtracting $1$ (corresponding to the target neuron) from the total number of locations stimulated. A larger value of $N_{act}$ indicates the stimulation was diffused. We chose different rectangular regions in the cortical layer at a depth of $1$mm from the cortical surface (roughly corresponding to the start of Layer 5 of the cortex) as the regions of interest for different simulation studies. The width and length of these regions of interest were chosen to ensure
good resolution and accuracy when determining the stimulation region. For each region of interest, we evaluate the neuron model's response at $800$ different spatial locations uniformly randomly chosen from the region of interest. The exact values of these regions are provided in Appendix D. 

\subsection{Implementing LCMV-E and \HP}
We use the \texttt{cvxpy} package~\cite{diamond2016cvxpy,agrawal2018rewriting} in \texttt{python}~\cite{10.5555/1593511} to implement the LCMV-E and \HP algorithm. We used the default tolerances to determine convergence. An initial study revealed that the best performance of \HP was typically achieved for $p=1,2$ or $3$ (see Appendix D). Hence, we only consider the value of $p=1,2$, and $3$. 

We used a nominal value of $E_{des}=1\text{Vm}^{-1}$ along the radial-in direction for studies utilizing the MRI head model. For the studies utilizing the sea of neurons model, we calculated the value of $E_{des}$ and the direction vector $\myVec{D}$ (see Sec.~\ref{sec:equiv-LCMV-EDM}) using an input-output curve (see Appendix D). Effectively, we chose the smallest value of $E_{des}$ for which we visually saw significant activation around the target region. The exact value of $E_{des}$ and  $\myVec{D}$ are as follows: $84.93$Vm$^{-1}$ along the direction $[\myminus0.68,0,0.73]$ in Sec.~\ref{sec:results:sea-of-neurons:c3-c4} and~\ref{sec:results:sea-of-neurons:neuron-type}, $84.94$Vm$^{-1}$ along the direction $[0,0,1]$ in Sec.~\ref{sec:results:sea-of-neurons:HD-tDCS}, and $70.27$Vm$^{-1}$ along the direction $ [\myminus0.75,\myminus0.41,0.51]$ in Sec.~\ref{sec:results:sea-of-neurons:opt-dir} and~\ref{sec:results:sea-of-neurons:diff-target}.  

All simulations studies compared \HP and LCMV-E across different values of $I_{safe}$ and $I_{tot}$. The smallest value for $I_{safe}$ was chosen as the lowest value for which the LCMV-E and \HP optimizations remain feasible. We parameterized $I_{tot}=I_{tot}^{mul}I_{safe}$ in both~\eqref{eq:algo:HP-main-opt} and~\eqref{eq:lcmv-e}. This particular parameterization allowed us to interpret $I_{tot}^{mul}$ as the maximum number of active electrodes that can pass $I_{safe}$ amount of current. We considered three different values of $I_{tot}^{mul}$: \{$2$, $4$, $8$\} for MRI head model studies and \{$2$, $4$, $6$\} for sea of neurons model studies. 

We chose $\myVec{E}_{tol}^+\myeq\myVec{E}_{tol}^{-}\myeq E_{tol}\myVec{1}_{|\mathcal{C}|}$ for MRI head model studies. Choosing a single $E_{tol}$ for all directions sufficed as we assumed that only the electric field along the radial-in direction stimulated neurons. We used a grid-search to determine the value of $E_{tol}$. For $p=1$ and $p=2,3$, we searched across the values $\{0.1\text{Vm}^{-1}$, $0.5\text{Vm}^{-1}$, $0.6\text{Vm}^{-1}$, $0.7\text{Vm}^{-1}\}$ and  $\{0.01\text{Vm}^{-1}$, $0.65\text{Vm}^{-1}$, $0.55\text{Vm}^{-1}$, $0.35\text{Vm}^{-1}\}$, respectively. We always present the result with the least $\mathcal{V}_{\mathrm{Th}}$.

Similarly, we chose $\myVec{E}_{tol}^{x+}\myeq\myVec{E}_{tol}^{x-}\myeq E_{tol}^x\myVec{1}_{|\mathcal{C}|}$, $\myVec{E}_{tol}^{y+}=\myVec{E}_{tol}^{y-}= E_{tol}^x\myVec{1}_{|\mathcal{C}|}$, and $\myVec{E}_{tol}^{z+}$ $\myeq\myVec{E}_{tol}^{z-}$ $\myeq E_{tol}^z\myVec{1}_{|\mathcal{C}|}$ for the sea of neurons model studies. We observe that allowing different tolerances along the different cardinal direction increases the performance of \HP for realistic neurons (see Sec.~\ref{sec:results:sea-of-neurons:c3-c4}). We determined the value of $[E_{tol}^x,E_{tol}^y,E_{tol}^z]$ using a random search across $30$ points for each sea of neurons model study. We chose $[E_{tol}^x,E_{tol}^y,E_{tol}^z]$ with the least $N_{act}$ value. The lower and upper limit of the random search were $0.1E_{des}$ and $0.7E_{des}$. Since, the sea of neurons model is computationally cumbersome, we did not perform the random search at all $I_{safe}$ and $I_{tot}$ values, but at a representative value of $I_{safe}$ and $I_{tot}$ (Appendix D has the exact values). Consequently, the chosen value of $[E_{tol}^x,E_{tol}^y,E_{tol}^z]$ is not optimal at all $I_{safe}$ and $I_{tot}$ values. If these chosen values of $[E_{tol}^x,E_{tol}^y,E_{tol}^z]$ were severely sub-optimal for some values of $I_{tot}$ and $I_{safe}$ leading to \HP performing worse than LCMV-E, then we re-tuned $[E_{tol}^x,E_{tol}^y,E_{tol}^z]$ at those values of $I_{safe}$ and $I_{tot}$ through the same random-search. For L1L1-norm optimization performed in Sec.~\ref{sec:results:sea-of-neurons:c3-c4}, we chose the value of $E_{tol}$ from the grid spanning the range $0.1E_{des}$-$0.9E_{des}$ (with increments of $0.1E_{des}$), which produced the least off-target stimulation at each $I_{safe}$ and $I_{tot}^{mul}$.

\subsection{Validation of the sub-sampling procedure discussed for MRI head models}
\begin{table}[ht]
    \centering
    \begin{tabular}{|c|c|c|c|}
    \hline
    \multicolumn{2}{|c|}{\HP}&    \multicolumn{2}{|c|}{LCMV-E}\\
    \hline
    Off-target matrix & Jaccard Index&Off-target matrix & Jaccard Index\\
    \hline
$\myVec{T}_c^{10}$ & $1$& $\myVec{T}_c^{10}$ & $0.95$ \\
$\myVec{T}_c^{25}$ & $1$& $\myVec{T}_c^{25}$ & $0.97$ \\
$\myVec{T}_c^{50}$ & $1$& $\myVec{T}_c^{50}$ & $0.98$ \\
\hline
    \end{tabular}
    \caption{The values of Jaccard index comparing the stimulated region generated using the sub-sampled off-target forward matrix with the stimulated region generated using the off-target matrix containing all voxels ($\myVec{T}_c$) for both \HP and LCMV-E algorithms.}
    \label{tab:appx:jaccard-index}
\end{table}
We harness the smoothness of electric field to reduce the size of the off-target matrix through the sub-sampling procedure discussed in the ``Realistic MRI head model'' section above. We validated our sub-sampling procedure by comparing the electric field and the stimulated region generated using the sub-sampled off-target matrix with the electric field and the stimulated region generated using the off-target matrix containing all off-target voxels. Briefly, we constructed $3$ different sub-sampled matrices $\myVec{T}_c^{10}$, $\myVec{T}_c^{25}$, and $\myVec{T}_c^{50}$ by using the sub-sampling procedure. The difference in $\myVec{T}_c^{10}$, $\myVec{T}_c^{25}$, and $\myVec{T}_c^{50}$ stems from sampling $10\%$, $25\%$, and $50\%$ of the off-target voxels, respectively, in the second step of the the sub-sampling procedure. Furthermore, we also generated the off-target matrix containing all off-target voxels $\myVec{T}_c$. Using $\myVec{T}_c^{10}$, $\myVec{T}_c^{25}$, $\myVec{T}_c^{50}$ and $\myVec{T}_c$, we solve the \HP algorithm (for $p=1$) and the LCMV-E algorithm for $I_{safe}=3.75$mA and $I_{tot}=I_{tot}^{mul}I_{safe}=4I_{safe}$. The value of $E_{des}$ was chosen to be $1\text{Vm}^{-1}$ and the value of $E_{tol}=0.5\text{Vm}^{-1}$ for the \HP algorithm. These values are taken from our study in Sec.~\ref{sec:results:mri:MC}. 
\begin{figure*}[t]
    \centering
    \includegraphics[width=0.8\textwidth]{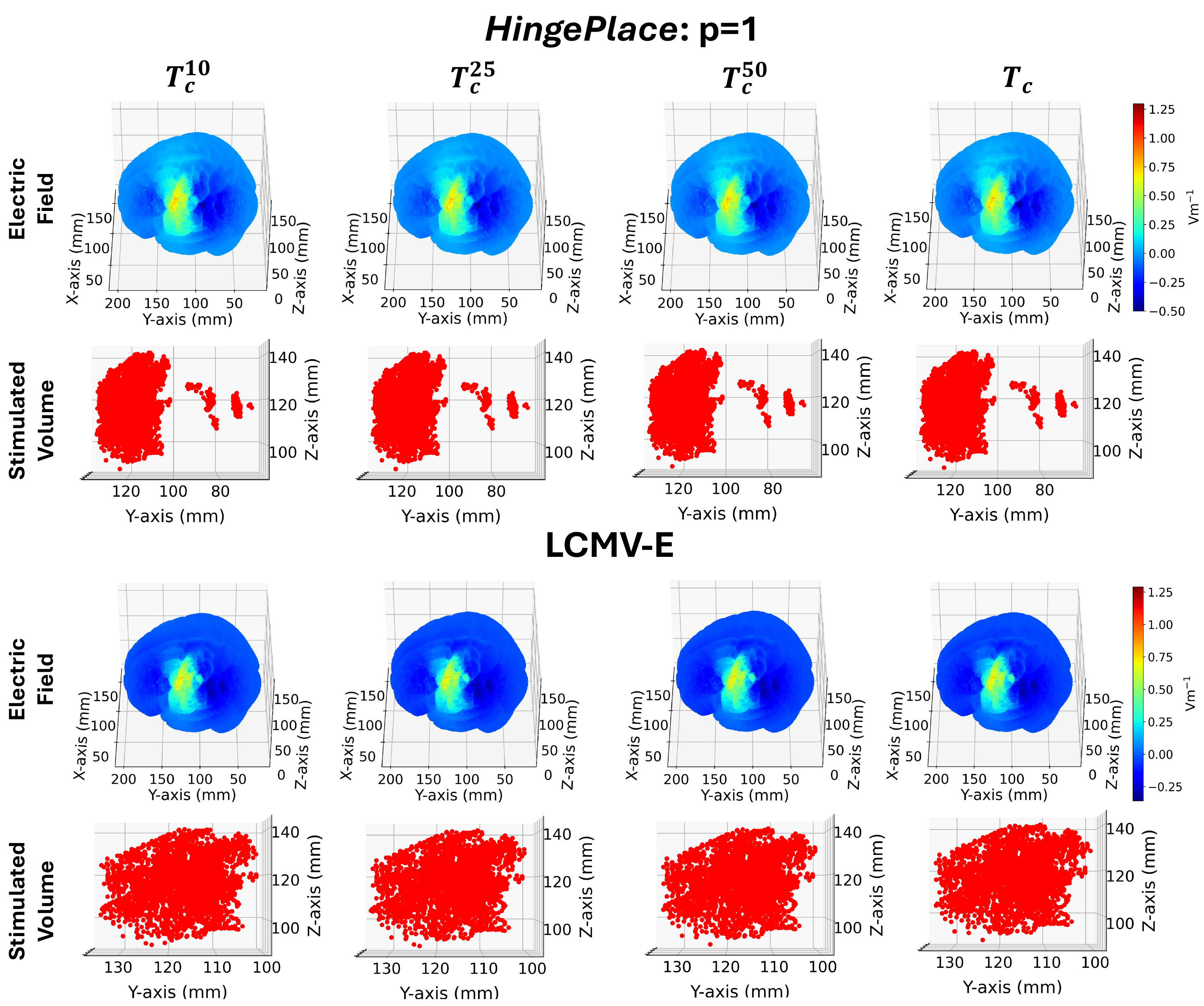}
    \caption{The plots of the electric fields and the corresponding stimulated regions generated by the \HP and LCMV-E algorithms when using the off-target matrices $\myVec{T}_c^{10}$, $\myVec{T}_c^{25}$, $\myVec{T}_c^{50}$, and $\myVec{T}_c$ (discussed in Appendix E).}
    \label{fig:appx:sub-sampling}
\end{figure*}
Fig.~\ref{fig:appx:sub-sampling} shows the electric field and the stimulated generated using the four off-target matrices, namely, $\myVec{T}_c^{10}$, $\myVec{T}_c^{25}$, $\myVec{T}_c^{50}$ and $\myVec{T}_c$, for both \HP and LCMV-E algorithms. We visually observe no significant differences between the electric field and the stimulated region generated using $\myVec{T}_c$ and those generated using $\myVec{T}_c^{10}$, $\myVec{T}_c^{10}$, and $\myVec{T}_{c}^{50}$. We quantify the similarity between the stimulated regions generated using the sub-sampled off-target forward matrix, namely, $\myVec{T}_c^{10}$, $\myVec{T}_c^{10}$, and $\myVec{T}_{c}^{50}$ with the stimulated region generated using $\myVec{T}_c$ with Jaccard index~\cite{costa2021further} in Table~\ref{tab:appx:jaccard-index}. We observe that Jaccard index is extremely close to $1$ in all cases, showing that the off-target stimulated regions generated using the sub-sampled off-target matrices are the same as the off-target stimulated regions generated using $\myVec{T}_c$. This empirical study provides evidence that we can reduce the number of voxels in the off-target matrix using our sub-sampling procedure while not significantly affecting the outcome of the \HP and LCMV-E algorithms. Furthermore, reducing the size of the off-target matrix provides a significant speed-up. The computation time using $\myVec{T}_c^{25}$ is $93$s whereas the computation time using $\myVec{T}_c$ is $403$s, providing a $4\times$ speed-up.
\subsection{The dimensions of the regions of interest used in the sea of neurons model}
We provide the exact dimensions of the rectangular regions chosen as the regions of interest for the sea of neurons model studies in Table~\ref{tab:sea-of-neurons:ROI}. The length and breadth of the rectangular regions are parallel to the $x$ and $y$ axis. Hence, we express these regions by presenting the $x$ and $y$ intervals over which they exist. All the rectangular regions were chosen at a depth of $1.3$cm from the scalp surface, roughly corresponding to the start of Layer $5$ of the cortex.  
\begin{table}[h]
    \centering
    \begin{tabular}{|c|c|c|}
    \hline
        Study & $x$-interval & $y$-interval  \\
        \hline
        Sec.~\ref{sec:results:sea-of-neurons:c3-c4}& [$-5$cm,$1$cm] &[$-1.5$cm,$2.5$cm]\\
        Sec.~\ref{sec:results:sea-of-neurons:HD-tDCS}& [$-2.5$cm,$2.5$cm] &[$-2.5$cm,$2.5$cm]\\
        Sec.~\ref{sec:results:sea-of-neurons:opt-dir}& [$-2.2$cm,$3$cm] &[$-2.5$cm,$2.5$cm]\\
        Sec.~\ref{sec:results:sea-of-neurons:neuron-type}& [$-5$cm,$1$cm] &[$-2$cm,$2.5$cm]\\
        Sec.~\ref{sec:results:sea-of-neurons:diff-target}: TP-1& [$-1.5$cm,$4.5$cm] &[$-2.5$cm,$2.5$cm]\\
        Sec.~\ref{sec:results:sea-of-neurons:diff-target}: TP-2& [$-4.5$cm,$3.8$cm] &[$-3.5$cm,$5.5$cm]\\
        Sec.~\ref{sec:results:sea-of-neurons:diff-target}: TP-3& [$-3$cm,$1.3$cm] &[$-1.3$cm,$1.8$cm]\\
        Sec.~\ref{sec:results:sea-of-neurons:diff-target}: TP-4& [$-3$cm,$4.3$cm] &[$-4$cm,$2.5$cm]\\
        \hline
    \end{tabular}
    \caption{The exact dimensions of the rectangular region chosen as the regions of interest for the sea of neurons model studies.}
    \label{tab:sea-of-neurons:ROI}
\end{table}
\subsection{The input-output curve used for determining the $E_{des}$ for the sea of neurons studies}
For the simulation studies described in Sec.~\ref{sec:results:sea-of-neurons:c3-c4} and Sec.~\ref{sec:results:sea-of-neurons:HD-tDCS}, we calculated the direction vector $\myVec{D}$ and the value $E_{des}$ by calculating an input-output curve. The intuition behind the input-output curve was to emulate how tES is typically performed in experimental settings, where an initial input-output curve is established to tune the parameters of electrical stimulation for each subject. 

\begin{figure*}
    \centering
    \includegraphics[width=\linewidth]{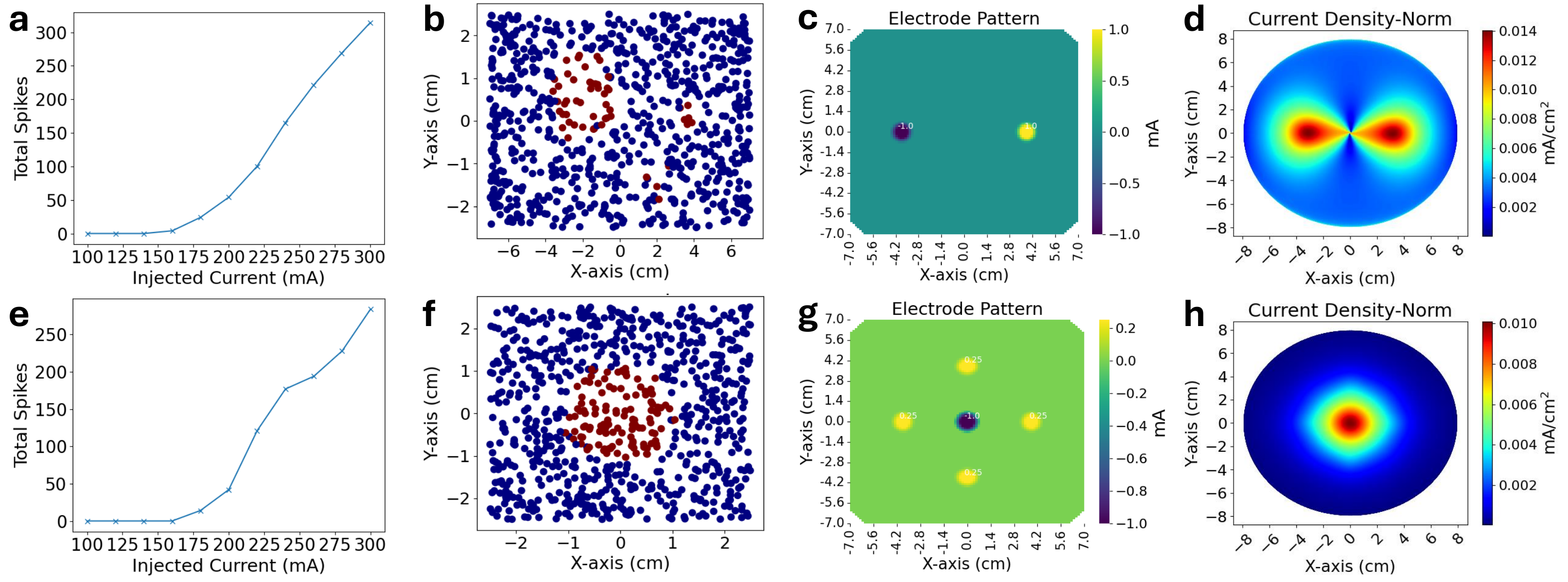}
    \caption{\textbf{a} and \textbf{d} show the total number of spikes induced in the region of interest as a function of the total input current for the C3-C4 and HD-tDCS baseline electrode patterns, respectively. \textbf{b} and \textbf{f} show the regions (in red dots) stimulated by the C3-C4 and HD-tDCS baseline electrodes at the injected current level of $200$mA and $220$mA, respectively. \textbf{c} and \textbf{g} show a schematic of the C3-C4 and HD-tDCS baseline electrode patterns, respectively, injecting $1$mA total current. \textbf{d} and \textbf{h} show the corresponding current density magnitude at a depth of $1.3$cm from the scalp surface (i.e., at the region of interest) for the C3-C4 and HD-tDCS baseline electrode patterns (injecting $1$mA total current), respectively.}
    \label{fig:appx:ip-op}
\end{figure*}

We used two electrode montages typically used in tES literature, namely, the C3-C4 arrangement shown in Fig.~\ref{fig:appx:ip-op}c and the HD-tDCS arrangement shown in Fig.~\ref{fig:appx:ip-op}g. The shape of the current density magnitude induced by these baseline electrode patterns is shown in Fig.~\ref{fig:appx:ip-op}d and h. The input-output curve was established by increasing the total injected current from $100$mA to $300$mA in increments of $25$mA. We quantified the neural response by $N_{act}$, i.e., the total number of locations where the neuron spiked. Fig.~\ref{fig:appx:ip-op}a and e show the input-output curve for the C3-C4 and HD-tDCS montages, respectively. 

By visual inspection, we observed significant neural activation (roughly a disc of radius of $1$cm) at $200$mA for C3-C4 montage and at $220$mA for HD-tDCS montage (see Fig.~\ref{fig:appx:ip-op}b and e). Hence, the reference electrode montages we used to determine the target location, $E_{des}$, and $\myVec{D}$ were calulated from the electric field induced by C3-C4 montage and HD-tDCS montages injecting $200$mA and $220$mA, respectively. Specifically, for the study in Sec.~\ref{sec:results:sea-of-neurons:c3-c4}, we chose the target point as the location where the electric field magnitude (induced by the C3-C4 montage at $200$mA injected current) was maximum in the region of interest, namely, the shell at a depth of $1.3$cm from the scalp. The value of $E_{des}$ and $\myVec{D}$ correspond to the electric field magnitude and direction at this chosen target point. A similar procedure determined the location of target point, the value of $E_{des}$, and $\myVec{D}$ for the study in Sec.~\ref{sec:results:sea-of-neurons:HD-tDCS}, where the baseline electrode pattern was the HD-tDCS pattern at $220$mA total injected current. For the study in Sec.~\ref{sec:results:sea-of-neurons:neuron-type}, we chose the same target location, $E_{des}$, and $\myVec{D}$ as the study in Sec.~\ref{sec:results:sea-of-neurons:c3-c4} due to Sec.~\ref{sec:results:sea-of-neurons:neuron-type} performing the same study as Sec.~\ref{sec:results:sea-of-neurons:c3-c4} with a L2/3 pyramidal neuron model instead of a L5 pyramidal neuron model. 

\subsection{Finding the optimal direction of electric field for stimulating Layer 5 neurons used in the sea of neurons model}
\begin{figure*}
    \centering
    \includegraphics[width=\linewidth]{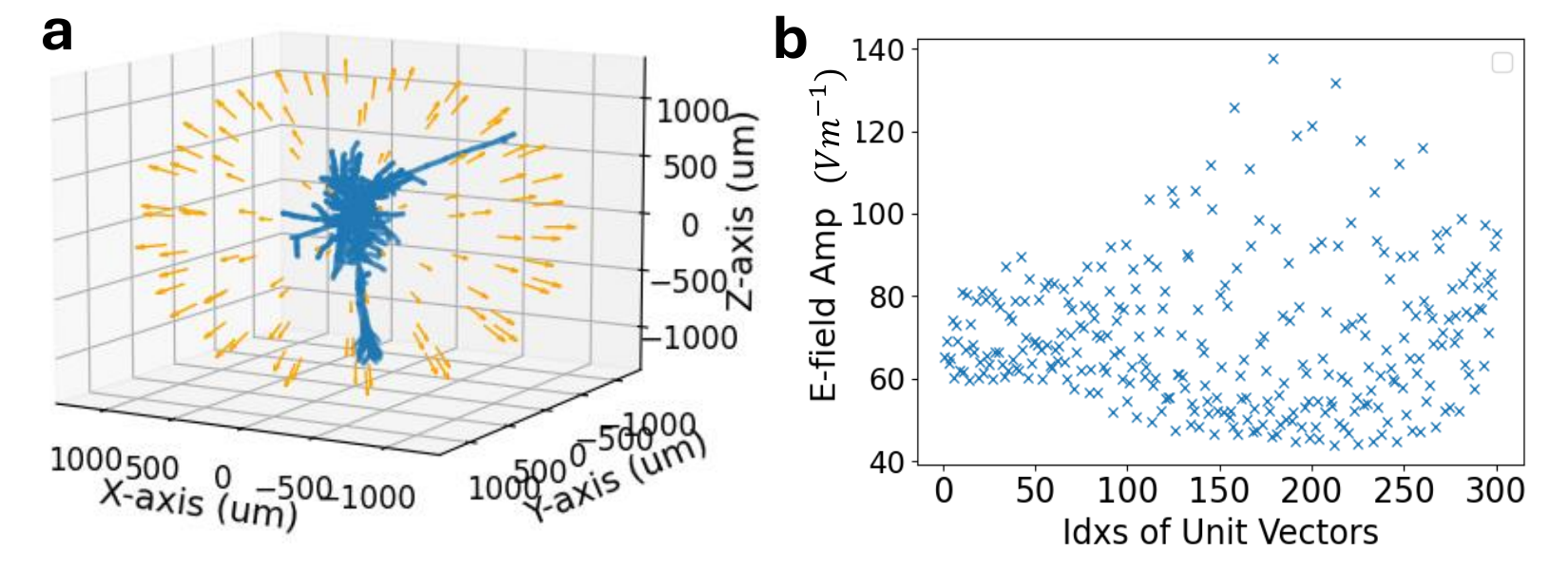}
    \caption{\textbf{a} provides a schematic of the different directions of the electric field relative to the neuron model tested to find the optimal direction of stimulation. \textbf{b} shows the corresponding threshold magnitude for the different directions tested to find the optimal direction }
    \label{fig:appx:opt-dir}
\end{figure*}

The stimulation studies employed in Sec.~\ref{sec:results:sea-of-neurons:opt-dir} and Sec.~\ref{sec:results:sea-of-neurons:diff-target} used the optimal direction of the electric field that required the smallest amplitude to make the neuron model spike. We find this direction by the following study: We stimulated the neuron model using uniform electric fields oriented along $300$ different directions (see Fig.~\ref{fig:appx:opt-dir}a). For each direction, we find the minimum value of the electric field amplitude required to stimulate the neuron model using binary search. Consequently, we obtain the threshold electric field amplitude of the electric field required to stimulate the neuron model for all the $300$ directions (see Fig.~\ref{fig:appx:opt-dir}b). We choose the direction having the least threshold amplitude. The value of $E_{des}$ was chosen to be $1.6\times$ the threshold amplitude to stimulate a significant area in the region of interest.  

\subsection{Finding the value of $p$ in \HP}
\begin{figure*}
    \centering
    \includegraphics[width=\linewidth]{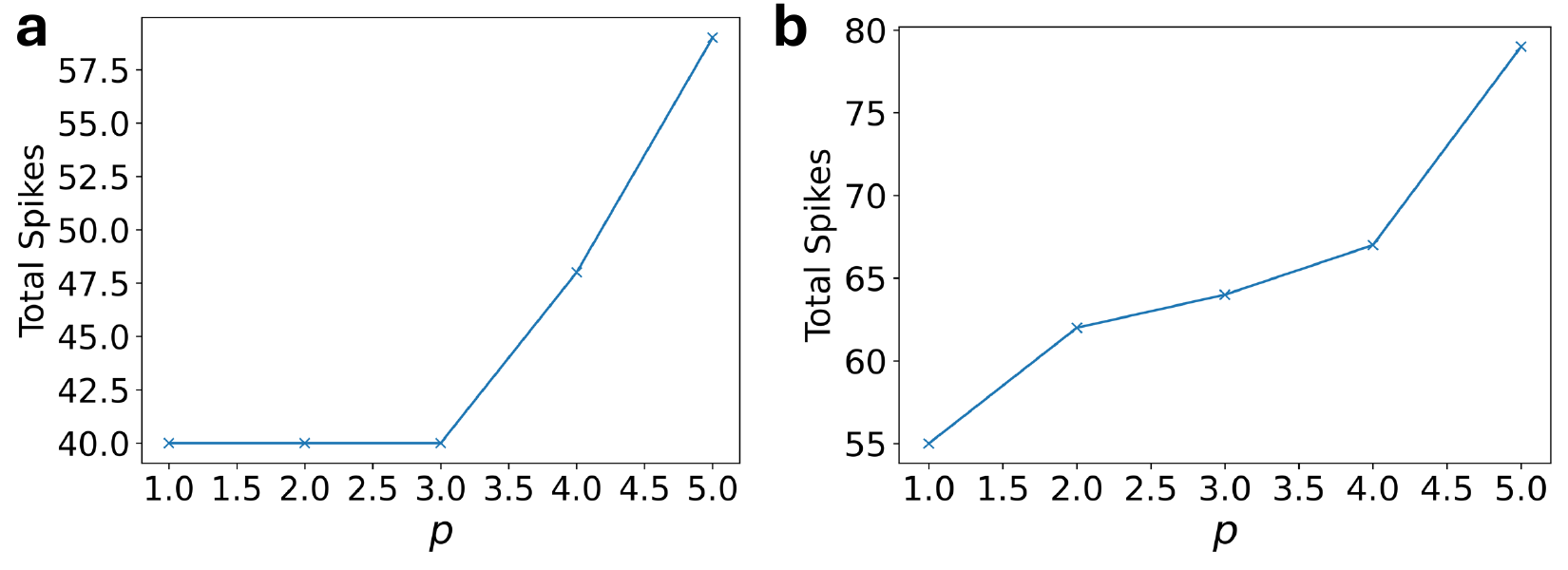}
    \caption{\textbf{a} and \textbf{b} show the value of $N_{act}$ as a function of $p$ in the \HP loss for the studies utilizing the hyperparameters of Sec.~\ref{sec:results:sea-of-neurons:opt-dir} and Sec.~\ref{sec:results:sea-of-neurons:neuron-type}, respectively, discussed in Appendix D. }
    \label{fig:appx:best-p}
\end{figure*}

We ran some preliminary studies to find which values of $p$ in the \HP loss produce the most focal neural response. We re-ran the simulation studies of Sec.~\ref{sec:results:sea-of-neurons:opt-dir} and Sec.~\ref{sec:results:sea-of-neurons:neuron-type} for only $(I_{safe},I_{tot}^{mul})=(130\text{mA},2)$ and $(I_{safe},I_{tot}^{mul})=(172\text{mA},4)$, respectively. We calculated the metric $N_{act}$ for the \HP algorithm with $p=1,2,3,4,$ and $5$ to asses which value of $p$ produced the most focal neural response in these two studies. The value of $[E_{tol}^{x},E_{tol}^{y},E_{tol}^{z}]$ were individually optimized by the random-search procedure discussed in the ``Sea of neurons model'' section above. Fig~\ref{fig:appx:best-p}a and b show the values of $N_{act}$ for the study utilizing the hyperparameters of Sec.~\ref{sec:results:sea-of-neurons:opt-dir} and Sec.~\ref{sec:results:sea-of-neurons:neuron-type}, respectively. We find that lower values of $p$ produced more focal neural responses compared to larger values of $p$. Hence, we decided to only test $p=1,2$ and $3$ in our studies in Sec.~\ref{sec:results}.
\newpage
\section*{}
\subsection{Additional representative plots for the simulation studies utilizing the sea of neurons model}
\begin{figure*}[htbb]
    \centering
\includegraphics[width=\textwidth]{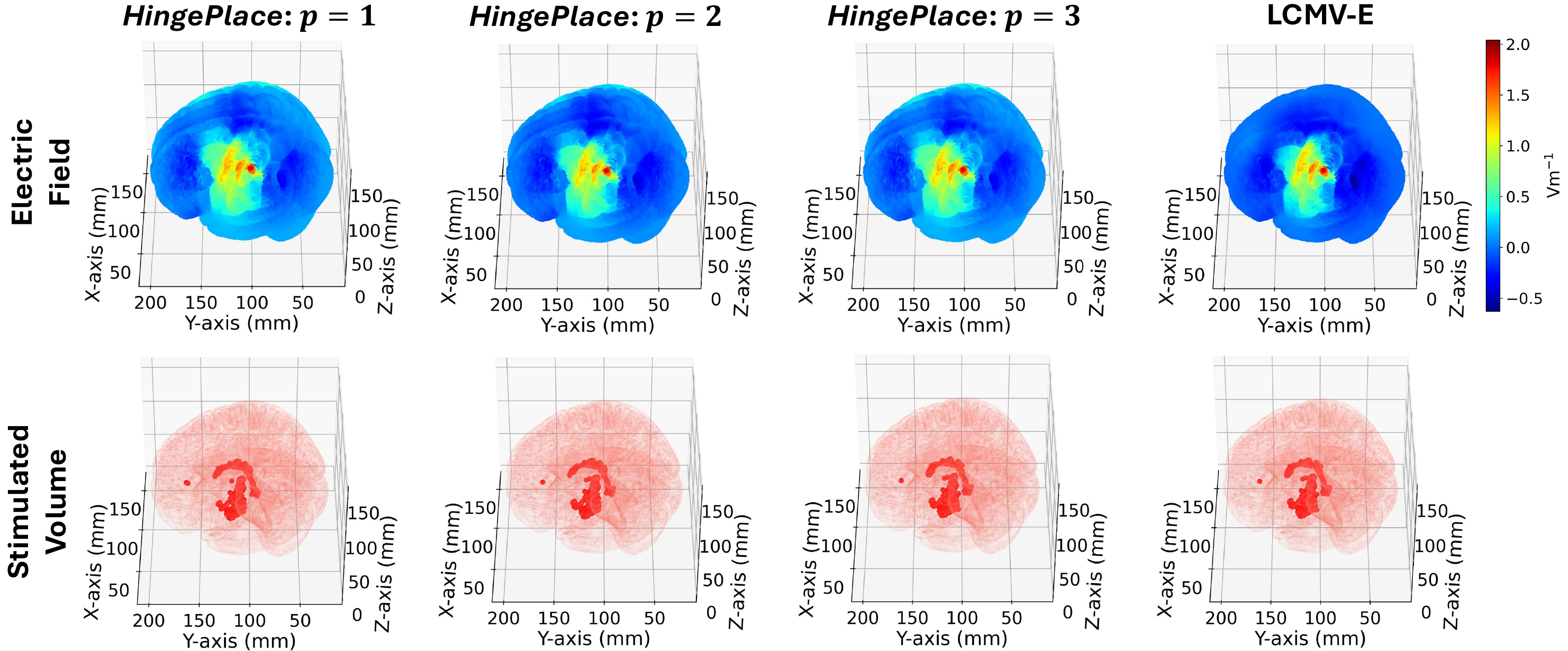}
    \caption{A representative plot of the electric field along the radial-in direction and the corresponding region having the radial-in electric field intensity above $0.8\text{Vm}^{-1}$ generated by electrode montages designed using the \HP and LCMV-E algorithms in Sec.~\ref{sec:results:mri:larger-focus}. The target location was the motor cortex and the values of $I_{safe}$ and $I_{tot}^{mul}$ were $5.2$mA and $8$, respectively.}
    \label{fig:mri-visualisation:larger-focus}
\end{figure*}

\begin{figure*}[htbp]
    \centering
\includegraphics[width=\textwidth]{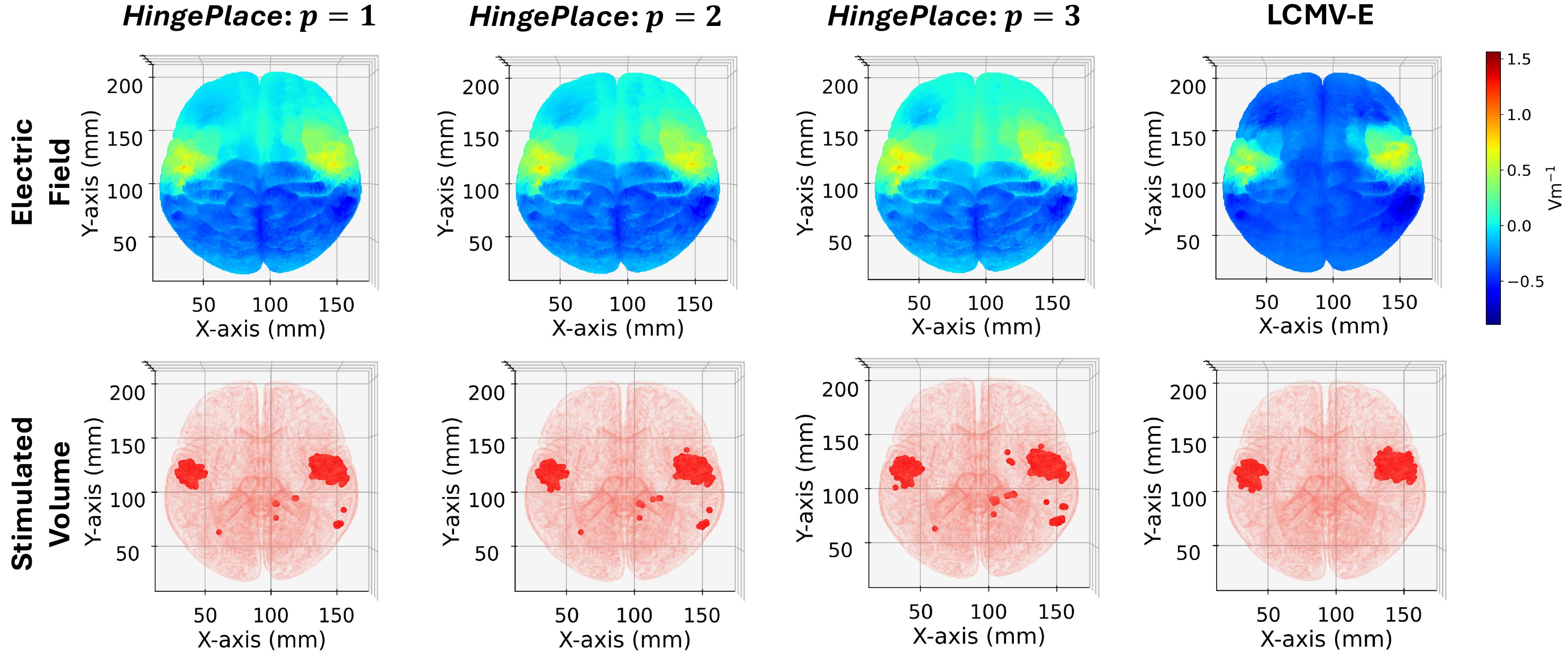}
    \caption{A representative plot of the electric field along the radial-in direction and the corresponding region having the radial-in electric field intensity above $0.8\text{Vm}^{-1}$ generated by electrode montages designed using the \HP and LCMV-E algorithms in Sec.~\ref{sec:results:mri:multi-site}. The target location was the motor cortex and the values of $I_{safe}$ and $I_{tot}^{mul}$ were $6.4$mA and $4$, respectively.}
    \label{fig:mri-visualisation:multisite}
\end{figure*}

Fig.~\ref{fig:mri-visualisation:larger-focus} and Fig.~\ref{fig:mri-visualisation:multisite} provide additional visualization of the electric field and the corresponding stimulated regions generated in the studies discussed in Sec.~\ref{sec:results:mri:larger-focus} and~\ref{sec:results:mri:multi-site}, respectively.

\subsection{Additional representative plots for the simulation studies utilizing the sea of neurons model}
\begin{figure*}[ht]
    \centering
   \includegraphics[width=\textwidth]{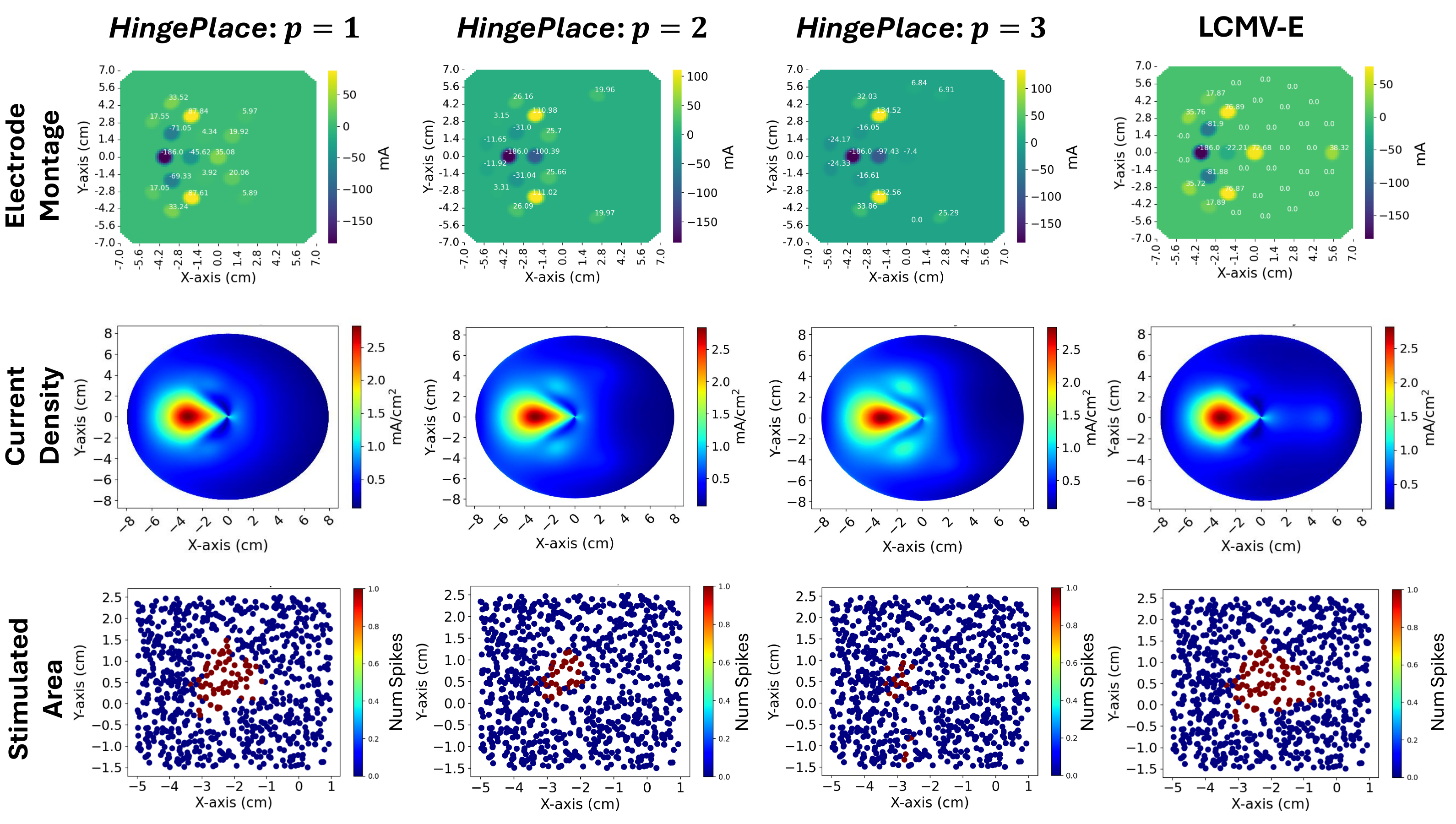}
    \caption{A representative plot of the electrode montage, the current density magnitude at the depth of $1$mm from the cortical surface (representing Layer 5), and the locations where the neurons fired in the Layer 5 of our sea of neurons model (denoted by the red dots) for the \HP and LCMV-E algorithms in Sec.~\ref{sec:results:sea-of-neurons:c3-c4}. The values of $I_{safe}$ and $I_{tot}^{mul}$ were $186$mA and $2$, respectively. The numbers in electrode montage denote the current injected at that electrode (in mA).}
    \label{fig:appx:neuron:visualisation:c3-c4}
\end{figure*}

\begin{figure*}[ht]
    \centering
   \includegraphics[width=\textwidth]{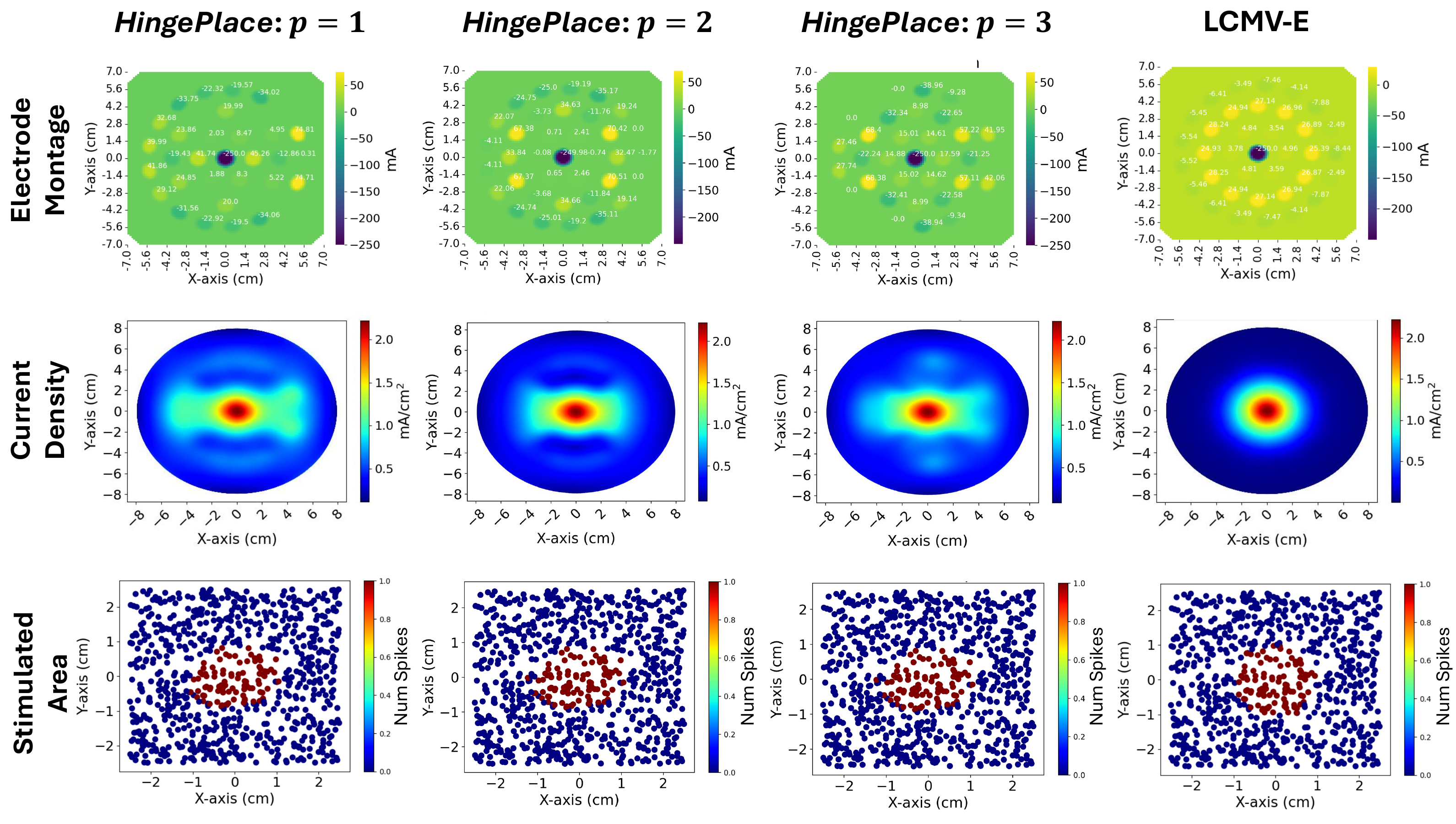}
    \caption{A representative plot of the electrode montage, the current density magnitude at the depth of $1$mm from the cortical surface (representing Layer 5), and the locations where the neurons fired in the Layer 5 of our sea of neurons model (denoted by the red dots) for the \HP and LCMV-E algorithms in Sec.~\ref{sec:results:sea-of-neurons:HD-tDCS}. The values of $I_{safe}$ and $I_{tot}^{mul}$ were $250$mA and $2$, respectively. The numbers in electrode montage denote the current injected at that electrode (in mA).}
    \label{fig:appx:neuron:visualisation:hd-tdcs}
\end{figure*}

\begin{figure*}[ht]
    \centering
   \includegraphics[width=\textwidth]{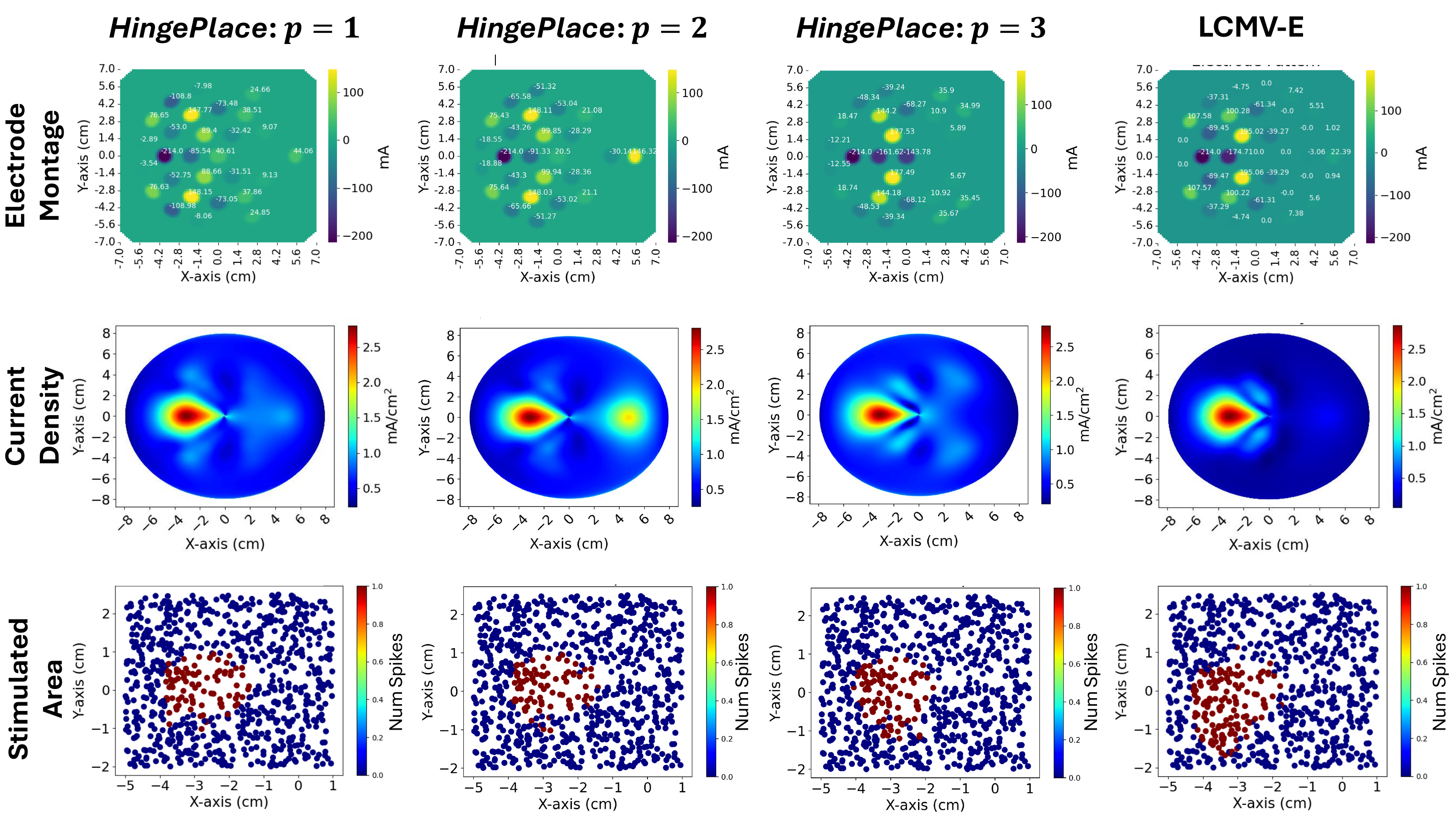}
    \caption{A representative plot of the electrode montage, the current density magnitude at the depth of $1$mm from the cortical surface (representing Layer 5), and the locations where the neurons fired in the Layer 5 of our sea of neurons model (denoted by the red dots) for the \HP and LCMV-E algorithms in Sec.~\ref{sec:results:sea-of-neurons:neuron-type}. The values of $I_{safe}$ and $I_{tot}^{mul}$ were $214$mA and $4$, respectively. The numbers in electrode montage denote the current injected at that electrode (in mA).}
    \label{fig:appx:neuron:visualisation:neuron-type}
\end{figure*}

Fig.~\ref{fig:appx:neuron:visualisation:c3-c4}, Fig.~\ref{fig:appx:neuron:visualisation:hd-tdcs}, and Fig.~\ref{fig:appx:neuron:visualisation:neuron-type} provide additional visualization of the electric field, electrode montage and the corresponding stimulated regions generated in the studies discussed in Sec.~\ref{sec:results:sea-of-neurons:c3-c4}, Sec.~\ref{sec:results:sea-of-neurons:HD-tDCS} and~\ref{sec:results:sea-of-neurons:neuron-type}, respectively.


\begin{thebibliography}{00}

\bibitem{kalu2012transcranial}
UG~Kalu, CE~Sexton, CK~Loo, and KP~Ebmeier.
\newblock Transcranial direct current stimulation in the treatment of major depression: a meta-analysis.
\newblock {\em Psychological medicine}, 42(9):1791, 2012.

\bibitem{mori2010effects}
Francesco Mori, Claudia Codec{\`a}, Hajime Kusayanagi, Fabrizia Monteleone, Fabio Buttari, Stefania Fiore, Giorgio Bernardi, Giacomo Koch, and Diego Centonze.
\newblock Effects of anodal transcranial direct current stimulation on chronic neuropathic pain in patients with multiple sclerosis.
\newblock {\em The Journal of Pain}, 11(5):436--442, 2010.

\bibitem{yook2011suppression}
Soon-Won Yook, Sung-Hee Park, Jeong-Hwan Seo, Sun-Jun Kim, and Myoung-Hwan Ko.
\newblock Suppression of seizure by cathodal transcranial direct current stimulation in an epileptic patient-a case report.
\newblock {\em Annals of rehabilitation medicine}, 35(4):579--582, 2011.

\bibitem{brunoni2013transcranial}
Andre~R Brunoni, Felipe Fregni, Alberto Priori, Roberta Ferrucci, and Paulo~S{\'e}rgio Boggio.
\newblock Transcranial direct current stimulation: challenges, opportunities, and impact on psychiatry and neurorehabilitation.
\newblock {\em Frontiers in psychiatry}, 4:19, 2013.

\bibitem{parkin2015non}
Beth~L Parkin, Hamed Ekhtiari, and Vincent~F Walsh.
\newblock Non-invasive human brain stimulation in cognitive neuroscience: a primer.
\newblock {\em Neuron}, 87(5):932--945, 2015.

\bibitem{kuo2013comparing}
Hsiao-I Kuo, Marom Bikson, Abhishek Datta, Preet Minhas, Walter Paulus, Min-Fang Kuo, and Michael~A Nitsche.
\newblock Comparing cortical plasticity induced by conventional and high-definition 4$\times$ 1 ring tdcs: a neurophysiological study.
\newblock {\em Brain stimulation}, 6(4):644--648, 2013.

\bibitem{fischer2017multifocal}
David~B Fischer, Peter~J Fried, Giulio Ruffini, Oscar Ripolles, Ricardo Salvador, Jaume Banus, William~Tyler Ketchabaw, Emiliano Santarnecchi, Alvaro Pascual-Leone, and Michael~D Fox.
\newblock Multifocal tdcs targeting the resting state motor network increases cortical excitability beyond traditional tdcs targeting unilateral motor cortex.
\newblock {\em Neuroimage}, 157:34--44, 2017.

\bibitem{dmochowski2011optimized}
Jacek~P. Dmochowski, Abhishek Dutta, Marom Bikson, Yuzhuo Su, and Lucas~C. Parra.
\newblock Optimized multi-electrode stimulation increases focality and intensity at target.
\newblock {\em Journal of Neural Engineering}, 8(4):046011, 2011.

\bibitem{horvath2014transcranial}
Jared~C Horvath, Olivia Carter, and Jason~D Forte.
\newblock Transcranial direct current stimulation: five important issues we aren't discussing (but probably should be).
\newblock {\em Frontiers in systems neuroscience}, 8:2, 2014.

\bibitem{horvath2015evidence}
Jared~Cooney Horvath, Jason~D Forte, and Olivia Carter.
\newblock Evidence that transcranial direct current stimulation (tdcs) generates little-to-no reliable neurophysiologic effect beyond mep amplitude modulation in healthy human subjects: a systematic review.
\newblock {\em Neuropsychologia}, 66:213--236, 2015.

\bibitem{antal2015conceptual}
A~Antal, D~Keeser, A~Priori, F~Padberg, and MA~Nitsche.
\newblock Conceptual and procedural shortcomings of the systematic review “evidence that transcranial direct current stimulation (tdcs) generates little-to-no reliable neurophysiologic effect beyond mep amplitude modulation in healthy human subjects: a systematic review” by horvath and co-workers.
\newblock {\em Brain Stimulation: Basic, Translational, and Clinical Research in Neuromodulation}, 8(4):846--849, 2015.

\bibitem{laakso2019can}
Ilkka Laakso, Marko Mikkonen, Soichiro Koyama, Akimasa Hirata, and Satoshi Tanaka.
\newblock Can electric fields explain inter-individual variability in transcranial direct current stimulation of the motor cortex?
\newblock {\em Scientific reports}, 9(1):626, 2019.

\bibitem{fernandez2020unification}
Mariano Fernández-Corazza, Sergei Turovets, and Carlos~Horacio Muravchik.
\newblock Unification of optimal targeting methods in transcranial electrical stimulation.
\newblock {\em NeuroImage}, 209:116403, 2020.

\bibitem{im2008determination}
Chang-Hwan Im, Hui-Hun Jung, Jung-Do Choi, Soo~Yeol Lee, and Ki-Young Jung.
\newblock Determination of optimal electrode positions for transcranial direct current stimulation ({tDCS}).
\newblock {\em Physics in Medicine \& Biology}, 53(11):N219, 2008.

\bibitem{park2011novel}
Ji-Hye Park, Seung~Bong Hong, Do-Won Kim, Minah Suh, and Chang-Hwan Im.
\newblock A novel array-type transcranial direct current stimulation (tdcs) system for accurate focusing on targeted brain areas.
\newblock {\em IEEE Transactions on Magnetics}, 47(5):882--885, 2011.

\bibitem{sadleir2012target}
Rosalind Sadleir, Tracy~D. Vannorsdall, David~J. Schretlen, and Barry Gordon.
\newblock Target optimization in transcranial direct current stimulation.
\newblock {\em Frontiers in Psychiatry}, 3, 2012.

\bibitem{dutta2013using}
Arindam Dutta and Anirban Dutta.
\newblock Using electromagnetic reciprocity and magnetic resonance current density imaging to fit multi-electrode montage for non-invasive brain stimulation.
\newblock In {\em 2013 6th International IEEE/EMBS Conference on Neural Engineering (NER)}, pages 447--451. IEEE, 2013.

\bibitem{ruffini2014optimization}
Giulio Ruffini, Michael~D. Fox, Oscar Ripolles, Pedro Cavaleiro~Miranda, and Alvaro Pascual-Leone.
\newblock Optimization of multifocal transcranial current stimulation for weighted cortical pattern targeting from realistic modeling of electric fields.
\newblock {\em NeuroImage}, 89:216--225, 2014.

\bibitem{guler2016optimization}
Seyhmus Guler, Moritz Dannhauer, Burak Erem, Rob Macleod, Don Tucker, Sergei Turovets, Phan Luu, Deniz Erdogmus, and Dana~H. Brooks.
\newblock Optimization of focality and direction in dense electrode array transcranial direct current stimulation {(tDCS)}.
\newblock {\em Journal of Neural Engineering}, 13(3):036020, May 2016.

\bibitem{salman2016concurrency}
Adnan Salman, Allen Malony, Sergei Turovets, Vasily Volkov, David Ozog, and Don Tucker.
\newblock Concurrency in electrical neuroinformatics: parallel computation for studying the volume conduction of brain electrical fields in human head tissues.
\newblock {\em Concurrency and Computation: Practice and Experience}, 28(7):2213--2236, 2016.

\bibitem{cancelli2016simple}
Andrea Cancelli, Carlo Cottone, Franca Tecchio, Dennis~Q Truong, Jacek Dmochowski, and Marom Bikson.
\newblock A simple method for {EEG} guided transcranial electrical stimulation without models.
\newblock {\em Journal of neural engineering}, 13(3):036022, 2016.

\bibitem{wagner2016optimization}
Sven Wagner, Martin Burger, and Carsten~H. Wolters.
\newblock An optimization approach for well-targeted transcranial direct current stimulation.
\newblock {\em SIAM Journal on Applied Mathematics}, 76(6):2154--2174, 2016.

\bibitem{fernandez2016transcranial}
Mariano Fernández-Corazza, Sergei Turovets, Phan Luu, Erik Anderson, and Don Tucker.
\newblock Transcranial electrical neuromodulation based on the reciprocity principle.
\newblock {\em Frontiers in Psychiatry}, 7, 2016.

\bibitem{dmochowski2017optimal}
Jacek~P Dmochowski, Laurent Koessler, Anthony~M Norcia, Marom Bikson, and Lucas~C Parra.
\newblock Optimal use of eeg recordings to target active brain areas with transcranial electrical stimulation.
\newblock {\em Neuroimage}, 157:69--80, 2017.

\bibitem{guler2018computationally}
Seyhmus Guler, Moritz Dannhauer, Biel Roig-Solvas, Alexis Gkogkidis, Rob Macleod, Tonio Ball, Jeffrey~G Ojemann, and Dana~H Brooks.
\newblock Computationally optimized {ECoG} stimulation with local safety constraints.
\newblock {\em Neuroimage}, 173:35--48, 2018.

\bibitem{saturnino2019accessibility}
Guilherme~Bicalho Saturnino, Hartwig~Roman Siebner, Axel Thielscher, and Kristoffer~Hougaard Madsen.
\newblock Accessibility of cortical regions to focal tes: Dependence on spatial position, safety, and practical constraints.
\newblock {\em NeuroImage}, 203:116183, 2019.

\bibitem{khan2022individually}
Asad Khan, Marios Antonakakis, Nikolas Vogenauer, Jens Haueisen, and Carsten~H. Wolters.
\newblock Individually optimized multi-channel {tDCS} for targeting somatosensory cortex.
\newblock {\em Clinical Neurophysiology}, 134:9--26, 2022.

\bibitem{prieto2022l1}
Fernando {Galaz Prieto}, Atena Rezaei, Maryam Samavaki, and Sampsa Pursiainen.
\newblock L1-norm vs. l2-norm fitting in optimizing focal multi-channel {tES} stimulation: linear and semidefinite programming vs. weighted least squares.
\newblock {\em Computer Methods and Programs in Biomedicine}, 226:107084, 2022.

\bibitem{wang2023multi}
Mo~Wang, Kexin Lou, Zeming Liu, Pengfei Wei, and Quanying Liu.
\newblock Multi-objective optimization via evolutionary algorithm (movea) for high-definition transcranial electrical stimulation of the human brain.
\newblock {\em NeuroImage}, 280:120331, 2023.

\bibitem{kandel2000principles}
Eric~R Kandel, James~H Schwartz, Thomas~M Jessell, Steven Siegelbaum, A~James Hudspeth, and Sarah Mack.
\newblock {\em Principles of neural science}, volume~4.
\newblock McGraw-hill New York, 2000.

\bibitem{hsu2023robust}
Gavin Hsu, A~Duke Shereen, Leonardo~G Cohen, and Lucas~C Parra.
\newblock Robust enhancement of motor sequence learning with 4 ma transcranial electric stimulation.
\newblock {\em Brain stimulation}, 16(1):56--67, 2023.

\bibitem{huang2019realistic}
Yu~Huang, Abhishek Datta, Marom Bikson, and Lucas~C Parra.
\newblock Realistic volumetric-approach to simulate transcranial electric stimulation—roast—a fully automated open-source pipeline.
\newblock {\em Journal of neural engineering}, 16(5):056006, 2019.

\bibitem{Caldas-Martinez2024}
Sara Caldas-Martinez, Chaitanya Goswami, Mats Forssell, Jiaming Cao, Alison~L. Barth, and Pulkit Grover.
\newblock Cell-specific effects of temporal interference stimulation on cortical function.
\newblock {\em Communications Biology}, 7(1):1076, Sep 2024.

\bibitem{griffiths2005introduction}
David~J Griffiths.
\newblock Introduction to electrodynamics, 2005.

\bibitem{goswami2021hingeplace}
Chaitanya Goswami and Pulkit Grover.
\newblock Hingeplace: Focused transcranial electrical current stimulation that allows subthreshold fields outside the stimulation target.
\newblock In {\em 2021 43rd Annual International Conference of the IEEE Engineering in Medicine \& Biology Society (EMBC)}, pages 1577--1583, 2021.

\bibitem{gomez2024perspectives}
Jose Gomez-Tames and Mariano Fern{á}ndez-Corazza.
\newblock Perspectives on optimized transcranial electrical stimulation based on spatial electric field modeling in humans.
\newblock {\em Journal of Clinical Medicine}, 13(11), 2024.

\bibitem{aberra2018biophysically}
Aman~S Aberra, Angel~V Peterchev, and Warren~M Grill.
\newblock Biophysically realistic neuron models for simulation of cortical stimulation.
\newblock {\em Journal of neural engineering}, 15(6):066023, 2018.

\bibitem{rattay1986analysis}
Frank Rattay.
\newblock Analysis of models for external stimulation of axons.
\newblock {\em IEEE Transactions on Biomedical Engineering}, BME-33(10):974--977, 1986.

\bibitem{rattay1999basic}
Frank Rattay.
\newblock The basic mechanism for the electrical stimulation of the nervous system.
\newblock {\em Neuroscience}, 89(2):335--346, 1999.

\bibitem{fellner2022finite}
Andreas Fellner, Amirreza Heshmat, Paul Werginz, and Frank Rattay.
\newblock A finite element method framework to model extracellular neural stimulation.
\newblock {\em Journal of Neural Engineering}, 19(2):022001, 2022.

\bibitem{Hermann2020}
Bertrand Hermann, Federico Raimondo, Lukas Hirsch, Yu~Huang, M{\'e}lanie Denis-Valente, Pauline P{\'e}rez, Denis Engemann, Fr{\'e}d{\'e}ric Faugeras, Nicolas Weiss, Sophie Demeret, Benjamin Rohaut, Lucas~C. Parra, Jacobo~D. Sitt, and Lionel Naccache.
\newblock Combined behavioral and electrophysiological evidence for a direct cortical effect of prefrontal tdcs on disorders of consciousness.
\newblock {\em Scientific Reports}, 10(1):4323, Mar 2020.

\bibitem{farahani2024transcranial}
Forouzan Farahani, Niranjan Khadka, Lucas~C Parra, Marom Bikson, and Mih{\'a}ly V{\"o}r{\"o}slakos.
\newblock Transcranial electric stimulation modulates firing rate at clinically relevant intensities.
\newblock {\em Brain Stimulation}, 17(3):561--571, 2024.

\bibitem{forssellner2021}
Mats Forssell, Vishal Jain, Chaitanya Goswami, Sara Caldas-Martinez, Pulkit Grover, and Maysam Chamanzar.
\newblock Effect of focality of transcranial currents on neural responses.
\newblock In {\em 2021 10th International IEEE/EMBS Conference on Neural Engineering (NER)}, pages 289--292, 2021.

\bibitem{ayache2012stroke}
Samar~S Ayache, Wassim~H Farhat, Hela~G Zouari, Hassan Hosseini, Veit Mylius, and Jean-Pascal Lefaucheur.
\newblock Stroke rehabilitation using noninvasive cortical stimulation: motor deficit.
\newblock {\em Expert review of neurotherapeutics}, 12(8):949--972, 2012.

\bibitem{sun2022repetitive}
Pingping Sun, Lei Fang, Jianzhong Zhang, Yang Liu, Guodong Wang, and Rui Qi.
\newblock Repetitive transcranial magnetic stimulation for patients with fibromyalgia: a systematic review with meta-analysis.
\newblock {\em Pain Medicine}, 23(3):499--514, 2022.

\bibitem{szelenyi2007transcranial}
Andrea Szel{\'e}nyi, Karl~F Kothbauer, and Vedran Deletis.
\newblock Transcranial electric stimulation for intraoperative motor evoked potential monitoring: stimulation parameters and electrode montages.
\newblock {\em Clinical neurophysiology}, 118(7):1586--1595, 2007.

\bibitem{alam2016spatial}
Mahtab Alam, Dennis~Q Truong, Niranjan Khadka, and Marom Bikson.
\newblock Spatial and polarity precision of concentric high-definition transcranial direct current stimulation (hd-tdcs).
\newblock {\em Physics in Medicine \& Biology}, 61(12):4506, 2016.

\bibitem{miranda2014predicting}
Pedro~C Miranda, Abeye Mekonnen, Ricardo Salvador, and Peter~J Basser.
\newblock Predicting the electric field distribution in the brain for the treatment of glioblastoma.
\newblock {\em Physics in Medicine \& Biology}, 59(15):4137, 2014.

\bibitem{boyd2004convex}
Stephen Boyd, Stephen~P Boyd, and Lieven Vandenberghe.
\newblock {\em Convex optimization}.
\newblock Cambridge university press, 2004.

\bibitem{gorski2007biconvex}
Jochen Gorski, Frank Pfeuffer, and Kathrin Klamroth.
\newblock Biconvex sets and optimization with biconvex functions: a survey and extensions.
\newblock {\em Mathematical methods of operations research}, 66(3):373--407, 2007.

\bibitem{hiriart1996convex}
Jean-Baptiste Hiriart-Urruty and Claude Lemar{\'e}chal.
\newblock {\em Convex analysis and minimization algorithms I: Fundamentals}, volume 305.
\newblock Springer science \& business media, 1996.

\bibitem{bertsekas2009convex}
Dimitri Bertsekas.
\newblock {\em Convex optimization theory}, volume~1.
\newblock Athena Scientific, 2009.

\bibitem{oostenveld2001five}
Robert Oostenveld and Peter Praamstra.
\newblock The five percent electrode system for high-resolution eeg and erp measurements.
\newblock {\em Clinical neurophysiology}, 112(4):713--719, 2001.

\bibitem{forssell2021effect}
Mats Forssell, Chaitanya Goswami, Ashwati Krishnan, Maysamreza Chamanzar, and Pulkit Grover.
\newblock Effect of skull thickness and conductivity on current propagation for noninvasively injected currents.
\newblock {\em Journal of Neural Engineering}, 2021.

\bibitem{markram2015reconstruction}
Henry Markram, Eilif Muller, Srikanth Ramaswamy, Michael~W Reimann, Marwan Abdellah, Carlos~Aguado Sanchez, Anastasia Ailamaki, Lidia Alonso-Nanclares, Nicolas Antille, Selim Arsever, et~al.
\newblock Reconstruction and simulation of neocortical microcircuitry.
\newblock {\em Cell}, 163(2):456--492, 2015.

\bibitem{carnevale2006neuron}
Nicholas~T Carnevale and Michael~L Hines.
\newblock The neuron book, 2006.

\bibitem{hines2009neuron}
Michael Hines, Andrew~P Davison, and Eilif Muller.
\newblock Neuron and python.
\newblock {\em Frontiers in neuroinformatics}, 3:391, 2009.

\bibitem{merton1980stimulation}
PA~Merton and HB~Morton.
\newblock Stimulation of the cerebral cortex in the intact human subject.
\newblock {\em Nature}, 285(5762):227--227, 1980.

\bibitem{diamond2016cvxpy}
Steven Diamond and Stephen Boyd.
\newblock {CVXPY}: {A} {P}ython-embedded modeling language for convex optimization.
\newblock {\em Journal of Machine Learning Research}, 17(83):1--5, 2016.

\bibitem{agrawal2018rewriting}
Akshay Agrawal, Robin Verschueren, Steven Diamond, and Stephen Boyd.
\newblock A rewriting system for convex optimization problems.
\newblock {\em Journal of Control and Decision}, 5(1):42--60, 2018.

\bibitem{10.5555/1593511}
Guido Van~Rossum and Fred~L. Drake.
\newblock Python 3 reference manual, 2009.

\bibitem{costa2021further}
Luciano da~F Costa.
\newblock Further generalizations of the jaccard index.
\newblock {\em arXiv preprint arXiv:2110.09619}, 2021.

\end{thebibliography}
\end{document}